\documentclass[aps, amsfonts,nofootinbib, showpacs, byrevtex,
preprintnumbers, showkeys, preprint]{revtex4}  
\usepackage{bm}
\usepackage{bbm}  
\usepackage{amsmath}
\usepackage{longtable}

\newcommand{\tr}{\mbox{tr}}
\newcommand{\Tr}{\mbox{Tr}}
\newcommand{\unit}{\mathbbm{1}}
\newcommand{\PP}{\mathsf{P}}
\newcommand{\BB}{\mathsf{B}}
\newcommand{\BBd}{\mathsf{B}^{\dagger}}

\def\be{\begin{equation}}
\def\ee{\end{equation}}
\def\bea{\begin{eqnarray}}
\def\eea{\end{eqnarray}}

\renewcommand{\vec}{\mathbf}
\begin{document}
\title{Generalized Random Phase Approximation and Gauge
Theories}
\author{Oliver Schr\"oder}
\email[Email:]{schroedo@lns.mit.edu}
\affiliation{Center for Theoretical Physics, Laboratory for Nuclear Science
  and Department of Physics, Massachusetts Institute of Technology, Cambridge,
Massachusetts 02139}
\author{Hugo Reinhardt}
\email[Email:]{hugo.reinhardt@uni-tuebingen.de}
\affiliation{Institut f\"ur Theoretische Physik, Eberhard-Karls Universit\"at
  zu T\"ubingen, D-72076 T\"ubingen, Federal Republic of Germany} 
\preprint{MIT-CTP-3355, UNITU-THEP-06/2003}
\begin{abstract}
Mean-field treatments of Yang-Mills theory face the problem of how
to treat the Gauss law constraint. In this paper we try to face this
problem by studying the excited states instead of the ground
state. For this purpose we extend the operator approach to the Random 
Phase Approximation (RPA) well-known from nuclear physics and recently also
employed in pion physics to general bosonic theories with a standard
kinetic term. We focus especially on conservation laws, and how they
are translated from the full to the approximated theories, demonstrate
that the operator approach has the same spectrum as the RPA derived
from the time-dependent variational principle, and give - for
Yang-Mills theory - a discussion of the moment of inertia connected
to the energy contribution of the zero modes to the RPA ground state
energy. We also indicate a line of thought that might be useful to
improve the results of the Random Phase Approximation.
\end{abstract}
\maketitle
\newpage
%
%
%
\section{Introduction and Motivation}
Recently, variational calculations based on Gaussian wave functionals
have stirred some interest when applied to Yang-Mills theory
\cite{Kerman:1989kb, Kogan:1995wf, Diakonov:1998ir, Zarembo:1998xq,
Heinemann:2000ja, Schroeder:2002b, completeref}. A
problem constantly encountered is the fact that Gaussian wave
functionals are not gauge invariant, i.e. they are not annihilated by
the Gauss law operator (except for the case of
electrodynamics). Therefore one faces the situation, that the
mean-field treatment (to which the Gaussian wave functionals
correspond) does not respect a symmetry of the Hamiltonian. This is a
situation commonly encountered in nuclear physics where one often even
wants to break as many symmetries as possible in order to store a
maximum amount of correlations in a wave function of a very simple
form. We therefore look to nuclear physics for a possible remedy of
this problem. \\
One possibility is the introduction of a projector which
projects the Gaussian wave functional onto the subspace of gauge
invariant functionals \cite{Kerman:1989kb, Kogan:1995wf,
Diakonov:1998ir, Zarembo:1998xq, Heinemann:2000ja,
Schroeder:2002b}. Another possibility is the \textit{Random Phase 
Approximation}. In this framework one studies not the ground state but
excited states, and one finds that - if the mean-field ground state
has a broken symmetry - spurious excitations exist. Under certain
conditions these spurious excitations decouple, however, and we have
the pleasant situation that our excitations are as good as they would
be if we had started from a gauge invariant mean-field state. \\
The computation of the excited states of Yang-Mills theory
is especially interesting for a variety of reasons: \textit{First}, one has an
alternative point of view from which to shed on light onto the
confinement problem, compared to the usual approaches that try to
compute the linear potential between static quarks. \textit{Second}, one has
quite accurate lattice data for some of the low lying glueballs for
both two and three spatial dimensions, e.g. \cite{Teper:1998te,
  Teper:1998kw}. This allows to judge the 
approximations that usually occur in the process of analytical
calculations. On the other hand, there is quite an amount of
information that is very hard to obtain from the lattice, like masses
of \textit{excited} glueballs, and also regularities and patterns in the
glueball spectrum which can only be observed but not explained within
the lattice framework. \textit{Third}, in recent years the glueball
spectrum has also become a matter of interest for theories that claim
to have a very non-trivial connection to Yang-Mills theory like
e.g. supergravity theories \cite{Csaki:1998qr, deMelloKoch:1998qs,
Minahan:1998tm}. \\ 
There are basically two approaches to the generalized Random Phase
Approximation (\textit{gRPA}). The first one starts from the
\textit{time-dependent variational principle} of Dirac
\cite{Kerman:1976yn, Kerman:1995uj, Kerman:1998vt}. The variational
parameters are decomposed into one 
part that solves the static equations of motion and one ``small''
fluctuating part. In this context, the nature of gRPA as a \textit{harmonic
approximation} becomes clear; the general procedure is outlined in
app. \ref{app_RPA_from_tdvp}, since the equivalence to the second approach
to gRPA has up to now only been shown for the fermionic case
\cite{Kerman:1976yn}. The second approach is based on the formalism of
creation and annihilation operators, well-known in nuclear physics
\cite{Ring:1980}, and has recently been applied also in the context of pion
physics \cite{Aouissat:1996va, Aouissat:1997be, Aouissat:1998nu}. In
contrast to these latter investigations, we consider here \textit{generic}
bosonic field theories; they are only restricted by the requirements
that the kinetic energy shall have a standard form, and the remainder
of the Hamiltonian shall be expressible as a polynomial in the field
operators. \\
The structure of the paper is as follows: We begin by introducing (in
sec. \ref{sec_Hamiltonian_formalism}) the Hamiltonian formulation of
generic bosonic theories which will be the subject of investigation in
the main part of this paper.
In sec. \ref{sec_manybodylanguage} we rewrite the usual canonical
formulation in terms of creation and annihilation operators thereby
bridging the gap between canonical and many-body treatment. Especially,
we will give an expression of the Hamiltonian for a generic bosonic
system (with a standard kinetic energy term) in terms of
creation/\-annihilation operators. This is followed by an alternative
form of the Schr\"odinger equation useful in the context of gRPA. We
then introduce the two crucial 
approximations needed  to obtain the gRPA equations from the
Schr\"odinger equation. The second of these approximations can - under
certain conditions - be rephrased in terms of the so-called
quasi-boson approximation which will allow for a simpler formulation. We
then discuss the normal mode form of the Hamiltonian, and see what
kind of difficulties appear if the gRPA equations have zero mode solutions.
In sec. \ref{sec_Conserv_laws} we will discuss how conservation laws
translate from the full theory to the theory approximated by gRPA,
when zero mode solutions are implied, and also the connection to symmetry
breaking in the mean-field treatment is indicated. We will also draw
attention to the fact that the character of the symmetry under
consideration can change from a non-Abelian to an Abelian symmetry. 
In sec. \ref{sec_on_PQ_formalism} we demonstrate the equivalence of the gRPA
formulation based on the time-dependent variational principle to the operator
approach employed in this paper; this equivalence has so far been
demonstrated only in the case of fermionic systems \cite{Kerman:1976yn}. In
sec. \ref{sec_mom_of_in} we discuss further how the normal mode form of the
Hamiltonian is altered if the gRPA equations have zero mode solutions,
and the special role that is played by the moment of inertia. We then
restrict the further discussion from generic bosonic theories to $SU(N)$
Yang-Mills theory, where we discuss both the moment of inertia in
general, and also give an explicit computation of the leading terms in
a perturbative expansion. We end the section with a discussion of the
possibility of 
interpreting the moment of inertia as the static quark potential.
In sec. \ref{sec_summary_and_conclusion} we give a critical evaluation
of the generalized Random Phase Approximation and give an outlook to
further applications. At the end of the paper a number of appendices
is given. In app. \ref{app_RPA_from_tdvp} we give a short account of
the gRPA as derived from the time-dependent variational principle in
so far as it is needed for the purposes of this paper. In
app. \ref{app_YMT} we discuss 
shortly Yang-Mills theory in Weyl gauge, and how it fits into the
general framework of bosonic theories as discussed in the main body of
the text. 
In app. \ref{app_explicit_expressions} a number of explicit
expressions and computations can be found, in app. \ref{gRPA_comm_rel}
the proof for bosonic commutation relations among gRPA excitation
operators is given, and in app. \ref{app_pot} the proof of a theorem used in
sec. \ref{sec_manybodylanguage} can be found. \\
The important issue of renormalization will not be addressed in this paper. 
This topic in the context of the gRPA is subject to further investigation. 
In the approach to the gRPA that is based on the time-dependent variational
principle a discussion of renormalization issues for the $\phi^4$ theory has
been given in \cite{Kerman:1995uj, Kerman:1998vt}.
\section{Hamiltonian Formalism \label{sec_Hamiltonian_formalism}} 
In this section the Hamiltonian formalism for fairly general bosonic
theories is introduced. Since there are good references to the
subject, e.g. \cite{Jackiw:1988sf, Yee:1992jc}, we will be very brief
here. \\ 
The basic variables in the Hamiltonian approach consist of the the
\textit{field operators} $\phi_i$ and the \textit{canonical momenta}
$\pi_i$. In order to allow for sufficient generality we will use
super-indices which can - besides the position coordinate $\vec{x}$ -
also contain spatial indices or internal (e.g. color) indices. The Einstein
summation convention is adopted, implying sums over all discrete and
integrals over all continuous variables.
These basic variables satisfy the basic commutation relations
\be
[ \phi_i , \pi_j] = i \delta_{ij}.
\ee
We will work in the Schr\"odinger representation, i.e. we will not
work with abstract states but with a field representation of the
states (analogous to the position representation in quantum
mechanics); the objects to be considered are thus wave functionals
$\psi[\phi]$, given by
\be
\psi[\phi] = \langle \phi | \psi \rangle,
\ee
for the state $|\psi\rangle$. The states $| \phi \rangle$ are
eigenstates of the field operators $\phi_i$.
The usage of $| \phi \rangle$ as basis states implies that we work in
the position representation, i.e. in the remainder of the paper we
will realize $\phi$ multiplicatively and $\pi$ as a
derivative operator:
\be
\langle \phi | \phi_i | \psi \rangle = \phi_i \psi[\phi] \text{
and } \langle \phi | \pi_i | \psi \rangle = \frac{1}{i}
\frac{\delta}{\delta \phi_i} \psi[\phi],
\ee
or more pragmatically we will read in all formulas where 
the field operator $\phi$ appears this only as a multiplicative $\phi$
and $\pi$ as $\delta/(i \delta \phi)$. \\ 
Since we work in the Schr\"odinger picture the operators considered
are all time-independent. The states, however, satisfy the
Schr\"odinger equation
\be
i \partial_t | \psi(t) \rangle = H | \psi(t) \rangle,
\ee
where $H$ is the Hamiltonian of the system. In the main section
we will be mostly interested in stationary states, i.e. states that
can be written as $|\psi(t) \rangle = e^{-i E t} | \psi(0)
\rangle$. These are then eigenstates of the Hamiltonian
\be
H | \psi \rangle = E | \psi \rangle.
\ee
\section{Formulation of Many-body Language for Generic Bosonic
Theories \label{sec_manybodylanguage}}
\subsection{Creation and Annihilation Operators
\label{sec_on_creat_annihil_op}}
In this section we will apply and generalize the operator approach
used in nuclear physics \cite{Ring:1980, MW69, Thouless:1960a,
Thouless:1960b, Brown:1961, Thouless:1962, MW70, Soloviev:1976,
Munchow:1985gt, Nesbet:1964} and recently also in pion physics
\cite{Aouissat:1996va, Aouissat:1997be, Aouissat:1998nu}. We draw the
connection between the canonical treatment and the many-body language
as was already briefly indicated in \cite{Kerman:1997xx}. 
Since for our purposes it is sufficient, and in order to allow 
close comparison to gRPA as obtained from the time-dependent
variational principle, cf. app. \ref{app_RPA_from_tdvp}, we will consider
Hamiltonians of the form
\be
H = \frac{1}{2}\pi_i^2 + V[\phi]. \label{gRPAHamiltonian}
\ee 
$V[\phi]$ is a functional of the field operator, in the following
referred to as 'potential'. \\
Since the basis of this approach is the {\it stationary} Schr\"odinger
equation, we start therefore from the most general {\it
time-independent} Gaussian state ($i,j$ are super-indices):
\be
\psi[\phi] = \mathcal{N} \exp{\left(-(\phi-\bar{\phi})_{i} (\frac{1}{4} G^{-1}
    - i \Sigma)_{ij}(\phi-\bar{\phi})_{j} + i \bar{\pi}_i
(\phi-\bar{\phi})_{i}\right)}, \label{RPA_most_general_TI_Gauss}
\ee
where $\mathcal{N}$ is a normalization constant.
A Gaussian state allows the explicit construction of  creation and annihilation
 operators as linear combinations of $\phi_i$ and $\pi_i$ that satisfy
 the basic relations
\be
a_i \psi[\phi] = 0 \ \text{and} \ [a_i, a^{\dagger}_j] = \delta_{ij},
\ee
where the latter relation fixes the normalization. Using
eq.\,(\ref{RPA_most_general_TI_Gauss}) as reference state to be
annihilated by $a_i$ one obtains as explicit expressions for
$a^{\dagger}, a$:
\bea
a_i^{\dagger} & = &  U_{ij}\left\{\left(\frac{1}{2} G^{-1}_{jk} + 2i
  \Sigma_{jk}\right)(\phi-\bar{\phi})_k - i
  \left(\pi-\bar{\pi}\right)_j  \right\}, \label{gRPAdefadagger} \\
a_i & = & U_{ij} \left\{\left(\frac{1}{2} G^{-1}_{jk} - 2i
  \Sigma_{jk}\right)(\phi-\bar{\phi})_k + i
  \left(\pi-\bar{\pi}\right)_j \right\},  \label{gRPAdefa} 
\eea
where $U$ is (implicitly) defined via the relation
\be
U_{ij} U_{jk} = G_{ik} \label{eq_def_of_U}
\ee
and could also be called the square root of G.
A short excursion on the existence and the implicit assumption of
reality of U is in order here: Since $G_{ik}$ is a real symmetric matrix
(matrix is used in the generalized sense s.t. also continuous
indices are allowed) one can always diagonalize it. Therefore one can
also always write down a $U$ as given above. However, $G$ has to
satisfy another condition, namely all of its eigenvalues have to be
strictly positive, since otherwise one will run into two kinds of
problems: If G has a zero eigenvalue $G^{-1}$ does not exist and the
matrix element of $\langle \pi^2 \rangle$ will be infinite. This is 
certainly undesirable, and will be assumed not to be the case in the
following. 
Moreover, if G has a negative eigenvalue, $\psi[\phi]$ is not even
normalizable, since in the direction of the negative eigenvalue $\psi[\phi]$
will increase exponentially for increasing values of $\phi$. Thus,
all eigenvalues of G must be strictly positive, therefore U can be chosen
to be real. \\
Since $a_i,a_i^{\dagger}$ are just given via linear combinations of
$\phi$ and $\pi$, one can invert these relations to obtain
$\phi,\pi$ 
in terms of $a, a^{\dagger}$ and the parameters of the Gaussian wave
functional we started with:
\bea
\phi_i  &=&  \bar{\phi}_{i} + U_{ij} \Big(a_{j} + a^{\dagger}_{j}\Big),
\label{gRPAdefphi} \\ 
\pi_{i} &=&  \bar{\pi}_{i} + 2i \Big\{ \Big(\frac{1}{4} G^{-1}_{ik} - i
    \Sigma_{ik} \Big) U_{kj} a^{\dagger}_j - 
    \Big(\frac{1}{4}     G^{-1}_{ik} + 
    i \Sigma_{ik} \Big) U_{kj} a_j \Big\}. \label{gRPAdefpi}
\eea
One should note that the dependence of the canonical operators on the
parameters  
of the wave functional we choose is only seeming, since the creation
and annihilation operators depend implicitly on these parameters as
well. If one inserts eqs.\,(\ref{gRPAdefadagger}), (\ref{gRPAdefa})  into
eqs.\,(\ref{gRPAdefphi}), (\ref{gRPAdefpi}) one obtains an identity
$\phi=\phi, \pi = \pi$. In a practical sense, however, we have
transferred information that is contained in the wave functional to the
operators, since in the following the only property of $\psi[\phi]$
that we will use for practical computations is that $a_i \psi[\phi]
= 0$. All parameter dependence that usually comes about by
calculating matrix elements now enters the formulas via normal
ordering. \\ Since we have now a representation of the 
canonical operators in terms of $a, a^{\dagger}$ (referred to in the following
as \textit{c/a representation}), {\it all} operators 
permissible in a canonical system can be expressed in terms of $a,
a^{\dagger}$, especially the Hamiltonian which for obvious reasons is
central to the following calculations. \\
\subsection{Form of Hamiltonian in the c/a Representation}
We have required the Hamiltonian to have a certain structure
[cf. eq.\,(\ref{gRPAHamiltonian})]. The Hamiltonian there resolves
naturally into a kinetic energy $T=\frac{1}{2} \pi^2$ and a potential term
$V[\phi]$  which is a functional of $\phi$ only.  It is
very useful to normal-order these expressions to make further
progress. In the course of this, one makes the useful observation that
one can write the kinetic energy as :
\bea
\frac{1}{2} \pi^2_i  & = & \quad \Bigg\{\frac{1}{2} \left(
\bar{\pi}_{i}^2 + \frac{1}{4} \Tr{(G^{-1})} + 4 \Tr{(\Sigma G
    \Sigma)} \right) \Bigg\} \nonumber \\ & &  +  \Bigg\{ \left(
    \frac{\delta}{\delta     \bar{\pi}_{k_1}} \langle  
\frac{1}{2} \pi^2_i \rangle \right)\left( \frac{i}{2} U^{-1}_{k_1 j_1}
  \left(a^{\dagger}_{j_1} - a_{j_1} \right) + 2 \Sigma_{k_1l_1} U_{l_1 j_1}
    \left(a^{\dagger}_{j_1} + a_{j_1}\right) \right) \Bigg\}
    \nonumber \\
  & & +  \Bigg\{ \left( \frac{\delta}{\delta G_{k_1 k_2}}  \langle \frac{1}{2}
    \pi^2_i \rangle \right) U_{k_1 j_1} U_{k_2 j_2} \left(
    a^{\dagger}_{j_1} a^{\dagger}_{j_2} + a_{j_1} a_{j_2} + 2
    a^{\dagger}_{j_1} a_{j_2} \right) \Bigg\} \nonumber \\
 & & +  \Bigg\{ \left(\frac{\delta}{\delta \Sigma_{k_1 k_2}}  \langle
    \frac{1}{2} \pi^2_i \rangle \right) U^{-1}_{k_1 j_1}
    U^{-1}_{k_2 j_2} \frac{i}{4} \left(a^{\dagger}_{j_1} a^{\dagger}_{j_2} -
    a_{j_1} a_{j_2} \right) \Bigg\}  \nonumber \\
 & & + \Bigg\{\left(\frac{1}{4} G^{-1}_{j_1 j_2} + \frac{i}{2} \left(U^{-1}
    \Sigma U - U \Sigma 
    U^{-1}\right)_{j_1 j_2} \right) 2 a^{\dagger}_{j_1} a_{j_2}\Bigg\}
    \hspace*{0.5em}, 
\label{kinetic_RPA}
\eea
where $\Tr$ is a trace over the super-indices.
A similar result can be found for the potential part (except that it
is independent of $\Sigma$ and $\bar{\pi}$). 
In app. \ref{app_pot} we will demonstrate that the
potential can be decomposed into c/a operators s.t. the prefactors can
be written as functional derivatives of the expectation value of the
potential between Gaussian states, and the c/a operators always
appear in a fixed structure\footnote{e.g. if in the potential there
are terms that 
contain two creation operators, or two annihilation operators, or one
creation/ one annihilation operator, they can always be written as
$\text{factor} \times (a^{\dagger}_i a^{\dagger}_j + a_i a_j + 2
a^{\dagger}_i a_j)$, and similarly for all other terms that contain a
fixed sum of creation and annihilation operators. For the kinetic
terms things are a bit different, but we have given the decomposition
of the only kinetic term allowed in eq.\,(\ref{kinetic_RPA}).}.
In the following we will restrict ourselves to the contributions to
$V$ with up to four c/a operators, since the terms containing higher
numbers of c/a operators do not contribute\footnote{That this is true
can be seen by usage of Wick's theorem and the so-called {\it second
gRPA approximation} $a|\rangle = 0$ that will be introduced in
sec.\,\ref{sec_ex_op}.} to the gRPA matrices that will be introduced in
eq.\,(\ref{defA-F}): 
\bea
V[\phi] & = & \quad  \langle V[\phi] \rangle \nonumber \\  
& & +  \Bigg(\frac{\delta}{\delta \bar{\phi}_{k_1}} \langle V[\phi]
\rangle \Bigg) 
U_{k_1 j_1} \bigg(a^{\dagger}_{j_1} + a_{j_1}\bigg)  \nonumber \\
 & & +  \Bigg( \frac{\delta}{\delta G_{k_1 k_2}}  \langle 
V[\phi] \rangle \Bigg)  U_{k_1 j_1} U_{k_2 j_2}
\bigg( 
a^{\dagger}_{j_1} a^{\dagger}_{j_2} + a_{j_1} a_{j_2} + 2 a^{\dagger}_{j_1}
a_{j_2} \bigg)  \nonumber  \\
& & +  \frac{1}{3} \Bigg( \frac{\delta}{\delta \bar{\phi}_{k_1}}
  \frac{\delta}{\delta G_{k_2 k_3}}  \langle  V[\phi] \rangle \Bigg) U_{k_1
  j_1} U_{k_2 j_2} U_{k_3 j_3} \nonumber \\ & &  
\quad \times \bigg(a^{\dagger}_{j_1} a^{\dagger}_{j_2} a^{\dagger}_{j_3} + 3
  a^{\dagger}_{j_1} a^{\dagger}_{j_2}  a_{j_3} 
+ 3 a^{\dagger}_{j_1}  a_{j_2} a_{j_3} +  a_{j_1}
  a_{j_2}  a_{j_3} \bigg) \label{potential_RPA}  \\ 
& & +  \frac{1}{6} \Bigg( \frac{\delta}{\delta G_{k_1 k_2}}
  \frac{\delta}{\delta G_{k_3 k_4}}  \langle V[\phi] \rangle \Bigg) U_{k_1
  j_1} U_{k_2 j_2} U_{k_3 j_3} U_{k_4   j_4}  \nonumber \\ & & \quad \times
\bigg( a^{\dagger}_{j_1} a^{\dagger}_{j_2} a^{\dagger}_{j_3} 
  a^{\dagger}_{j_4} + 4 a^{\dagger}_{j_1} a^{\dagger}_{j_2} a^{\dagger}_{j_3}
  a_{j_4} + 6 a^{\dagger}_{j_1}
  a^{\dagger}_{j_2} a_{j_3} a_{j_4}    
 +\, 4 a^{\dagger}_{j_1} a_{j_2}  a_{j_3} a_{j_4} + a_{j_1} a_{j_2} a_{j_3
}   a_{j_4}\bigg). \nonumber  
\eea
By adding the expressions eq.\,(\ref{kinetic_RPA},
\ref{potential_RPA}) together, one observes that the Hamiltonian has a
very simple schematic structure:
\bea
H & = & \quad \langle H \rangle \nonumber \\
& & +  \left(\frac{\delta}{\delta \bar{\phi}_{k_1}} \langle H
\rangle \right) 
U_{{k_1} j_1} \left(a^{\dagger}_{j_1} + a_{j_1}\right)
\nonumber \\
 & & +  \left(\frac{\delta}{\delta \bar{\pi}_{k_1}} \langle H \rangle
\right) \left( \frac{i}{2} U^{-1}_{{k_1} j_1} 
  \left(a^{\dagger}_{j_1} - a_{j_1} \right) + 2 \Sigma_{k_1 l} U_{l j_1}
  \left(a^{\dagger}_{j_1} 
    + a_{j_1}\right) \right) 
\nonumber \\
 & & +  \left( \frac{\delta}{\delta G_{k_1 k_2}}  \langle 
H \rangle \right)  U_{k_1 j_1} U_{k_2 j_2}
\left( 
a^{\dagger}_{j_1} a^{\dagger}_{j_2} + a_{j_1} a_{j_2} + 2 a^{\dagger}_{j_1}
a_{j_2} \right)  
\nonumber \\
& & +  \left(\frac{\delta}{\delta \Sigma_{k_1 k_2}}  \langle H \rangle
\right)   U^{-1}_{k_1 j_1} U^{-1}_{k_2 j_2} \frac{i}{4} \left(
a^{\dagger}_{j_1} a^{\dagger}_{j_2} - a_{j_1} a_{j_2} \right) 
\nonumber \\ 
& & +  \left( \frac{1}{4}G^{-1}_{j_1 j_2} + \frac{i}{2} \left(U^{-1}
    \Sigma U - U \Sigma 
    U^{-1}\right)_{j_1 j_2} \right) 2 a^{\dagger}_{j_1} a_{j_2}
\nonumber \\ 
& & +  \frac{1}{3} \Bigg( \frac{\delta}{\delta \bar{\phi}_{k_1}}
  \frac{\delta}{\delta G_{k_2 k_3}}  \langle  H \rangle \Bigg) U_{k_1
  j_1} 
U_{k_2 j_2} U_{k_3 j_3} \nonumber \\ && \quad \times
\bigg(a^{\dagger}_{j_1} a^{\dagger}_{j_2} a^{\dagger}_{j_3} + 3
  a^{\dagger}_{j_1} a^{\dagger}_{j_2}  a_{j_3} 
+ 3 a^{\dagger}_{j_1}  a_{j_2}   a_{j_3} +  a_{j_1}  a_{j_2}  a_{j_3}
\bigg) 
\nonumber \\ 
& & +  \frac{1}{6} \left( \frac{\delta}{\delta G_{k_1 k_2}}
  \frac{\delta}{\delta G_{k_3 k_4}}  \langle V \rangle
 \right) U_{k_1
  j_1} U_{k_2 j_2} U_{k_3 j_3} U_{k_4 j_4} \label{H_in_ca_rep}  \\ & & \quad
\times  
\bigg( a^{\dagger}_{j_1} 
  a^{\dagger}_{j_2} a^{\dagger}_{j_3} 
  a^{\dagger}_{j_4} + 4 a^{\dagger}_{j_1} a^{\dagger}_{j_2} a^{\dagger}_{j_3}
  a_{j_4}  
  + 6 a^{\dagger}_{j_1}   a^{\dagger}_{j_2} a_{j_3} a_{j_4}   
+ 4 a^{\dagger}_{j_1} a_{j_2}  a_{j_3} a_{j_4} + a_{j_1} a_{j_2} a_{j_3}
  a_{j_4} \bigg) \, . \nonumber
\eea
One should note that in all save the
last term, the derivatives are taken of $\langle H \rangle$
whereas in the last term the derivative is taken of $\langle V
\rangle$. The importance of this will become clear in
sec.\,\ref{pqformulationofRPA}.
Up to now, we have taken an arbitrary Gaussian as a
reference state: we have defined our creation/annihilation operators
relative to that state - nothing else. Eq.\,(\ref{H_in_ca_rep}) is 
exactly the same as $H$ given in eq.\,(\ref{gRPAHamiltonian}).
However, it is obvious from eq.\,(\ref{H_in_ca_rep}) that the
Hamiltonian greatly simplifies if we 
choose a specific reference state: a state that is a stationary point
of the energy functional $\langle H \rangle$ under variation of the parameters
$\bar{\phi}, \bar{\pi}, G, \Sigma$ - in other words, \textit{a state
that satisfies the Rayleigh-Ritz variational principle}\footnote{ 
One should note that we require the same in the
time-dependent approach, when we decompose the time-dependent
parameters $\bar{\phi}(t), \bar{\pi}(t), G(t), \Sigma(t)$ into a
static part and small fluctuations,
cf. app. \ref{app_RPA_from_tdvp}. The equations for the static part
are identical to the Rayleigh-Ritz equations.}. 
\subsection{Alternative Form of Schr\"odinger equation}
In the following we want to generalize the RPA known for many-body physics to
the quantum field theory described by the Hamiltonian of 
eq.\,(\ref{gRPAHamiltonian}). For this purpose we assume that the exact
(excited) state $| \nu \rangle$ can be created from the exact vacuum $| 0
\rangle$ by an operator $Q^{\dagger}_{\nu}$, i.e.
\be
| \nu \rangle = Q^{\dagger}_{\nu} | 0 \rangle.
\ee 
If $E_{\nu}$ denotes the corresponding eigenvalue of $H$ we have
\be
H Q^{\dagger}_{\nu} | 0 \rangle = E_{\nu} Q^{\dagger}_{\nu} |0 \rangle.
\ee
If we denote the vacuum energy by $E_0$, we can write this
equivalently with a commutator:
\be
\ [H, Q^{\dagger}_{\nu}] | 0 \rangle = (E_{\nu} - E_0)
Q^{\dagger}_{\nu} | 0 \rangle. \label{Schroedinger_comm}
\ee
We may multiply both sides of the equation with an arbitrary operator
$\delta Q$, and obtain an expectation value by multiplying from the
left with $\langle 0 |$:
\be
\langle 0 | \delta Q [H, Q^{\dagger}_{\nu}] | 0 \rangle = (E_{\nu} -
E_0) \langle 0 | \delta Q Q^{\dagger}_{\nu}  | 0 \rangle.
\ee
By subtracting zero on both sides we obtain an equation that only contains
expectation values of commutators: 
\be 
\langle 0 | [\delta Q,[H,Q^{\dagger}_{\nu}]] | 0 \rangle = (E_{\nu} -
E_0) \langle 0 | [\delta Q, Q^{\dagger}_{\nu}] | 0 \rangle.
\label{equationofmotion} 
\ee
Using this equation as a starting point to derive the (generalized) RPA
is known in nuclear physics as the {\it equations of motion method}
\cite{Rowe:1968,Ring:1980}. 
\subsection{First and Second gRPA Approximations \label{sec_ex_op}}
The generalized Random Phase Approximation consists now of
approximating eq.\,(\ref{equationofmotion}) in order to obtain a
solvable set of equations. Two obvious candidates for approximations
suggest themselves: \textit{first}, the excitation operator
$Q_{\nu}^{\dagger}$, \textit{second}, the vacuum state. In the
remainder of the paper, we will call the approximation concerning the
first topic \textit{the first gRPA approximation}, the approximation
concerning the vacuum state \textit{the second gRPA approximation}
(even though this may be doubtful linguistically). The second gRPA
approximation is actually easier to state, therefore we start with it:
\textit{approximate  in all expressions involving vacuum expectation values
of commutators the true vacuum state by the reference state of the 
creation/annihilation operators}, which in this context we will
usually refer to as \textit{mean-field vacuum} and which will be denoted by
$|\, \rangle$. Later on, we will be even more 
restrictive and require the reference state to really be a stationary
state of the Rayleigh-Ritz principle as was indicated before, but we
will state explicitly from when on this additional restriction will be
necessary.  As already noted, the \textit{first gRPA approximation}
 deals with the class
of allowed operators. In nuclear physics it is quite reasonable to
assume that the lowest excited state above a Hartree-Fock ground
state consists of a particle-hole excitation. This results in the
so-called Tamm-Dancoff approximation. In the generalized RPA one
 assumes that one has a correlated ground state, s.t. not only the
creation of a hole and a particle, but also their destruction is a possible
excitation. In Yang-Mills theory - the theory we will be
ultimately interested in - it is by far not so clear what
structure the lowest excitation will have; we have thus taken the two
following guiding principles (an argument similar to our second
principle can be found in \cite{Aouissat:1996va, Aouissat:1998nu}) 
\begin{enumerate}
\item one of the main differences between fermionic and bosonic
systems is that in the latter there exist single-particle condensates;
thus, one has at least to extend the \textit{ansatz} for the
excitation operator by linear terms allowing for fluctuations. 
Furthermore, since we can compare to the gRPA as derived from the
time-dependent variational principle, we will see that the ansatz for
$Q^{\dagger}_{\nu}$ to be proposed leads to equations of motion that are
identical to those derived from the time-dependent variational principle,
thereby verifying the ansatz to be the correct one. 
\item The principal goal of the Random Phase Approximation is the
restoration of symmetries that are violated at the mean-field level. 
We give here a short outline, drawing from concepts that will be
introduced further below, in order to motivate from this property of
symmetry restoration the form to be allowed for the excitation
operators. \\ The first point is that the second gRPA approximation can
be replaced (under certain conditions) by the so-called
\textit{quasi-boson approximation} (QBA). There, it will turn out that
gRPA equations can be written as 
\be
\ [H_B, Q^{\dagger}_{B \nu}] = (E_{\nu} - E_0) Q^{\dagger}_{B \nu}
\ee
where $\mathcal{O}_B$ indicates that the QBA has been used. 
One the other hand, for a symmetry generator (in our case the Gauss
law operator) $\Gamma$ which is a one-body operator (i.e. an operator
that can be written as a linear combination of the operators $a,
a^{\dagger}, a a, a^{\dagger} a^{\dagger}, a^{\dagger} a$), one can
derive that  
\be
\ [H, \Gamma] = 0 \to [H_B, \Gamma_B] = 0.
\ee
If the mean-field vacuum is annihilated by $\Gamma$ this is a trivial
statement, since in this case $\Gamma_B \equiv 0$. However, if the
symmetry is violated on the mean-field level, one obtains a
non-trivial result, namely that $\Gamma_B$ is a
solution of the gRPA equations with zero excitation energy
\textit{provided that the class of excitation operators contains}
$\Gamma_B$. Since all gRPA excitations are orthogonal this would be a
very desirable state of affairs: effectively the spurious
excitations caused by the symmetry violation on the mean-field level
would not affect the physical excitations we are interested in. 
\end{enumerate}
To put it all in a nut-shell: one looks at the Gauss law operator, determines
its structure (cf. app. \ref{app_Gauss_law}), and keeps the class
of excitation operators so large that the Gauss law operator belongs
to this class. All the concepts and claims will become clear and will be
proved in what follows. After this rather long motivation, we will
consider excitation operators of the following form:
\be
Q_{\nu}^{\dagger} = \frac{1}{2} 
\left(X^{\nu}_{mi} a^{\dagger}_m a^{\dagger}_i - Y^{\nu}_{mi} a_m
  a_i \right) +
\left(\tilde{Z}^{\nu}_m a^{\dagger}_m - Z^{\nu}_m a_m \right),
\label{Ansatz_Q_nu}
\ee
where $X, Y, Z,\tilde{Z}$ are called \textit{amplitudes}.
From the form of the Gauss law operator one would have expected a few
more terms; the reasons for leaving them out will be explained in
sec. \ref{sec_restr_on_ex_op_II}.
With the ansatz (\ref{Ansatz_Q_nu}) we derive from
eq.\,(\ref{equationofmotion}) a closed set of  
equations for the amplitudes $X^{\nu}, Y^{\nu}, Z^{\nu},\tilde{Z}^{\nu}$.
To this end, the
(up to now arbitrary) operators $\delta Q$ will be appropriately chosen,
together with the {\it second gRPA approximation}. 
We choose $\delta Q$ in such a way that the different amplitudes are
extracted individually on the RHS of eq.\,(\ref{equationofmotion}). It
will be useful to consider $\delta Q \in \{a_i, a_i^{\dagger}, a_i^{\dagger}
a_j, a_i a_j, a_i^{\dagger} a_j^{\dagger} \}$, and compute the
commutators $[\delta Q, Q^{\dagger}_{\nu}]$, approximating the real
vacuum by the mean-field vacuum according to the second gRPA approximation.
We obtain:
\be
\begin{array}{lclclcl}
\langle [ a_j, Q^{\dagger}_{\nu} ] \rangle & = & \hspace*{0.8em}
\tilde{Z}^{\nu}_j & \hspace*{0.5em} & 
\langle [ a^{\dagger}_j, Q^{\dagger}_{\nu} ] \rangle& = & 
\hspace*{0.8em} Z^{\nu}_j \\ 
\langle [ a^{\dagger}_n a^{\dagger}_j, Q^{\dagger}_{\nu} ] \rangle&
= & \frac{1}{2} Y^{\nu}_{\{jn\}} & \hspace*{0.5em} & \langle [ a_n a_j,
Q^{\dagger}_{\nu} ] \rangle & = & \frac{1}{2} X^{\nu}_{\{jn\}} \\
\langle [ a^{\dagger}_n a_j, Q^{\dagger}_{\nu} ] \rangle& = &
\hspace*{0.8em} 0, \\
\end{array} \label{coeff_of_Q_sim_dras}
\ee
where we have used the notation $\langle | \ldots | \rangle = \langle
\ldots \rangle$ and introduced the abbreviation
$X_{\{ij\}}$ defined as $X_{\{ij\}}= X_{ij}+X_{ji}$, and
correspondingly for $Y_{\{ij\}}$.
With the $\delta Q$s defined above we
now have to compute the LHS of eq.\,(\ref{equationofmotion}). For this
purpose we introduce a number of matrices
\be
\begin{array}{rclrcl}
A_{njmi} & = & \langle [ a^{\dagger}_n a^{\dagger}_j, [H, a^{\dagger}_m
a^{\dagger}_i]] \rangle, & 
D_{njm} & = & \langle [ a^{\dagger}_n a^{\dagger}_j, [H, a_m ]]
\rangle, \\
B_{njmi} & = & \langle [ a^{\dagger}_n a^{\dagger}_j, [H, a_m a_i]] \rangle
, &
E_{nm} & = & \langle [ a^{\dagger}_n, [H, a^{\dagger}_m ]] \rangle, \\
C_{njm} & = & \langle [ a^{\dagger}_n a^{\dagger}_j, [H, a^{\dagger}_m ]]
\rangle,&
F_{nm} & = & \langle [ a^{\dagger}_n, [H, a_m ]] \rangle, \\
\end{array} \label{defA-F}
\ee
and study their properties under interchange of labels. This will
allow to reduce the number of independent entries of the LHS of
eq.\,(\ref{equationofmotion}). The basic tools for this study are
the Jacobi identity and the second gRPA approximation, i.e. $a |
\rangle = 0$. We also use frequently that $[a, a] = [a^{\dagger},
a^{\dagger}] = 0$.
It turns out that the matrices $A$ and $E$ are symmetric in all
indices, whereas $F$ is a hermitean matrix,  $F^{*}_{nm} = F_{mn}$.
In order to establish that $B$ is also hermitean in the sense
$B_{njmi}^{*} = B_{minj}$ (whereas it is symmetric under interchange
$n \leftrightarrow j, m \leftrightarrow i$) we need the second gRPA
approximation since only upon usage of this approximation $\langle [
H, [a^{\dagger}_m a^{\dagger}_i, a_n a_j]] \rangle$ will be zero
generally. \\ 
Apart from one matrix, all other matrices that arise from inserting
the different $\delta Q$s into the LHS of eq.\,(\ref{equationofmotion}) 
are trivially related to the
matrices $A,\ldots,F$ introduced above; this one non-trivial matrix is 
\be
\langle [a_n^{\dagger}, [H, a_m a_i]] \rangle \label{Dnontriv}.
\ee
It is related to the matrix D, cf. eq.\,(\ref{defA-F}), via
\be
\langle [a_n^{\dagger}, [H,a_m a_i]] \rangle  =  \delta_{in} \Lambda_m
+ \delta_{nm} \Lambda_i + D^{*}_{min}, 
\ee
with $\Lambda_{i,m} = \langle [H, a_{i,m}] \rangle$. 
In sec.\,\ref{subsec_the_quasi_boson_approx}, the second gRPA
approximation will be replaced by the so-called {\it quasi-boson
approximation}. In that formulation, it 
will be obvious that we have to neglect the terms $\Lambda$. From the
explicit expression for the  expectation values, given further below in
eq.\,(\ref{eq_express_lambda}), we see that, at the stationary
mean-field point\footnote{i.e., that point 
in parameter space where the energy expectation value is stationary
with respect to variations of $\bar{\phi}, \bar{\pi}, G, \Sigma$; for
$\Lambda$ to be zero it is necessary that $\frac{\delta}{\delta
\bar{\phi}_i} \langle H \rangle = 0$, and $\frac{\delta}{\delta
\bar{\pi}_i} \langle H \rangle = 0$.},
$\Lambda$ is indeed zero. We note two points:
\begin{itemize}
\item The gRPA matrix [essentially the LHS of
eq.\,(\ref{equationofmotion})] is hermitian iff $\Lambda = 0$.
\item The second gRPA approximation and the quasi-boson approximation
to be introduced below are compatible iff $\Lambda = 0$;  such a constraint
does not appear in nuclear physics as is easily comprehensible in
two different ways: First, the matrix under consideration (as well as
the matrices $C,D$) are non-zero only if the Hamiltonian contains
terms with in total three creation/annihilation operators,
i.e. terms that violate particle-number conservation. Such terms are
not allowed in the usual nuclear physics framework, and thus there is
no possibility for the above consistency condition to arise\footnote{In
the bosonic theories under consideration here, these terms with three
creation/annihilation operators appear in the Hamiltonian usually due
to a condensate.}. Second, if we replace (as we will do in
sec. \ref{subsec_the_quasi_boson_approx}) the operators containing two
ordinary boson creation operators by a new boson operator, in the case of
bosonic theories we have still to keep the original boson, in
contrast to the RPA treatment of fermion systems, where it is not necessary to
retain the original fermions once one has the formulation in terms of bosonic 
operators at hand;  
\end{itemize}
This whole discussion allows us now to write down a first form of the
gRPA equations:
\be
\left( \begin{array}{rrrr} \frac{1}{2} B^*_{ij;kl} & -
\frac{1}{2}A^*_{ij;kl} & D^*_{ij;k} & -C^*_{ij;k} \\
-\frac{1}{2}A_{ij;kl} & B_{ij;kl}  & -C_{ij;k} & D_{ij;k}
\\ \frac{1}{2} D_{kl;i} & -\frac{1}{2} C^*_{kl;i} & F^*_{ik} &
-E^*_{ik} \\ -\frac{1}{2} C_{kl;i} & 
\frac{1}{2} D^*_{kl;i} & -E_{ik}  & F_{ik} \end{array} \right)
\left( \begin{array}{c} \frac{1}{2} X^{\nu}_{\{kl\}} \\ \frac{1}{2}
Y^{\nu}_{\{kl\}}  \\ \tilde{Z}^{\nu}_k \\ Z^{\nu}_k
\end{array} \right) = \Omega_{\nu} \left( \begin{array}{cccc} 1
& & & \\ & -1 & & \\ & & 1 & \\ & & & -1 \end{array} \right)
\left( \begin{array}{c} \frac{1}{2} X^{\nu}_{\{ij\}} \\ \frac{1}{2}
Y^{\nu} _{\{ij\}} \\ \tilde{Z}^{\nu}_i \\ Z^{\nu}_i
\end{array} \right)   
\label{RPAeq1}
\ee
with $\Omega_{\nu} = E_{\nu} - E_0$.
The factors $\frac{1}{2}$ associated with some of the matrices $C,D$
shouldn't lead one to conclude that the gRPA matrix on the LHS is not
hermitian. In fact, as we will see later on, we can write the gRPA
equations in a very compact form resulting from a hermitian
Hamiltonian (at the stationary point of the mean-field equations
$\frac{\delta}{\delta \bar{\phi}_i} \langle H \rangle \stackrel{!}{=} 0$, and
$\frac{\delta}{\delta \bar{\pi}_i} \langle H \rangle \stackrel{!}{=} 0$).
\subsection{The Quasi-Boson
Approximation\label{subsec_the_quasi_boson_approx}}
Sometimes, it is useful to have the second gRPA approximation at hand
without having to take expectation values. This can be done with the
help of the {\it quasi-boson approximation}. The name has its roots in
nuclear physics \cite{MW69, Ring:1980}, where two fermions are
combined into one boson. This 
works only approximately\footnote{i.e. the ordinary bosonic commutation
relations are only fulfilled if we take the mean-field expectation
value of the commutator of the boson pairs, and thus in some sense
employ the second gRPA approximation, by using the mean-field instead of
the exact vacuum state.}, so that the result was christened a {\it
quasi-boson}. Interestingly enough, we can do the same in a bosonic
system: we replace a two boson operator (like $aa$ or $a^{\dagger}
a^{\dagger}$) by a new operator that again has bosonic commutation
relations. \\
We construct the new boson-pair operators $\BB, \BBd$ s.t. the
commutation relations are identical to the mean-field expectation
values of commutators containing still the pair of original boson
operators: 
\bea
\begin{array}{lclclclclcl}
\,[ a_m a_i, a_j ]\, & = & 0 & & & & &  \rightarrow & \, [ \BB_{mi},a_j] & = &
0    \\ 
\,[ a_m a_i, a^{\dagger}_j]\,   & = & \delta_{ij} a_m + \delta_{mj} a_i &
\rightarrow &  \langle [ a_m a_i, a^{\dagger}_j ] \rangle & = & 0 
& \rightarrow & \, [ \BB_{mi},a^{\dagger}_j] & = & 0  \\
\,[ a^{\dagger}_m a^{\dagger}_i, a_j]\,   & = & -\delta_{ij} a^{\dagger}_m -
\delta_{mj} a^{\dagger}_i & \rightarrow &  \langle [ a^{\dagger}_m
a^{\dagger}_i, a_j ] \rangle &= & 0  & \rightarrow & \,
[\BBd_{mi},a_j] & = & 0.   
\end{array} \nonumber
\eea
The only non-vanishing commutator involving $\BB, \BBd$
originates from $[ a_m a_i, a^{\dagger}_n a^{\dagger}_j]$: 
\bea
 \,[\frac{1}{\sqrt{2}} a_m a_i, \frac{1}{\sqrt{2}} a^{\dagger}_n
 a^{\dagger}_j]\,  &=& \frac{1}{2} \left( \delta_{in} 
\delta_{mj} + \delta_{ij} \delta_{mn} +  \delta_{in}
a^{\dagger}_j a_m 
+ \delta_{mn} a^{\dagger}_j a_i +  \delta_{ij} a^{\dagger}_n a_m +
\delta_{mj} a^{\dagger}_n a_i \right) \nonumber 
\\ \rightarrow [ \BB_{mi},
\BBd_{nj}]  
&=& \frac{1}{2} \left( \delta_{in} \delta_{mj} + \delta_{ij}
\delta_{mn} \right). \nonumber 
\nonumber
\eea
We therefore replace
\be
\begin{array}{lcl}
\frac{1}{\sqrt{2}} \big(a_m a_i \big) & \leftrightarrow & \BB_{mi} \\
\frac{1}{\sqrt{2}} \big(a^{\dagger}_m a^{\dagger}_i \big) & \leftrightarrow &
\BBd_{mi}. 
\end{array}
\label{replace_aa_adad}
\ee
The operators $\BB, \BBd$ have the following commutation relations
\bea
\,[a_j, \BB_{mi}] = [a_j, \BBd_{mi} ] &=&
[\BB_{mi}, \BB_{nj}] = 0 \\ 
\mbox{and} \hspace*{1.5em} \,[\BB_{mi}, \BBd_{nj}] &=& \frac{1}{2}
(\delta_{in} \delta_{mj} + \delta_{ij} \delta_{mn}). 
\eea
With these new operators, the excitation operators 
\be
Q^{\dagger}_{\nu} = \frac{1}{2} 
\Big( X^{\nu}_{mi} a^{\dagger}_m a^{\dagger}_i -  Y^{\nu}_{mi} a_m a_i \Big)
+ \left(\tilde{Z}^{\nu}_m a^{\dagger}_m - Z^{\nu}_m a_m \right)
\ee
become: 
\be
Q^{\dagger}_{B \nu} = \frac{1}{\sqrt{2}} 
\left(X^{\nu}_{mi}  \BBd_{mi} -  Y^{\nu}_{mi} \BB_{mi} \right) + 
\left(\tilde{Z}^{\nu}_m a^{\dagger}_m - Z^{\nu}_m a_m \right).
\label{Qdaggerdef}
\ee
At this point a short comment on how one constructs quasi-boson
approximations is in order, cf. also eq.\,(\ref{qb_approx_gen_obo}) below. 
Usually we will have to consider only one- 
and two-body operators, but the procedure should work for
higher-body operators as well: 
\begin{itemize}
\item[(i)] first, one writes down the operator
$\mathcal{O}$ under consideration as a polynomial of $a, a^{\dagger}$
operators. Then one 
tries to extract the coefficients of the different powers 
in analogy to
eqs.\,(\ref{coeff_of_Q_sim_dras}), (\ref{defA-F}) via (a) taking
multiple commutators of $\mathcal{O}$ with up to two $a, a^{\dagger}$
operators, and then (b)  taking vacuum expectation values of these
multiple commutators employing the second gRPA approximation. 
\item[(ii)] The quasi-boson approximation $\mathcal{O}_B $ of this
operator $\mathcal{O}$ is then given as a polynomial of $\BB, \BBd, a,
a^{\dagger}$ operators. The coefficients of the different powers are
extracted by taking multiple commutators of  $\mathcal{O}_B $ with $a,
a^{\dagger}, \BB, \BBd$ {\it without} taking expectation values. One then
requires that the coefficients of $\mathcal{O}_B$ determined in this way
are identical to those of $\mathcal{O}$ determined in (i) if one
replaces in the multiple commutators of (i) $\mathcal{O}$ by $\mathcal{O}_B$
and terms of the structure $a a$ by
$\sqrt{2} \BB$ and $a^{\dagger} a^{\dagger}$ by $\sqrt{2} \BBd$ as indicated in
eq.\,(\ref{replace_aa_adad}). 
\end{itemize}
This is precisely the procedure that led to eq.\,(\ref{Qdaggerdef}).
Using the procedure outlined, one can in fact express the QBA $\mathcal{
O}_B$ of a general one-body operator $\mathcal{O}$ in terms of mean-field
vacuum expectation values of commutators: 
\bea
\mathcal{O}_B = \langle \mathcal{O} \rangle &-& \frac{1}{\sqrt{2}} \langle
  [a_i^{\dagger} a_j^{\dagger}, 
  \mathcal{O}] \rangle \BB_{ij} + \frac{1}{\sqrt{2}} \langle [a_i
  a_j,\mathcal{O}] \rangle \BBd_{ij} 
- \langle [a_i^{\dagger},  \mathcal{
  O}] \rangle a_i + \langle [a_i,  \mathcal{O}] \rangle
  a_i^{\dagger}. \label{qb_approx_gen_obo} 
\eea
Incidentally, one can at this point easily verify -by writing out the
commutators - the claim made
above: the QBA of a symmetry generator (which is generally hermitian) of
  one-body type that annihilates the mean-field vacuum \textit{vanishes}. \\
Now we will follow the above procedure to obtain the Hamiltonian in
quasi-boson approximation. The coefficients mentioned are nothing but
the matrices $A,...,F$. We therefore have the requirements
\be \begin{array}{rclcl}
A_{njmi} & = & \langle [ a^{\dagger}_n a^{\dagger}_j, [H, a^{\dagger}_m
a^{\dagger}_i]] \rangle & \stackrel{!}{=} &  [ \sqrt{2} \BB^{\dagger}_{nj},
[H_B,  \sqrt{2} \BB^{\dagger}_{mi} ]], \\
B_{njmi} & = &  \langle [ a^{\dagger}_n a^{\dagger}_j, [H, a_m a_i]] \rangle &
\stackrel{!}{=} &    [\sqrt{2} \BB^{\dagger}_{nj}, [H_B, \sqrt{2} \BB_{mi} ]],
\\ 
C_{njm} & = & \langle [ a^{\dagger}_n a^{\dagger}_j, [H, a^{\dagger}_m ]]
\rangle  & \stackrel{!}{=} &  [ \sqrt{2} \BB^{\dagger}_{nj} , [H_B,
a^{\dagger}_m ]], \\  
D_{njm} & = & \langle [ a^{\dagger}_n a^{\dagger}_j, [H, a_m ]] \rangle  &
\stackrel{!}{=} &   [ \sqrt{2} \BB^{\dagger}_{nj} , [H_B, a_m ]], \\ 
E_{nm} & = & \langle [ a^{\dagger}_n, [H, a^{\dagger}_m ]] \rangle  &
\stackrel{!}{=} &   [ a^{\dagger}_n, [H_B, a^{\dagger}_m ]], \\ 
F_{nm} & = & \langle [ a^{\dagger}_n, [H, a_m ]] \rangle  & \stackrel{!}{=} &
[ a^{\dagger}_n, [H_B, a_m ]], 
\end{array}
\label{eq_fix_coeff_HB}
\ee 
where $H_B$ denotes the Hamiltonian in quasi-boson approximation.
At this point it seems in order to discuss a question that turned up
in connection with eq.\,(\ref{Dnontriv}): There we saw that one of the
double commutators was not directly connected to one of the matrices
$A,...,F$ but that some extra terms appeared, the
'$\Lambda$-terms'. If we consider the same double commutator in the
quasi-boson approximation, we obtain
\be
\langle [H,[a^{\dagger}_n, a_m a_i]] \rangle \rightarrow 
[H_B,\underbrace{[a^{\dagger}_n, \sqrt{2} \BB_{mi}]}_{= 0}] = 0.
\ee
Thus the quasi-boson approximation implies the {\it vanishing of } $\Lambda$.
An explicit expression can be given for $\Lambda$ as well
(cf. app. \ref{app_expl_A-F})
\be
\Lambda_i = 
\left(\frac{\delta}{\delta \bar{\phi}_k} \langle H \rangle \right)
U_{k i}  + \left(\frac{\delta}{\delta \bar{\pi}_k} \langle H \rangle
   \right) \left(\frac{1}{2i} U^{-1}_{k i} + 2 \Sigma_{kl} U_{l i} \right), 
\label{eq_express_lambda}
\ee
which vanishes if we choose the parameters of the reference state
s.t. the energy is stationary under their variation (at least
w.r.t. $\bar{\phi}, \bar{\pi}$). 
We obtain a consistency condition between the second gRPA approximation
and the quasi-boson approximation: the latter is equivalent to the
former {\it only at the stationary point} - at least w.r.t the
condensates $\bar{\phi}, \bar{\pi}$ - of $\langle H \rangle$.
Thus, in the following we will always assume that the state (which we
often call the {\it mean-field
vacuum})  relative to which our creation/annihilation operators are
defined [cf. eq.\,(\ref{RPA_most_general_TI_Gauss})] has its parameters chosen
s.t. it satisfies the Rayleigh-Ritz 
principle - that this restriction would become necessary was already
indicated when the second gRPA approximation was introduced. 
With this qualification, we can give the (now hermitian) Hamiltonian
in quasi-boson approximation:
\be
\begin{array}{lccl} H_B &=& E_{MF} & - \frac{1}{4}\left(A_{n_1 j_1 n_2 j_2} \BB_{n_1
      j_1}\BB_{n_2 j_2} + A^*_{n_1 j_1 n_2 j_2} \BB^{\dagger}_{n_1 j_1}
    \BB^{\dagger}_{n_2 j_2} \right) + \frac{1}{2} B_{n_2 j_2 n_1 j_1}
    \BB^{\dagger}_{n_1 j_1}\BB_{n_2 j_2} \\ & & & - 
    \frac{1}{\sqrt{2}}\left(C_{n_1 j_1 n_2} \BB_{n_1 j_1} a_{n_2} + C^*_{n_1
    j_1 n_2} \BB^{\dagger}_{n_1 j_1} a^{\dagger}_{n_2} \right)  \\ & &
    & + 
    \frac{1}{\sqrt{2}}\left(D_{n_1 j_1 n_2} \BB_{n_1 j_1} a^{\dagger}_{n_2} +
    D^*_{n_1 j_1 n_2} \BB^{\dagger}_{n_1 j_1} a_{n_2} \right) \\ & & &
- \frac{1}{2} \left(E_{n_1 n_2} a_{n_1} a_{n_2} + E^*_{n_1 n_2}
    a^{\dagger}_{n_1} a^{\dagger}_{n_2} \right) + F_{n_2 n_1}
    a^{\dagger}_{n_1} a_{n_2} \hspace*{0.5em},
\end{array}
\label{Hboson}
\ee
with $E_{MF} = \langle H \rangle$. We have achieved writing the
approximated Hamiltonian as a quadratic form, which can always be diagonalized.
Using this Hamiltonian $H_B$ one can see that the gRPA
equations eq.\,(\ref{RPAeq1}) can be written in the transparent
form\footnote{One inserts eq.\,(\ref{Hboson}) and
eq.\,(\ref{Qdaggerdef}) into eq.\,(\ref{RPA-SHO}) and compares the
coefficients of the different operators $a, a^{\dagger}, \BB,
\BBd$. This then reproduces eq.\,(\ref{RPAeq1}). }
\be
\ [ H_B, Q^{\dagger}_{B\nu} ] = \Omega_{\nu}
Q^{\dagger}_{B\nu}. \label{RPA-SHO} 
\ee
This form of the gRPA equations demonstrates that we have performed the
approximation to the dynamics (Hamiltonian) and to the excitation
operators consistently, since the form of the equations is identical to the
form of the Schr\"odinger equation eq.\,(\ref{Schroedinger_comm}) with
the {\it exact} Hamiltonian and the {\it exact} excitation
operators. \\ 
\subsection{Restriction on Excitation Operators, part II
\label{sec_restr_on_ex_op_II}} 
Having introduced all the concepts of gRPA, it is time to look again
at some terms that appear neither in $Q^{\dagger}_{\mu}$ nor in
$Q^{\dagger}_{B\mu}$. 
\begin{enumerate}
\item $Q_{\nu}^{\dagger}$ does not contain an $a^{\dagger} a$
term. The superficial reason for this is that the corresponding
amplitude cannot be extracted in a way similar to the other amplitudes
$X, Y, Z, \tilde{Z}$, since due to the second gRPA approximation
\be
\langle | [\delta Q, a^{\dagger} a ] |  \rangle = 
\langle  | \delta Q a^{\dagger} \underbrace{ a | \rangle}_{=0} - 
\underbrace{\langle  | a^{\dagger}}_{= 0} a \delta Q | 
\rangle = 0 . \label{noextract}
\ee
This point will become clearer in the context of the QBA, cf. item
\ref{second_footnote_on_absence_aadagger} below.
\item $Q_{\nu}^{\dagger}$ does not contain an $a
a^{\dagger}$ term. This is due to the fact that $a a^{\dagger}$ is
distinguished from $a^{\dagger} a$ only by a
constant. However, this constant would lead to a non-vanishing
expectation value of $Q_{\nu}^{\dagger}$ between mean-field vacua, in
view of the second gRPA approximation a situation certainly not
desirable for an excitation operator (all other terms contained in
$Q_{\nu}^{\dagger}$ vanish between mean-field vacua !). One could also
argue on a formal level that the way the generalized RPA is derived here does
not allow to determine any constant parts of $Q_{\nu}^{\dagger}$, so
if we cannot determine a constant, we shouldn't put it into our ansatz
in the first place. 
\item \label{second_footnote_on_absence_aadagger}Here we want to have
a look at the fact that $a^{\dagger} a$ does not appear in our excitation
operator from the perspective of the QBA. The excitation operator is
linear in all the different boson creation/annihilation operators that
we construct as one can see in 
eq.\,(\ref{Qdaggerdef}).  Thus, the question arises whether we can
construct a boson operator from $a^{\dagger}a$ similar to how we
construct one from $a^{\dagger} a^{\dagger}$. The problem is
immediately apparent: if we consider $C_{mi} = a_m^{\dagger} a_i$,
then its adjoint has the same structure as $C_{mi}$, since
$C_{mi}^{\dagger} = a_i^{\dagger} a_m = C_{im}$; if we compute the
commutator $[C_{ij}, C^{\dagger}_{kl}] = \delta_{jk} a^{\dagger}_i a_l
- \delta_{li} a^{\dagger}_k a_j$ we see that its mean-field
expectation value is zero. Therefore we cannot construct boson
operators with the correct commutation relations from $a^{\dagger}a$
in the same way as we did for $a^{\dagger} a^{\dagger}$, and thus they
don't appear in the excitation operators $Q^{\dagger}_{\nu}$. In
nuclear physics this is well-known. There one can see that the
quasi-boson operators correspond to a creation of a particle-hole
pair (or its annihilation), whereas the operator that creates and
annihilates a particle is translated into an operator that creates and
annihilates a quasi-boson \cite{MW69}.
\end{enumerate}
\subsection{Normal Mode Form \label{sec_on_normal_mode_form}}
In app. \ref{gRPA_comm_rel} we consider the commutation relations
of the normal mode operators. There we show that if {\it all}
excitation energies $\Omega_{\nu}$ are distinct and \textit{non-zero}, the
normal modes have the usual bosonic commutation relations: 
\be
[Q_{B\mu}, Q^{\dagger}_{B\nu}] = \delta_{\mu \nu}.
\ee
Using this fact, we can conclude from eq.\,(\ref{RPA-SHO}) 
\be
H_B = E_{RPA} + \sum_{\nu} \Omega_{\nu} Q^{\dagger}_{B\nu}
Q_{B\nu}. \label{HBdiagonal}  
\ee
The sum over $\nu$ extends over all positive semi-definite $\Omega_{\nu}$.
The constant $E_{RPA}$ can be determined as usual,
cf. \cite{Ring:1980}, namely by requiring
\be
\langle H_B \rangle = E_{MF}.
\ee
Using
\be
\langle Q^{\dagger}_{B\nu} Q_{B\nu} \rangle = \frac{1}{2} \sum_{mi}
   |\frac{1}{2}Y^{\nu}_{\{mi\}}|^2 + \sum_i |Z^{\nu}_i|^2, 
\ee
with $Y^{\nu}_{\{mi\}} = Y^{\nu}_{mi} + Y^{\nu}_{im}$ this gives
\be
E_{RPA}  = E_{MF} - \sum_{\nu} \Omega_{\nu}
  \left(\frac{1}{2} \sum_{mi} |\frac{1}{2}Y^{\nu}_{\{mi\}}|^2 + \sum_i
  |Z^{\nu}_i|^2\right). \label{firstRPAenergy}
\ee
A similar expression can be obtained in nuclear physics
\cite{Ring:1980} but without the appearance of $|Z^{\nu}_i|^2$. 
The changes brought about by excitations with zero excitation energy
will be considered in sec. \ref{sec_mom_of_in}.
\section{Conservation Laws \label{sec_Conserv_laws}}
In this section we want to demonstrate that - under certain conditions -
conservation laws from the full theory translate to conservation laws in the
gRPA treatment. The proof follows the lines of the fermionic case as
given in \cite{MW69}. 
For the conservation laws to hold in the approximated theory, it is
mandatory that the reference state minimizes the expectation value of
the Hamiltonian, since then the terms of the Hamiltonian linear in
$a, a^{\dagger}, aa, a^{\dagger} a^{\dagger}$ vanish,
cf. eq.\,(\ref{H_in_ca_rep}). One should also keep in mind that the
terms of $H$ linear in $a, a^{\dagger}$ have to vanish anyway, since
otherwise a treatment using the QBA is not valid as was discussed before. \\
The fact that conservation laws translate into the approximated theory
has to be taken with a grain of salt, however, since it turns out
(and will be discussed in sec. \ref{sec_change_of_char}) that the character of
symmetries may change. Non-Abelian symmetries are usually reduced to
Abelian symmetries which is not surprising since the gRPA is basically
a small-fluctuation approximation which does not probe the group
manifold. 
\subsection{General Observation \label{subsec_conservlaws_GO}}
With these qualifications, let us now compute the commutator of $H_B$,
cf. eq.\,(\ref{Hboson}), with a general one-body operator $\Gamma_B$
\be
\Gamma_B = \Gamma_0 + \left(\Gamma^{10}_{n_1} a^{\dagger}_{n_1} +
  \Gamma^{01}_{n_1} a_{n_1} \right) \label{express_for_Gamma_B} + \sqrt{2} 
\left(\Gamma^{20}_{n_1 n_2}
\BB^{\dagger}_{n_1 n_2} + \Gamma^{02}_{n_1 n_2} \BB_{n_1 n_2} \right),
\ee
where the coefficients have been chosen, s.t. the full operator before
QBA reads 
\be
 \Gamma = \Gamma_0 + \Gamma^{10}_{n_1} a^{\dagger}_{n_1} + 
  \Gamma^{01}_{n_1} a_{n_1} + \Gamma^{20}_{n_1 n_2}
a^{\dagger}_{n_1} a^{\dagger}_{n_2} + \Gamma^{02}_{n_1 n_2} a_{n_1}
a_{n_2} + \Gamma^{11}_{n_1 n_2} a^{\dagger}_{n_1}
  a_{n_2}. \label{eq_full_gamma} 
\ee
Then the commutator gives
\bea 
[ H_B, \Gamma_B] &=&  -\frac{1}{\sqrt{2}} \BB_{n_1 j_1}  \big( 
  C_{n_1 j_1 m_1}   \Gamma^{10}_{m_1} +  D_{n_1 j_1 m_1}  \Gamma^{01}_{m_1}
 + A_{n_1 j_1 m_1 m_2}  \Gamma^{20}_{m_1  m_2} + B_{n_1 j_1 m_1 m_2}
  \Gamma^{02}_{m_1 m_2} \big) \nonumber \\  
 & & + \frac{1}{\sqrt{2}} \BB_{n_1 j_1}^{\dagger} \big( 
  C^{*}_{n_1 j_1 m_1}   \Gamma^{01}_{m_1} +  D^{*}_{n_1 j_1 m_1}
 \Gamma^{10}_{m_1} 
+ A^{*}_{n_1 j_1 m_1 m_2}
  \Gamma^{02}_{m_1 m_2} + B^{*}_{n_1 j_1 m_1 m_2}   \Gamma^{20}_{m_1 m_2}
  \big) \nonumber \\   
& & - a_{n_1} \big( E_{n_1 m_1} \Gamma^{10}_{m_1}   + F_{n_1 m_1}
\Gamma^{01}_{m_1}   
+ \Gamma^{20}_{m_1 m_2} C_{m_1 m_2 n_1}  
+  \Gamma^{02}_{m_1 m_2} D^{*}_{m_1 m_2 n_1}  \big) \nonumber \\ 
& & + a^{\dagger}_{n_1} \big( 
E^{*}_{n_1 m_1} \Gamma^{01}_{m_1} + F^{*}_{n_1 m_1}   
  \Gamma^{10}_{m_1}  
+ \Gamma^{02}_{m_1 m_2} C^{*}_{m_1 m_2 n_1}  
+ \Gamma^{20}_{m_1 m_2} D_{m_1 m_2 n_1}  \big). 
\label{gl1}
\eea
Let us now concentrate on the first line to make the principle clear;
we use the definition of $A,B,C,D$ as mean-field expectation values of
double commutators involving the \textit{full Hamiltonian},
cf. eq.\,(\ref{eq_fix_coeff_HB}). Then it is clear that - using also
eq.\,(\ref{eq_full_gamma}) - one can rewrite  
it as 
\bea
& & \BB_{n_1 j_1} \left(\langle [ a^{\dagger}_{n_1} a^{\dagger}_{j_1}, [H,
  a^{\dagger}_{m_1} ]] \rangle   \Gamma^{10}_{m_1} +  
\langle [ a^{\dagger}_{n_1} a^{\dagger}_{j_1}, [H, a_{m_1} ]] \rangle
\Gamma^{01}_{m_1}  \right. \nonumber \\ && \hspace*{3.5em} \left.
 + \langle [ a^{\dagger}_{n_1} a^{\dagger}_{j_1}, [H, a^{\dagger}_{m_1}
a^{\dagger}_{m_2} ]] \rangle  \Gamma^{20}_{m_1 m_2} 
+  \langle [
a^{\dagger}_{n_1} a^{\dagger}_{j_1}, [H, a_{m_1} a_{m_2}]] \rangle 
  \Gamma^{02}_{m_1 m_2} \right)
 \nonumber \\
& = &  \BB_{n_1 j_1} \big(  \langle [ a^{\dagger}_{n_1} a^{\dagger}_{j_1}, [H,
  \Gamma ]]  \rangle -  \Gamma^{11}_{m_1 m_2} \underbrace{\langle [
  a^{\dagger}_{n_1} a^{\dagger}_{j_1}, [H,  a_{m_1}^{\dagger} a_{m_2} ]]
  \rangle}_{(*)} \big).   
\eea
The last term $(*)$ is proportional to $H^{02}$, i.e. that term in the
Hamiltonian that multiplies two annihilation operators, and is zero in
the case we are considering here, i.e. our reference state minimizes
the expectation value of the Hamiltonian. The procedure can be
repeated for the other three terms in $[H_B, \Gamma_B]$. Eventually,
one finds
\bea [H_B, \Gamma_B] &=& - \frac{1}{\sqrt{2}}
    \langle [a^{\dagger}_{n_1} a^{\dagger}_{j_1}, [H, \Gamma]] \rangle
    \BB_{n_1 j_1}  +  \frac{1}{\sqrt{2}} \langle 
  [a_{n_1} a_{j_1}, [H, \Gamma]] \rangle \BB^{\dagger}_{n_1 j_1}
    \label{conserv1} \\ & & -
   \langle [a^{\dagger}_{n_1}, [H, \Gamma]] \rangle a_{n_1} +
  \langle [a_{n_1} , [H, \Gamma]] \rangle a^{\dagger}_{n_1}
    + \text{terms  that vanish at the stationary point}. \nonumber 
\eea
Two points should be noted:
\begin{enumerate}
\item In eq.\,(\ref{qb_approx_gen_obo}) we have given a formula of the
QBA of a general one-body operator; if $[H, \Gamma]$ is a one-body
operator, then by comparing eq.(\ref{conserv1}) to
eq.\,(\ref{qb_approx_gen_obo}) we conclude that $[H_B, \Gamma_B]$ is
the QBA of this operator (apart from a possible mean-field expectation value). 
\item  The stationarity condition ($H^{20}= H^{10} = 0$) has been used
several times; if we are not at the stationary point, some of the
terms neglected above do not vanish, i.e. $[H, \Gamma]=0$ {\it does
not imply} $[H_B, \Gamma_B]=0$ away from the stationary point.
\end{enumerate}
We want to assume that we have chosen the reference state
s.t. indeed $H^{20}= H^{10} = 0$. Then eq.\,(\ref{conserv1})
simplifies to 
\be
\ [H_B, \Gamma_B] = 0.
\ee
Comparing this with the gRPA equations as given in eq.(\ref{RPA-SHO}) we see
that - since $\Gamma_B$ is exactly of the form of $Q_{B \nu}^{\dagger}$
- that $\Gamma_B$ is a solution of the gRPA equations with excitation
energy zero, i.e. it is a zero mode.
\subsection{Change of Symmetry Character\label{sec_change_of_char}}
It seems that by this construction we have precisely obtained what we
wanted: the spurious excitation made possible by the deformed
mean-field state is a solution by itself, and does not influence the
other - physical - solutions inappropriately. However, this result has
to be taken with a bit of caution: as has been indicated before, the
character of the symmetry may change. This can be seen as follows:
Compute the commutator of the QBA $Q_B, P_B$ of two arbitrary one-body
operators $Q,P$. Using eq.\,(\ref{qb_approx_gen_obo}) it is very
simple to obtain
\be
\ [Q_B, P_B] = \langle [ Q, P ] \rangle.
\ee
In the case of Yang-Mills theory, we obtain for the commutator of the Gauss
law operators 
\be
~[\Gamma^a_B, \Gamma^b_B] = \langle [\Gamma^a, \Gamma^b] \rangle = i
f^{abc} \langle \Gamma^c \rangle. 
\ee
At the stationary point, however, we have for the bosonic theories
under consideration 
$\bar{\bf \pi} = \Sigma = 0$, since $\langle \pi \pi \rangle$ is
quadratic in both 
$\bar{\pi}$ and $\Sigma$, cf. eq.\,(\ref{eq_expect_pi_squared}). Thus,
in the case of Yang-Mills theory, we obtain  
\be
~[\Gamma^a_B, \Gamma^b_B] =0, 
\ee
using the expression given in eq.\,(\ref{eq_Gauss_law_c-a-decomp}) for
$\langle \Gamma^a \rangle$. In other words, the QBA has reduced the
non-Abelian $SU(N)$ symmetry to an Abelian $U(1)^{N^2-1}$ symmetry. 
This is a phenomenon also well known from perturbation theory.
\subsection{Possibilities for Improvement}
One main drawback of the gRPA is the fact that non-Abelian symmetries
may be reduced to Abelian symmetries. There have been a couple of
investigations in nuclear physics, e.g. \cite{ Ring:1980, MW69}, and other
quantum field theories (where the fields were in the fundamental
representation) 
\cite{Aouissat:1998nu, Aouissat:1996va, Aouissat:1997be} whether one
can regard the gRPA as a certain order in a systematic
expansion\footnote{Within the path integral approach one can show that the
  mean-field approximation and the RPA correspond to the first and second,
  respectively, order in the loop expansion \cite{Reinhardt:1980}.},
symbolically
\be
\text{complete result} = \text{mean field} + \text{gRPA} +
\text{higher orders}. 
\ee
A first hint on how one may construct such higher orders for one-body
operators like the Gauss law operator can be obtained from
the following observation: if we calculate (as we have done above) the
commutator of two quasi-boson approximated Gauss law operators, we
obtain the mean-field value of the full commutator; apparently, in the
expansion alluded to above, taking a commutator reduces the order in
the expansion by one (for a similar observation in nuclear physics,
cf. e.g. \cite{Ma74, Ring:1980}). 
If we construct a next order of say the Gauss law operator, taking the
appropriate commutator should give the quasi-boson approximation of the
commutator, the latter being again a one-body operator. In formulas,
if we write\footnote{One should note that this is a slight deviation from the
  conventions in the rest of the paper, e.g. in
  eq.\,(\ref{qb_approx_gen_obo}) the mean-field order $\langle
  \mathcal{O} \rangle$ is assigned to $\mathcal{O}_B$.}  
\be
\underbrace{\Gamma}_{\text{full operator}} = \underbrace{\langle
\Gamma \rangle}_{\text{mean-field}} +
\underbrace{\Gamma_B}_{\text{conventional gRPA}} +
\underbrace{\Gamma_{2B}}_{\text{next term}} + \text{higher orders}
\ee 
we have 
\be
\ [\Gamma^a_B, \Gamma^b_B] = \langle [\Gamma^a, \Gamma^b] \rangle
\ee
and $\Gamma_{2B}$ would be correct if it would reproduce
\be
 [\Gamma^a_B, \Gamma^b_{2B}] +  [\Gamma^a_{2B}, \Gamma^b_B] =  [\Gamma^a,
 \Gamma^b]_B. 
\ee
Such a $\Gamma_{2B}$ can indeed be constructed. If one requires that
$\Gamma_{2B}$ reproduces the same commutators as $H_B$ in
eq.\,(\ref{eq_fix_coeff_HB}) [with $H$ obviously replaced by $\Gamma$
on the LHS of eq.\,(\ref{eq_fix_coeff_HB})] one obtains $\Gamma_{2B}$ as 
\be
\Gamma_{2B} = 2 \Gamma^{11}_{kj} \BB^{\dagger}_{kl} \BB_{lj} +
\Gamma^{20}_{j_1 j_2} a^{\dagger}_{j_1} a^{\dagger}_{j_2} + 
\Gamma^{02}_{j_1 j_2} a_{j_1} a_{j_2} + \Gamma^{11}_{kj} a^{\dagger}_k a_j
\ee
where the coefficients $\Gamma^{ab}_{ij}$ are defined in
eq.\,(\ref{eq_full_gamma}). \\ 
For the calculation of the commutator $[\Gamma^a_B, \Gamma^b_{2B}]$ we
use a technique very similar to the one used in
sec. \ref{subsec_conservlaws_GO} by rewriting the coefficients of
$\Gamma^b_{2B}$ as double commutators (note that the coefficients of
$\Gamma^b_{2B}$ are called ${}^b\Gamma^{11}_{jk}$ etc; note also that - as in
the remainder of the paper - we use $\Gamma_{\{ij\}} = \Gamma_{ij} +
\Gamma_{ji}$)
\be
{}^b\Gamma^{11}_{kj} = \langle [a^{\dagger}_j, [\Gamma^b, a_k]]
\rangle \quad;\quad {}^b\Gamma^{02}_{\{j_1 j_2\}} = - \langle
[a^{\dagger}_{j_1}, [\Gamma^b, a^{\dagger}_{j_2}]] \rangle \quad;\quad 
{}^b\Gamma^{20}_{\{j_1 j_2\}} = - \langle [a_{j_1}, [\Gamma^b,
a_{j_2}]] 
\rangle
\ee
Using these results, one can rewrite
\begin{subequations}
\bea
\ [\Gamma^a_B, \Gamma^b_{2B}] &=& a_{j_1} \langle [a_{j_1}^{\dagger},
[\Gamma^b, ({}^a \Gamma^{10}_{j_2} a^{\dagger}_{j_2} + 
{}^a \Gamma^{01}_{j_2} a_{j_2})]] \rangle - a_{j_1}^{\dagger} \langle [a_{j_1},
[\Gamma^b, ({}^a \Gamma^{10}_{j_2} a^{\dagger}_{j_2} + 
{}^a \Gamma^{01}_{j_2} a_{j_2})]] \rangle \nonumber \\ 
& & + \frac{1}{\sqrt{2}} \BB_{j_1 j_2} \langle [a_{j_1}^{\dagger}
a_{j_2}^{\dagger}, 
[\Gamma^b, ({}^a \Gamma^{02}_{k_1 k_2} a_{k_1} a_{k_2})]] \rangle 
- \frac{1}{\sqrt{2}} \BBd_{j_1 j_2} \langle [a_{j_1} a_{j_2}, 
[\Gamma^b, ({}^a \Gamma^{20}_{k_1 k_2} a^{\dagger}_{k_1}
a^{\dagger}_{k_2})]] \rangle \nonumber \\
& = & [\Gamma^b_{B}, \Gamma^a_{2B}] \\ & & -
a_{j_1} \langle [ 
a^{\dagger}_{j_1}, [ \Gamma^a, \Gamma^b]] \rangle + a^{\dagger}_{j_1} \langle [
a_{j_1}, [ \Gamma^a, \Gamma^b]] \rangle    \label{eq_commut_pb} \\ & & -
\frac{1}{\sqrt{2}} \BB_{j_1 
j_2} \langle [ a^{\dagger}_{j_1} a^{\dagger}_{j_2} , [\Gamma^a,
\Gamma^b]] \rangle + \frac{1}{\sqrt{2}} \BBd_{j_1 j_2} \langle [
a_{j_1} a_{j_2}, [ \Gamma^a, \Gamma^b]] \rangle \label{eq_commut_pc}
\eea
\end{subequations}
Since $[\Gamma^a, \Gamma^b]$ just again gives a one-body operator,
we can compare eqs.\,(\ref{eq_commut_pb}), (\ref{eq_commut_pc}) to
eq.\,(\ref{qb_approx_gen_obo}), and see that it is indeed the QBA
of\footnote{Note again that in this section the expectation value $\langle [
  \Gamma^a, \Gamma^b] \rangle$ is not incorporated into $[\Gamma^a,
  \Gamma^b]_B$.} 
$[\Gamma^a, \Gamma^b]$, i.e. $[\Gamma^a, \Gamma^b]_B$. Thus we have
fulfilled our goal: we have constructed an extension of $\Gamma^a_B$
which improves the commutation relations as was desired. \\
At this point one should keep in mind the following: We have
constructed $\Gamma_{2B}$ by requiring that $\langle [ a^{\dagger}_{i}
a^{\dagger}_{j}, [ \Gamma, a_k ]] \rangle \stackrel{!}{=} [\sqrt{2} \BBd_{ij}
, [ \Gamma_{2B}, a_k ]]$. This gave \textit{zero} as the coefficient
of $a^{\dagger} \BB$. Had we chosen instead to require $\langle [
a_k , [ \Gamma, a^{\dagger}_{i} a^{\dagger}_{j}]] \rangle
\stackrel{!}{=} [a_k , [ \Gamma_{2B}, \sqrt{2} \BBd_{ij}]]$  we would
have obtained a non-zero coefficient for $a^{\dagger} \BB$. 
This incompatibility of the two construction principles goes back
basically to the problem that the commutation relations for $a, \BB$
are constructed to work with one-body operators (as was noted
in the context of $H_B$ where both coefficients are required to be
equal. This can be achieved by going to the stationary point since the
offending terms all vanish there; here we cannot follow the same route
since the state is already specified). 
The ambiguity is here in some sense resolved, since the latter construction
does not lead to the aspired commutation relations\footnote{There can be
circumstances where even in the latter construction the correct commutation
relations can be produced. We have investigated the case where the symmetry
generators are the Gauss law operators, and found two instances where the
correct commutation relations were produced even in the presence of the $a
\BBd, a^{\dagger} \BB$-terms. These examples were: $G^{-1} \propto \unit$ -
this is not a very likely mean-field vacuum since there is no correlation
between different points in space - and the case where $\hat{\bar{\vec{D}}}
U^{-1} = 0$. For the definition of $U$, cf. eq.\,(\ref{eq_def_of_U}). But this
requirement on $U^{-1}$ implies that $G^{-1}$ has zero modes, and is thus
excluded in the gRPA framework, cf. sec.\,\ref{sec_on_creat_annihil_op}. Thus,
for practical purposes, at least in Yang-Mills theory, the required
commutation relations resolve all ambiguities in the construction of
$\Gamma_{2B}$.}.  
\section{P-Q Formalism \label{sec_on_PQ_formalism}}
In sec. \ref{sec_on_normal_mode_form} we have learned that the usual
gRPA procedure encounters problems when the gRPA equations have zero
mode solutions; 
in sec. \ref{sec_Conserv_laws} we have learned that in the case we are
interested in - namely a mean-field vacuum that breaks certain
symmetries - zero modes appear necessarily. Therefore, it seems that we
may have a problem here. This problem, however, was already
encountered in nuclear physics, and solved in \cite{MW69}. 
We present the problem and its solution for the simplest case
available, namely the simple harmonic oscillator. We then give an
alternative formulation of the gRPA in QBA using - instead of
creation/annihilation operators - operators with canonical commutation
relations. This formulation will furthermore allow a simple comparison between
the gRPA equations derived from the time-dependent variational principle
and the operator approach. 
\subsection{Resolution of Zero-Mode Problem \label{morefunwithzeromodes}}
In the nuclear physics literature it is well-known \cite{MW69} that
the appearance of zero-modes, despite being welcome in our case,
leads to all sorts of problems, especially that the solutions of the
gRPA equations no longer
form a complete set\footnote{In nuclear physics this is a problem
since one there cannot prove that the Hamiltonian can be written as a
sum of $Q^{\dagger}_{\nu} Q_{\nu}$. Completeness is lost in the
following way: We see - in eq.\,(\ref{RPA-SHOadj}) -  that from every solution
of the gRPA
equations we can form another solution by considering the adjoint
operator, therefore the solutions to the gRPA equations always come in
pairs. In the case of zero modes, this is not necessarily true any more,
since now the 'excitation operators' that we find may be hermitian;
thus, by considering the adjoint, we do not get a new solution. In fact
this situation is the usual case for zero modes, as we have seen in
sec.\,\ref{sec_Conserv_laws} the excitation operators
for the zero modes are just the symmetry generators, which are usually
hermitian.
In our discussion in app.\,\ref{gRPA_comm_rel}, we hit upon the
problem from a different direction; 
we see that only if there are no zero modes in the spectrum we can
show that the excitation operators can be considered as independent
oscillators. 
}. Marshallek {\it et al.} \cite{MW69} however pointed out that 
the problems one has are actually an artifact of the
creation/annihilation operator formalism. We want to give a very primitive
illustration of their point: One can write down a quantum mechanical
oscillator in two equivalent forms (we have set the mass equal to one
for simplicity): 
\bea
H & = & \frac{p^2}{2} + \frac{\omega^2}{2} x^2 \\
  & = & \omega(a^{\dagger} a + \frac{1}{2}),
\eea
where $\omega$ denotes the frequency of the harmonic oscillator. If
we now send the frequency to zero, something seemingly strange happens:
\bea
 \frac{p^2}{2} + \frac{\omega^2}{2} x^2 & \stackrel{\omega \rightarrow
 0}{=}&  \frac{p^2}{2}, \\
 \omega(a^{\dagger} a + \frac{1}{2}) & \stackrel{\omega \rightarrow
 0}{=}& 0. 
\eea
If one now looks at the transformation from canonical to c/a
operators, one begins to see more clearly:
\bea
a =  \sqrt{\frac{\omega}{2}} x + \frac{i}{\sqrt{2 \omega}} p &
\stackrel{\omega  \rightarrow
 0}{=}& \frac{i}{\sqrt{2 \omega}} p, \\
a^{\dagger} =   \sqrt{\frac{\omega}{2}} x - \frac{i}{\sqrt{2 \omega}} p
 &\stackrel{\omega \rightarrow 0}{=}& -\frac{i}{\sqrt{2 \omega}} p 
\eea
and if we take this small-$\omega$ behavior into account, there is
actually no paradox\footnote{One should note that, in this limit $\omega
\rightarrow 0$, no new information can be obtained by considering the
adjoint of $a^{\dagger}$.}, since
\be
\frac{\omega}{2} a^{\dagger} a \stackrel{\omega \rightarrow 0}{=}
\frac{\omega}{2} (\frac{1}{\sqrt{\omega}})^2 p^2 = \frac{p^2}{2}.
\ee
This is actually the solution of the problems involving zero
modes: one puts the system into an external field, s.t. there are no
zero modes any more, since the external field\footnote{The external
field can in fact be engineered just to do that, and this works as
follows: We solve the gRPA equations and determine all modes
available. The operator belonging to the zero mode will be known
beforehand - it is just the quasi-boson approximation of the (known)
symmetry generator, which is usually a hermitian operator, and so will
be the operator associated with the zero mode.
One then constructs a canonically conjugate variable, s.t. $[q,p] = i$ and $q$
commutes with all other normal mode creation/annihilation operators, and adds a
term $\omega^2 q^2$ to the Hamiltonian; this will lift the zero mode
to a finite frequency. By this construction, only the zero mode will
be lifted and nothing strange will happen to the other modes.
Two points should be mentioned: First, the construction of a
canonically conjugate coordinate to a such a generator is only
possible in the gRPA context, since we have seen in
sec.\,\ref{sec_Conserv_laws} that the 
symmetry we have in mind is usually reduced to an Abelian symmetry; if we
would still have the non-Abelian symmetry this would not be possible
(the problems with the construction of 'angle operators' in nuclear
physics are well-known, for a recent discussion cf. e.g.
\cite{Torre:1998}). As a second point, one should mention that 
in a quantum field theory there might be problems with breaking gauge
symmetries in intermediate steps. Since we have indicated here only a
formal construction, it is not easy to see where these kinds of problems
might surface.}
 breaks all
symmetries. One diagonalizes the Hamiltonian into a set of harmonic
oscillators using canonical coordinates, and at the end sends the
external field to zero. Then the would-be zero modes indeed become zero modes
with the correct operator representation, namely $p^2$, and the
non-zero modes can be rewritten using creation and annihilation
operators. As a last comment on this topic, one should keep in mind the
following: since we already have some prejudice about the zero mode
operators (as the symmetry generators), we cannot just choose their
mass to be one as we did in the example. We will rather have to allow for a
mass tensor, and this will be dealt with in section
\ref{sec_mom_of_in}.  
\subsection{Equivalence of Operator and TDVP Approach to gRPA
\label{pqformulationofRPA}} 
We have seen that at least three formulations (using the tdvp, 
the two gRPA approximations, and the quasi-boson approximation) of the
generalized Random Phase Approximation of bosonic systems are available. 
In this
paragraph we want to add one more, since it simplifies the
treatment of those Hamiltonians we set out to consider,
cf. eq.\,(\ref{gRPAHamiltonian}). Since we start from the gRPA
formulation using quasi-boson operators, the whole treatment is only
valid for the mean-field vacuum representing a solution of the Rayleigh-Ritz
variational principle (at least w.r.t. variations of $\bar{\phi}$ and
$\bar{\pi}$). In this section we assume that the mean-field vacuum
is a solution of the Rayleigh-Ritz variational principle w.r.t. all
parameters $\bar{\phi}, \bar{\pi}, \Sigma, G$.
We consider linear combinations of $a, a^{\dagger}$ and $\BB, \BBd$,
s.t.
\be
\begin{array}{lclclcl}
q_m & = & \frac{1}{\sqrt{2}}(a^{\dagger}_m + a_m) & \hspace*{2em} &
p_m & = & \frac{i}{\sqrt{2}}(a^{\dagger}_m - a_m) \\
Q_{mi} & = & \frac{1}{\sqrt{2}}(\BBd_{mi} + \BB_{mi}) & \hspace*{2em} &
P_{mi} & = & \frac{i}{\sqrt{2}}(\BBd_{mi} - \BB_{mi}).
\end{array}
\ee
They fulfill the usual canonical commutation relations:
\be
[q_m, p_n] = i \delta_{nm} \hspace*{0.5em} ; \hspace*{0.5em}
[Q_{mi}, P_{nj}] = \frac{i}{2}(\delta_{mn} \delta_{ij} + \delta_{mj}
\delta_{ni})  \label{canon_comm}  
\ee
and all other commutators vanishing.  
If we now rewrite the Hamiltonian $H_B$ in terms of these new
operators, we obtain a complicated expression where the real and imaginary
parts of the matrices $A,...,F$ are separated. We can now put to use 
that we have computed the matrices $A,...,F$ in
app.\,\ref{app_expl_A-F} ; we have found that 
\begin{enumerate}
\item $A,C,D$ are always real
\item $C=-D$
\item $B,E,F$ may be imaginary; however, as this imaginary part is
due to $\Sigma$, and in the theories under consideration\footnote{This
excludes the cranking Hamiltonian $H_{cr} = H - \int d^3x \, 
\omega^a(\vec{x}) \Gamma^a(\vec{x})$ - where $H$ is the Yang-Mills
Hamiltonian and $\omega^a(\vec{x})$ is a Lagrange multiplier,
cf. e.g. \cite{Heinemann:2000ja} - here,  
since it is not of the form given in eq.\,(\ref{gRPAHamiltonian}).} at
the stationary point $\Sigma = 0$, the imaginary parts of $B,E,F$ vanish.
\item As a matter of fact, $E$ vanishes altogether at the stationary
point.
\item A last simplification is that one may decompose $B=
B^{11}+B^{22}$ and, using this notation, we have $A = - B^{22}$.
\end{enumerate}
Using this information, the Hamiltonian simplifies considerably:
\be
\begin{array}{ccccl}
H_B & = & E' & + & Q_{n_1 j_1} Q_{n_2 j_2} \frac{1}{4}(B_{n_1 j_1 n_2
j_2} - A_{n_1 j_1 n_2 j_2}) \vspace*{0.5em} \\ & & & + & P_{n_1 j_1}
P_{n_2 j_2} \frac{1}{4}(B_{n_1 j_1 n_2 j_2} + A_{n_1 j_1 n_2 j_2})
\vspace*{0.5em} \\ 
& & & + & \sqrt{2} Q_{n_1 j_1} q_{n_2} D_{n_1 j_1 n_2} \vspace*{0.5em} \\ 
& & & + & \frac{1}{2}
F_{n_2 n_1} (q_{n_2} q_{n_1} + p_{n_2} p_{n_1})
\end{array}
 \label{pqHamiltonianatstatpoint}
\ee
and can even be written in a nice matrix form:
\bea
H = E' &+& 
  \frac{1}{2}( P_{n_1 j_1} \hspace*{0.8em} p_{n_1} ) \left( \begin{array}{cc}
\frac{1}{2}(B_{n_1 j_1 n_2 j_2} + A_{n_1 j_1 n_2 j_2}) & 0 \\ 0 &
  F_{n_2 n_1} \end{array} \right) \left( \begin{array}{c} P_{n_2 j_2}
\\ p_{n_2} \end{array} \right) \nonumber \\
&+&  \frac{1}{2}( Q_{n_1 j_1} \hspace*{0.8em} q_{n_1} ) \left(
\begin{array}{cc} 
\frac{1}{2}(B_{n_1 j_1 n_2 j_2} - A_{n_1 j_1 n_2 j_2}) & \sqrt{2}
D_{n_1 j_1 n_2} \\ \sqrt{2} D_{n_2 j_2 n_1} & F_{n_2 n_1} \end{array}
\right) \left( \begin{array}{c} Q_{n_2 j_2} \\ q_{n_2} \end{array} \right),
\label{pqHamiltonian1}
\eea
where $E'$ is a constant that has to be chosen s.t. $\langle H_B
\rangle = E_{MF}$. 
This is a viable starting point for proving the equivalence between the
gRPA formulation originating from the time-dependent variational
principle and the one using the quasi-boson approximation.
We will rewrite eq.\,(\ref{pqHamiltonian1}) into a Hamiltonian that
leads to the same eigenvalue equations as does the Hamiltonian of the
small-fluctuation approach given in eq.\,(\ref{tdvpHamiltonian1}).
From this we will conclude - since we have both times the same matrix
to diagonalize - that the spectra of the Hamiltonians of the two
approaches are identical, and we will see a correspondence in the
eigenvectors.  
For this purpose we note the following:
\begin{enumerate}
\item We start with the term quadratic in $p,P$; there we have
\bea
& & \frac{1}{2}(B+A)_{n_1 j_1 n_2 j_2} \nonumber \\ & = & (\frac{1}{2} U^{-1}_{n_1 m_1} U^{-1}_{j_1
i_1}) (\frac{1}{2}  
U^{-1}_{n_2 m_2} U^{-1}_{j_2 i_2}) (G \unit)_{m_1 i_1; m_2 i_2}  
\eea
and
\be
F_{n_2 n_1} = (\frac{1}{\sqrt{2}} U^{-1}_{n_2 m_2}) (\frac{1}{\sqrt{2}}
U^{-1}_{n_1 m_1}) \delta_{m_2 m_1},  
\ee
where we have used the abbreviation 
\bea
(G \unit)_{ij;kl} = G_{ki}
\delta_{lj} +G_{kj} \delta_{li} + G_{jl} \delta_{ki} + G_{li}
\delta_{kj}. 
\eea
This suggests that one should introduce new momentum operators
\bea
{\bf P}_{m_1 i_1} & = &  \frac{1}{2} U^{-1}_{n_1 m_1} U^{-1}_{j_1 i_1}
P_{n_1 j_1}, \\ 
{\bf p}_{m_1}  & = & \frac{1}{\sqrt{2}} U^{-1}_{n_1 m_1} p_{n_1}.
\eea
\item Now we consider the term quadratic in $q,Q$; there the
observation 
\bea
B^{11}_{minj}
& = & 8 U_{m k_1} U_{i k_2} U_{n k_3} U_{j k_4}
\frac{\delta}{\delta G_{k_1 k_2}}\frac{\delta}{\delta G_{k_3 k_4}}
\langle T \rangle \\
B^{22}_{minj} - A_{minj}
& = & 8 U_{m k_1} U_{i k_2} U_{n k_3} U_{j k_4}
\frac{\delta}{\delta G_{k_1 k_2}}\frac{\delta}{\delta G_{k_3 k_4}}
\langle V \rangle 
\eea
(where $T$ denotes the kinetic energy $\frac{1}{2} \pi_i^2$) is
useful. Then one can rewrite 
\bea
(B-A)_{minj} 
& = & 8 U_{m k_1} U_{i k_2} U_{n k_3} U_{j k_4}
\frac{\delta}{\delta G_{k_1 k_2}}\frac{\delta}{\delta G_{k_3 k_4}}
\langle H \rangle.  \nonumber \\
\eea
Next, one can put to use the fact that the mean-field vacuum is a stationary
point (s.p.) and thus $\delta \langle H \rangle/ \delta G =0$, the identity
eq.\,(\ref{trade_phiphi_forG}) from app. \ref{app_pot} and the
fact that the kinetic energy $\langle \frac{1}{2} \pi_i^2 \rangle$ is
independent of $\bar{\phi}$: 
\bea
F_{n_2 n_1} & = & \frac{1}{2} G^{-1}_{n_2 n_1} \nonumber \\
& = & - 4 \left(\frac{\delta}{\delta G_{k_1 k_2}} \langle T \rangle \right)
U_{k_1 n_1} U_{k_2 n_2} \nonumber \\ 
& \stackrel{s.p.}{=} & 4 \left(\frac{\delta}{\delta G_{k_1 k_2}} \langle V 
\rangle \right) U_{k_1 n_1} U_{k_2 n_2} \nonumber \\
&\stackrel{eq.\,(\ref{trade_phiphi_forG})}{=}& 2 \left(\frac{\delta^2 }{\delta 
\bar{\phi}_{k_1} \delta \bar{\phi}_{k_2}} \langle V \rangle \right) U_{k_1 n_1}
U_{k_2 n_2} \nonumber \\
& = & 2 \left(\frac{\delta^2 }{\delta \bar{\phi}_{k_1}
\delta \bar{\phi}_{k_2}} \langle H \rangle \right) U_{k_1 n_1} U_{k_2 n_2}.
\eea
This now suggests introducing new coordinates as well:
\bea
{\bf Q}_{m_1 i_1} & = &  2 U_{n_1 m_1} U_{j_1 i_1} Q_{n_1 j_1}, \\
{\bf q}_{m_1}  & = & \sqrt{2} U_{n_1 m_1} q_{n_1}.
\eea
\item One should note that the newly defined coordinates and momenta
still fulfill the canonical commutation relations
eq.\,(\ref{canon_comm}); thus, the new definitions amount to a
canonical transformation that leaves the dynamics unchanged.
\end{enumerate}
We therefore end up with a Hamiltonian where the matrices in the
quadratic forms are identical to those appearing in the Hamiltonian of 
the small-fluctuation approach, eq.\,(\ref{tdvpHamiltonian1}),
\bea
H = E' &+& 
  \frac{1}{2}({\bf p} \hspace*{0.8em} {\bf P} ) 
\left( \begin{array}{cc}
\unit & 0 \\ 0 & (G \unit) \end{array} \right)
\left( \begin{array}{c} {\bf p} \\ {\bf P} \end{array} \right)
 \label{pqHamiltonian2} \\ 
&+&  \frac{1}{2}( {\bf q} \hspace*{0.8em} {\bf Q}) \left( \begin{array}{cc}
\frac{\delta^2 \mathcal{H}}{\delta \bar{\phi} \delta \bar{\phi}} &
\frac{\delta^2 \mathcal{H}}{\delta \bar{\phi} \delta G}
 \\\frac{\delta^2 \mathcal{H}}{\delta G \delta \bar{\phi}}  &
 \frac{\delta^2 \mathcal{H}}{\delta G \delta G}   \end{array} \right)
\left( \begin{array}{c}{\bf q} \\ {\bf Q} \end{array} \right), 
\nonumber 
\eea
since $\mathcal{H} = \langle H \rangle$ as defined in eqs.\,(\ref{cal_H}),
(\ref{tdvpRPA_smallfluct1}). As a last point, we want to show that the
eigenvalue equations that result from the Hamiltonian given in
eq.\,(\ref{pqHamiltonian2}) are identical to
eqs.\,(\ref{tdvp_final0}), (\ref{tdvp_final1}). This is in fact quite
simple; since the 
excitation operator $Q^{\dagger}_{B\,\nu}$ is a linear combination of
$\BB, \BBd, a, a^{\dagger}$ one can write it just as well as a linear
combination of $q, Q, p, P$ or ${\bf q}, {\bf Q}, {\bf p}, {\bf P}$. Thus,
we may write
\be
Q_{B\,\nu}^{\dagger} = \sum_{mi} (\tilde{Q}^{\nu}_{mi} {\bf Q}_{mi} +
\tilde{P}^{\nu}_{mi} {\bf P}_{mi}) +  \sum_{m} (\tilde{q}^{\nu}_{m}
{\bf q}_{m} + \tilde{p}^{\nu}_{m} {\bf p}_{m}). \label{pq_excit_ansatz}
\ee
We can now use the gRPA equations eq.\,(\ref{RPA-SHO}) to derive the
eigenvalue equations: we insert eq.\,(\ref{pq_excit_ansatz}) and
eq.\,(\ref{pqHamiltonian2}) into $[H_B, Q^{\dagger}_{B\,\nu}] =
\Omega_{\nu} Q^{\dagger}_{B\,\nu}$ and compare the coefficients of the
various operators ${\bf q}, {\bf p}, {\bf Q}, {\bf P}$. The resulting
equations are then easily expressed as
\be
\left( \begin{array}{cc} \frac{\delta^2 \mathcal{H}}{\delta
\bar{\phi} \delta \bar{\phi}} & \frac{\delta^2 \mathcal{H}}{\delta
\bar{\phi} \delta G} (G \unit) \\
\frac{\delta^2 \mathcal{H}}{\delta G \delta \bar{\phi}} &
\frac{\delta^2 \mathcal{H}}{\delta G \delta G} (G
\unit) \end{array} \right) \left(\begin{array}{c}
\tilde{q}^{\nu} \\ \frac{1}{2} \tilde{Q}^{\nu}_{s}
\end{array} \right)  = \Omega_{\nu}^2 \left(\begin{array}{c}
\tilde{q}^{\nu} \\ \frac{1}{2} \tilde{Q}^{\nu}_s
\end{array} \right)
\ee
and
\be
\left( \begin{array}{rr} \frac{\delta^2 \mathcal{H}}{\delta
\bar{\phi} \delta \bar{\phi}} & \frac{\delta^2 \mathcal{H}}{\delta
\bar{\phi} \delta G}  \\
(G \unit) \frac{\delta^2 \mathcal{H}}{\delta G \delta \bar{\phi}} & 
(G \unit) \frac{\delta^2 \mathcal{H}}{\delta G
\delta G}  \end{array} \right)
\left(\begin{array}{c} 
\tilde{p}^{\nu} \\ \frac{1}{2} \tilde{P}^{\nu}_{s}
\end{array} \right)  = \Omega_{\nu}^2 \left(\begin{array}{c}
\tilde{p}^{\nu} \\ \frac{1}{2} \tilde{P}^{\nu}_{s}
\end{array} \right),
\ee
where $\tilde{Q}^{\nu}_s, \tilde{P}^{\nu}_s$ denote the symmetric part
of $\tilde{Q}^{\nu}, \tilde{P}^{\nu}$ respectively, e.g.
$(\tilde{Q}^{\nu}_{s})_{ij} = \tilde{Q}^{\nu}_{ij} +
\tilde{Q}^{\nu}_{ji} = \tilde{Q}^{\nu}_{\{ij\}}$. 
These equations are now {\it identical} to eqs.\,(\ref{tdvp_final0}), 
(\ref{tdvp_final1}), 
only the components of the vectors have acquired {\it new names}: 
\be
\delta \bar{\phi}  \rightarrow \tilde{p}^{\nu}
\hspace*{0.5em};\hspace*{0.5em}  \delta G
\rightarrow \frac{1}{2} \tilde{P}^{\nu}_{s} \hspace*{0.5em}
\mbox{and} \hspace*{0.5em} 
\delta \bar{\pi}  \rightarrow \tilde{q}^{\nu}
\hspace*{0.5em};\hspace*{0.5em}  \delta \Sigma 
\rightarrow \frac{1}{2}\tilde{Q}^{\nu}_{s}.
\ee
Thus, we have proven that the two different approaches
to the generalized RPA, namely the small-fluctuation approach from the
time-dependent variational principle, and
the operator approach, give in fact the same spectrum of possible
excitations. Now the comments made at the end of
app.\,\ref{app_RPA_from_tdvp} carry over to the operator formulation of
gRPA: If the stability matrix of the mean-field problem is positive,
all 'eigenvalues' $\Omega^2_{\nu}$ are larger than zero, and thus all
$\Omega_{\nu}$s 
are real; if we are not at a minimum of the energy with our choice of
the mean-field vacuum, the stability matrix will also have negative
eigenvalues, and thus we will obtain complex conjugate pairs $\pm
i|\Omega_{\nu}|$. As long as we have real 'eigenvalues' $\Omega_{\nu}$
in the gRPA problem, the mean-field vacuum under consideration will be
stable at least w.r.t. small fluctuations. One should note that this
relation was of some importance in proving the bosonic commutation
relations of the excitation operators.
\section{The Moment of Inertia \label{sec_mom_of_in}}
\subsection{General Form of Kinetic Energy}
From the discussion in sec.\,\ref{morefunwithzeromodes} we saw that we
can in general write the Hamiltonian in quasi-boson approximation as
\be
H_B = E_{\text{RPA}} + \sum_{\nu \in \nu_{+}} \Omega_{\nu}
Q^{\dagger}_{\nu} Q_{\nu} + \sum_{\nu \in \nu_0} \frac{1}{2} P^2_{\nu},
\label{RPA_gen_diag_HB_incl_zero_modes} 
\ee
where $\{\nu_+\}$ denotes the set of modes with positive $\Omega_{\nu}$
and $\{\nu_0\}$ denotes the set of zero modes. 
However, quite often an alternative expression to
eq.\,(\ref{RPA_gen_diag_HB_incl_zero_modes}) in terms of symmetry
generators is useful.
For this we have to recall from sec.\,\ref{sec_Conserv_laws} that if our
reference state is a solution of the Rayleigh-Ritz equations, and if the
symmetry generators under consideration are one-body operators, then the
quasi-boson approximations $\Gamma^a_B$ of the symmetry generators commute
with the quasi-boson 
approximation of the Hamiltonian. This implies that they are zero mode
solutions of the gRPA equations. If we now assume that there are no
'accidental zero modes', i.e. the whole space of zero modes is spanned
by the $\Gamma^a_B$s, then we can write the $P_{\nu}$s, $\nu \in
\{\nu_0\}$, as linear combinations of the $\Gamma^a_B$s, and
$\sum_{\nu \in \{ \nu_0\}} P^2_{\nu}$ becomes a quadratic form in
terms of the $\Gamma^a_B$s. The advantage of this alternative
expression is two-fold: First, in general the quasi-boson
approximations of the symmetry operators are known. Second, the
expression one obtains is (physically) more transparent, and can be more
easily compared to other frameworks like e.g. the Thouless-Valatin method
\cite{Heinemann:1998cx, Heinemann:2000ja} or the Kamlah expansion
\cite{Schroeder:2002b}. The general dependence of $H_B$ on the generators
$\Gamma_B^a$ will then be\footnote{$H_{B,zm}$ means
'that part of the Hamiltonian in quasi-boson approximation that only
contains the {\it z}ero {\it m}ode operators'.}
\be
H_{B,zm} = \frac{1}{2} \Gamma_B^a (\mathcal{M}^{-1})^{ab} \Gamma_B^b, 
\ee
where $\mathcal{M}$ is the {\it moment-of-inertia} tensor, and Einstein's
summation convention is also used for the indices $a,b$ which are employed to
label the different symmetry generators $\Gamma_B^a$. We cannot (as we
have done in preceding discussions where masses were set to 1)
'normalize' $\mathcal{M}$ 'away', since the normalization of $\Gamma_B^a$ is
fixed by its very nature of a known operator. Furthermore, if the original
symmetry generators (before the quasi-boson approximation) are
generators of a non-Abelian symmetry, their normalization is fixed by
the commutation relations. Thus, in this section, we will 
see how one can actually compute the moment-of-inertia tensor,
cf. \cite{Ring:1980}. 
We have seen in sec.\,\ref{morefunwithzeromodes} that in cases of
appearances of zero modes one has to pass to a description of the
oscillators in terms of canonical coordinates\footnote{Let us once
again mention that the reduction of the non-Abelian to an Abelian
symmetry (in the Yang-Mills case) in the quasi-boson approximation
seems to be essential since otherwise one cannot construct canonically
conjugate 'angle' operators fulfilling the canonical commutation
relations \cite{Torre:1998}. } and momenta (which we
have done above). This actually allows to determine the
moment-of-inertia tensor. 
We assume that we can construct a set of
coordinates $\Theta_B^a$, s.t.
\be
 \ [\Theta_B^a, \Gamma_B^b ] = i \delta^{ab}, \label{Thetadet1}
\ee
and which commute with all other normal modes. Since we already know
the form of the part of $H_B$ that is not supposed to commute with $\Theta_B$,
we also require\footnote{We assume here that the moment-of-inertia
tensor is symmetric, which is at least true for the case of Yang-Mills
theory, since there $[\Gamma^a_B, \Gamma^b_B] = 0$.} 
\be
 \ [\Theta_B^a, H_B ] = i \mathcal{M}^{ab} \Gamma_B^b. \label{Thetadet2}
\ee
We start by making an {\it ansatz} of
$\Theta_B^a$ as a generalized one-body operator in quasi-boson
approximation\footnote{The usage of the p-q formulation as in section\,
\ref{pqformulationofRPA} will be useful in this context.}:
\be
\Theta_B^a = \Theta^a_{Q;mi} Q_{mi} + \Theta^a_{P;mi} P_{mi}
+\Theta^a_{q;m} q_{m} + \Theta^a_{p;m} p_{m}   \label{Thetadefgeneral}
\ee 
whereas in general $\Gamma_B$ reads
\be
\Gamma_B^a = \Gamma^a_{Q;mi} Q_{mi} + \Gamma^a_{P;mi} P_{mi}
+\Gamma^a_{q;m} q_{m} + \Gamma^a_{p;m} p_{m} \label{Gammadefgeneral}.
\ee 
In the latter case the coefficients are obviously known.
We now insert eqs.\,(\ref{Thetadefgeneral}), (\ref{Gammadefgeneral}) into
eq.\,(\ref{Thetadet2}), and compare coefficients. This gives in general
four equations, and allows to determine $\Theta^a_{Q;mi},
\Theta^a_{P;mi}, \Theta^a_{q;m}, \Theta^a_{p;m}$ in terms of $\mathcal{
M}^{-1}$ and the coefficients of $\Gamma_B$. If we now insert the thus
determined coefficients of $\Theta_B$ into the equation  that
results from eq.\,(\ref{Thetadet1}) we obtain an equation of the type
\be
(\mathcal{M}^{-1})^{ac} \mathcal{N}^{cb} = i \delta^{ab},
\ee
where $\mathcal{N}$ is a matrix given entirely in terms of the matrices
$A,...,F$ and the coefficients of $\Gamma_B$. Thus, one has to
invert the matrix $\mathcal{N}$ in order to obtain the correct kinetic
term for the zero modes. One should note that the logic is just the other way
around from what one would expect, namely that $\Theta_B$ is defined via
its commutator with $\Gamma_B$. However, for a proper definition of
$\Theta_B$ we also need the fact that it commutes with all the other
normal modes. 
These general considerations find their application to the specific
case of Yang-Mills theory in secs. \ref{subsec_YMT},
\ref{subsec_explicit_calc} below. 
\subsection{Energy Contributions of the Zero Modes}
We have discussed in some detail that zero modes cannot be treated via
the ordinary creation/annihilation operator formalism; they have to be
treated with the help of canonical coordinates and momenta. This
changes their contribution to the gRPA vacuum energy. The gRPA energy
in eq.\,(\ref{HBdiagonal}) was fixed s.t. 
\be
\langle H_B \rangle = E_{MF}.
\ee
Since now $H_B$ no longer contains oscillator modes only but rather
has the form 
\be
H_B = E_{{\rm RPA}} + \frac{1}{2} \Gamma_B^a (\mathcal{M}^{-1})^{ab}
\Gamma_B^b + \sum_{\nu \in \{\nu_+\}} \Omega^{\nu}
Q^{\dagger}_{\nu} Q_{\nu}, 
\ee
where $\{\nu_+\}$ denotes the set of modes with positive
$\Omega_{\nu}$, eq.\,(\ref{firstRPAenergy}) 
becomes in this context
\bea
E_{{\rm RPA}} &=& E_{{\rm MF}} -  \frac{1}{2}  
\langle \Gamma_B^a  (\mathcal{M}^{-1})^{ab} \Gamma_B^b \rangle 
\\ &-&
\sum_{\nu \in \{\nu_+\} } 
\Omega_{\nu}    \left(\frac{1}{2} \sum_{mi}
|\frac{1}{2}Y^{\nu}_{\{mi\}}|^2 + \sum_i |Z^{\nu}_i|^2\right), \nonumber
\eea
which is again similar to the result obtained in nuclear physics
\cite{Ring:1980}. In the special case of Yang-Mills theory to leading
order in perturbation theory,
we obtain an energy correction (see below) to the mean-field energy due to the
zero modes\footnote{In this context, one should note that $\langle
\Gamma^a_B \Gamma^b_B \rangle = \langle \Gamma^a \Gamma^b \rangle$.}  
\be
\Delta E = \frac{1}{2} G^{ab}_{\Delta}({\bf x},{\bf y}) \langle
\Gamma^a({\bf x}) \Gamma^b({\bf y}) \rangle
\ee
that is essentially identical in form
to what one obtains in a second order Kamlah expansion \cite{Schroeder:2002b}
and also to the Thouless-Valatin correction proposed in 
\cite{Heinemann:1998cx, Heinemann:2000ja}, but 
also with an important difference: the energy contribution which is
due to the zero modes (and thus ultimately due to the deformation) is
subtracted  off {\it after variation} of the mean-field vacuum wave
functional, and not before, as in the cases of \cite{Schroeder:2002b,
Heinemann:1998cx, Heinemann:2000ja}; therefore, the determination of the  
parameters of the mean-field vacuum is \textit{not influenced} by the
correction. 
\subsection{Special Considerations for Yang-Mills Theory \label{subsec_YMT}}
We have repeatedly emphasized that the quasi-boson formulation works
properly only if the reference state fulfills the Rayleigh-Ritz
equations (the parameters are such that 'the energy functional is at
its stationary point'). In Yang-Mills theory without a cranking
term\footnote{As mentioned before the cranking Hamiltonian differs from the
  ordinary Yang-Mills Hamiltonian by the term $\int d^3x\,\omega^a(\vec{x})
  \Gamma^{a}(\vec{x})$, where $\omega^a(\vec{x})$ is a Lagrange multiplier.}
this leads automatically to
\be
\bar{\bm{\pi}} = 0 \hspace*{1em} \mbox{and} \hspace*{1em} \Sigma = 0.
\ee
This simplifies the expression for the Gauss law operator
considerably, since, when we insert these results into the expression
given in App. \ref{app_Gauss_law}, we
obtain\footnote{\label{footnoteOnNoMoreSuperIndex} In this
section we have to give up the super-index notation used until now in
some places;
instead of one super-index, the operators $p$ will carry three indices
(color, spatial, and position), i.e. $p_m \rightarrow p^b_i({\bf x})$
where $b$ is the color, $i$ the spatial, and ${\bf x}$ the position index,
and correspondingly for $P_{mi}$. In this context, also the
index $a$ carried by $\Gamma^a_B$ has to be re-examined; in fact it
is also a super-index, consisting of a color index $a$ and a position
index ${\bf x}$: $\Gamma_B^a \rightarrow \Gamma^a_B({\bf x})$; the
same applies to the 
super-indices that are carried by the moment-of-inertia tensor: $\mathcal{
M}^{ab} \rightarrow \mathcal{M}^{ab}({\bf x},{\bf y})$. Having
clarified this, 
in the more formal parts of this section, we will still stick to the
super-index notation, since otherwise the formulas will become unreadable.} 
\bea
\Gamma_B^a &=& \Gamma^a_{P,mi} P_{mi} + \Gamma^a_{p,m} p_{m} \nonumber  \\
&=& (\Gamma^a_{P}({\bf x}))^{bc}_{ij}({\bf x}_1,{\bf x}_2)
P^{bc}_{ij}({\bf x}_1,{\bf x}_2) + 
(\Gamma^a_{p}({\bf x}))^{b}_{i}({\bf x}_1) p^{b}_{i}({\bf x}_1)
\eea
with
\bea
(\Gamma^a_{P}({\bf x}))^{b_1 b_2}_{n_1 n_2}({\bf z}_1,{\bf z}_2) & = &
\frac{1}{2} \left( \frac{\delta}{\delta 
\Sigma^{a_1 a_2,l_1 l_2}_{{\bf x}_1 {\bf x}_2}} \langle \hat{\bf
D}^{ab}_{{\bf x},i} 
{\bf \Pi}^{b}_{i {\bf x}} \rangle   \right) \nonumber \\
&\times&(U^{-1})^{a_1 b_1, l_1 n_1}_{{\bf x}_1 {\bf z}_1}
(U^{-1})^{a_2 b_2, l_2 n_2}_{{\bf x}_2 {\bf z}_2}
\hspace*{2em}\label{expr_Gamma_P}\\ 
(\Gamma^a_{p}({\bf x}))^{b_1}_{n_1}({\bf z}_1) & = & \frac{1}{\sqrt{2}} \left(
\frac{\delta}{\delta \bar{\bm{\pi}}^{a_1}_{l_1 {\bf y}_1}} \langle
\hat{\bf D}^{ab}_{{\bf x},i} {\bf \Pi}^{b}_{i {\bf x}} \rangle
\right) \nonumber \\ &\times&
(U^{-1})^{a_1 b_1,l_1 n_1}_{{\bf y}_1 {\bf
z}_1}, \label{expr_Gamma_p} 
\eea
where we don't integrate over ${\bf x}$. Further remarks on notation
can be found in footnote\,\ref{footnoteOnNoMoreSuperIndex}.
One should note that all double indices are summed over {\it except
for {\bf x}} ! Inserting eq.\,(\ref{Thetadefgeneral}) into
eq.\,(\ref{Thetadet2}), 
with the Hamiltonian given in eq.\,(\ref{pqHamiltonianatstatpoint}), we
obtain (apart from $\Theta_P = \Theta_p = 0$)  
\be\begin{array}{lcl}
\Theta^a_{q,m} F_{mi} & = & (\mathcal{M}^{-1})^{ab} \Gamma^b_{p,i}, \\
\Theta^a_{Q,mi} B^{11}_{minj} & = & (\mathcal{M}^{-1})^{ab}
\Gamma^b_{P,nj},
\end{array}
\ee
where we have used the properties of the stationary point,
i.e. the matrices $A,...,F$ are all real,
etc. [cf. sec.\,\ref{pqformulationofRPA} and
eqs.\,(\ref{eq_A_at_stationary_point}) - 
(\ref{eq_F_at_stationary_point})].  
We invert the second line in a way that will be discussed below in
some detail. Let us only mention at this point that $(B^{11})^{-1}$
is defined by
\bea
&& ((B^{11})^{-1})_{i_1 m_1 j_1 n_1} (B^{11})_{j_1 n_1 i_2 m_2}
\nonumber \\ &=&
\frac{1}{2} ( \delta_{i_1 i_2} \delta_{m_1 m_2} + \delta_{i_1 m_2}
\delta_{m_1 i_2}). \label{B11inv1}
\eea
We insert the result into the normalization eq.\,(\ref{Thetadet1})
and obtain
\bea
(\mathcal{M}^{-1})^{ac} \Big( \Gamma^c_{p,m} (F^{-1})_{mi}
\Gamma^b_{p,i} + 2\Gamma^c_{P,mi}
((B^{11})^{-1})_{minj} \Gamma^b_{P,nj} \Big) = \delta^{ab},
\label{eq_for_momofin}  
\eea
which bears some resemblance to the expression obtained in nuclear physics
\cite{Ring:1980}, especially if one notes that $(B^{11})_{minj}
\propto  (A+B)_{minj}$ and $F_{ij} \propto (E+F)_{ij}$ at the
stationary point.
\subsection{Explicit Calculations \label{subsec_explicit_calc}}
Unfortunately, we cannot read off an explicit expression for $(\mathcal{
M}^{-1})^{ab}$ from this in general. However, we 
can do two things: \textit{First}, we can try to evaluate the terms
\be
\Gamma^c_{p,m} (F^{-1})_{mi} \Gamma^b_{p,i} \text{ and }
2\Gamma^c_{P,mi} ((B^{11})^{-1})_{minj} \Gamma^b_{P,nj}  
\ee
further. This will provide us with the moment of inertia - though for
the energy correction due to zero modes we will need its inverse and
this cannot be given in a general form. \textit{Second}, we can go
back to perturbation theory, where we in fact can give the inverse
for the leading and next-to leading part in $g^2$ for the moment of
inertia. 
\subsubsection{$\Gamma^c_{p,m} (F^{-1})_{mi} \Gamma^b_{p,i}$}
It is a simple exercise to take $\Gamma^b_{p,i}$ and $(F^{-1})_{mi}$
to calculate
\be
(\mathcal{M}^{-1})^{ac}({\bf x}, {\bf y}) \left(  \left(\Gamma^c_{p}({\bf
y})\right)^{b_1}_{n_1}({\bf z}_1)  (F^{-1})^{b_1 b_2}_{n_1 n_2}({\bf
z}_1, {\bf z}_2) (\Gamma^b_{p}({\bf z}))^{b_2}_{n_2}({\bf
z}_2) \right) =  - ({\hat{\bar{\bf D}}} {\hat{\bar{\bf D}}})^{bc}({\bf
  z}) (\mathcal{M}^{-1})^{ca}({\bf z}, {\bf x}). 
\ee
\subsubsection{$2\Gamma^c_{P,mi} ((B^{11})^{-1})_{minj} \Gamma^b_{P,nj}$}
The evaluation of this expression needs a bit more thought.
The first step is to give a practical expression for
$((B^{11})^{-1})_{minj}$ beyond its implicit definition in
eq.\,(\ref{B11inv1}). For this purpose, it is useful to rewrite
$G^{-1}$, cf. eq.\,(\ref{RPA_most_general_TI_Gauss}), as 
\be
G^{-1}_{ij} = \sum_A \lambda_A \PP^A_{ij}
\ee
where $\lambda_A$ are the eigenvalues of $G^{-1}$ and $\PP^A$ are the
projectors onto the corresponding eigenspaces. The projectors are
complete in the sense that
\be
\sum_A \PP^A_{ij} = \delta_{ij}.
\ee
Then
\bea
B^{11}_{i_1 i_2 m_1 m_2} &=& \frac{1}{2} 
(G^{-1}_{i_1 m_1} \delta_{i_2 m_2} + 
 G^{-1}_{i_1 m_2} \delta_{i_2 m_1} + 
 G^{-1}_{i_2 m_1} \delta_{i_1 m_2} + 
 G^{-1}_{i_2 m_2} \delta_{i_1 m_1}) \\
&=&  \frac{1}{2} \sum_{A,B} (\lambda_A + \lambda_B) (\PP^{A}_{i_1 m_1}
\PP^B_{i_2 m_2} + \PP^{A}_{i_1 m_2} \PP^B_{i_2 m_1}).  
\eea
Using this expression for $B^{11}$ it is simple to verify that 
\be
((B^{11})^{-1})_{i_1 i_2 m_1 m_2} = \frac{1}{2} \sum_{A,B}
\frac{1}{\lambda_A + \lambda_B} (\PP^{A}_{i_1 m_1}
\PP^B_{i_2 m_2} + \PP^{A}_{i_1 m_2} \PP^B_{i_2 m_1}).
\ee
In order to give a compact expression, using super-indices, for
$\Gamma^b_{P,nj}$ we introduce $\hat{T}^a_{\vec{x}}$: with $n, m$ being
super-indices, where $m$ stands for $(\vec{x}_1, b, i)$ and $n$ stands for
$(\vec{x}_2, c, j)$ , we define (no integration over $\vec{x}$ !) 
\be
(\hat{T}^a)^{bc} \delta_{\vec{x}_1 \vec{x}} \delta_{\vec{x} \vec{x}_2}
\delta_{ij} = (\hat{T}^a_{\vec{x}})_{mn},
\ee
which inherits the property of anti-symmetry from $(\hat{T}^a)^{bc} = f^{bac}$
where $f^{bac}$ are the $SU(N)$ structure constants. 
Then we have
\be
(\Gamma^a_P(\vec{x}))_{n_1 n_2} = \frac{g}{2}((U \hat{T}^a_{\vec{x}}
U^{-1})_{n_2 n_1} + (U \hat{T}^a_{\vec{x}} U^{-1})_{n_1 n_2}).
\ee
This expression can now be cast in the same form as $(B^{11})^{-1}$ by
using the eigenvalue decomposition:
\be
U^{-1}_{ij} = \sum_A \sqrt{\lambda_A} \PP^A_{ij} \text{  and  } U_{ij} =
\sum_A \frac{1}{\sqrt{\lambda_A}} \PP^A_{ij},
\ee
namely one obtains
\be
(\Gamma^a_{P}(\vec{x}))_{n_1 n_2} = \frac{g}{2} \sum_{A,B} \frac{\lambda_B -
\lambda_A}{\sqrt{\lambda_A \lambda_B}} (\PP^A \hat{T}^a_{\vec{x}}
\PP^B)_{n_2 n_1}. 
\label{eq_eigenval_expression_for_GammaP}
\ee
Thus, we obtain overall
\be
2 (\Gamma^c_P(\vec{x}))_{i_1 i_2} ((B^{11})^{-1})_{i_1 i_2 m_1 m_2}
(\Gamma^b_P(\vec{y}))_{m_1 m_2} = - \frac{g^2}{2} \sum_{A,B} \left[
\frac{(\lambda_B - \lambda_A)^2}{\lambda_A \lambda_B (\lambda_A +
\lambda_B)} \Tr{(\PP^A \hat{T}^c_{\vec{x}} \PP^B \hat{T}^b_{\vec{y}})}
\right]. \label{eq_2nd_term_moi} 
\ee
Of course, the moment-of-inertia calculations ultimately have the goal
to be connected to the static quark potential. It turns out to be
interesting to compare the structure presented in
eq.\,(\ref{eq_2nd_term_moi}) to the structures one obtains in a
perturbative Coulomb gauge investigation into the static quark
potential \cite{Lee:1981}.
\subsubsection{A Short Excursion into Coulomb Gauge Perturbation Theory}
In Coulomb gauge, the charge density consists of two parts, namely the
external part $\rho^a_{ext}$ and the gauge part $\rho^a_{gauge} =
f^{abc} \vec{A}^b_i (\vec{\Pi}^{tr})^c_i$, where $\vec{\Pi}^{tr}$
denotes the transversal components of the momentum (the longitudinal
component is eliminated using the Gauss law constraint). We then
decompose the Coulomb gauge Hamiltonian into a free part, independent
of the coupling, and a rest. In Coulomb gauge, the Hamiltonian
contains arbitrary powers of $g$. The free part (i.e. that part of the
Hamiltonian which is left if we set $g = 0$) defines what in the
following is to be called \textit{gluons}. Since we are in Coulomb
gauge these are given in terms of the transversal fields and the
corresponding momenta. \\
In order to compute the energy corrections to the energy of the
perturbative ground state by the presence of external charges to
$\mathcal{O}(g^4)$ in Rayleigh-Schr\"odinger perturbation theory, we need on
the one hand the $\mathcal{O}(g^4)$ contribution to the interaction
Hamiltonian in order to compute 
\be
\langle 0 | H_{int} | 0 \rangle
\ee
and on the other hand the  $\mathcal{O}(g^2)$ contribution to the
interaction Hamiltonian in order to compute
\be
\sum_{N \neq 0} \frac{| \langle 0 | H_{int} | N \rangle |^2}{-E^0_N} 
\label{eq_sec_order_contrib}
\ee
where the sum runs over all states except for the (perturbative)
vacuum, and $-E^0_N$ is the energy of this state (the energy of the
perturbative vacuum has been set to zero). Since we are interested
only in that part of the interaction energy that is proportional to
$\rho_{ext}(\vec{x}_1) 
\times \rho_{ext} (\vec{x}_2)$, we drop everything from $H_{int}$
which is not quadratic in $\rho_{ext}$ for the first term, and not linear
in $\rho_{ext}$ for the second. Since the term of the Hamiltonian
which contains $\rho$ \textit{at all} is
\be
\rho^a \left(\frac{1}{\vec{D}\cdot \vec{\nabla}} \Delta
\frac{1}{\vec{D}\cdot \vec{\nabla}} \right)^{ab} \rho^b = \rho^a
\mathcal{O}^{ab} \rho^b
\ee 
(which consequently has to be expanded in powers of $g$), for the
second term we need $\rho^a_{ext} (\mathcal{O}^{ab} +
\mathcal{O}^{ba}) \rho^b_{gauge}$. 
Whereas the terms stemming from the first order contribution give the
anti-screening part of the one-loop $\beta$-function, the second order
contribution gives the \textit{screening} part of the one-loop
$\beta$-function (which is given by $-\frac{1}{12}$ times the anti-screening
part). We will go into a little bit
more detail of this latter contribution. What goes really into the
computation of eq.\,(\ref{eq_sec_order_contrib}) is
\be
\langle 0 | \rho^a_{gauge} | N \rangle = f^{abc} \langle 0 |
\vec{A}^b_i (\vec{\Pi}^{tr}_i)^c | N \rangle.
\ee
Using the representation in terms of creation and annihilation
operators it becomes clear that the matrix element $\langle 0 |
\rho^a_{gauge} | N \rangle$ is non-zero only if $|N\rangle$ contains two
gluons. If the gluons carry energy 
$\omega_{k_1}, \omega_{k_2}$ (and color index $m_1, m_2$) one obtains
\be
\langle N |\rho^a_{gauge} |0 \rangle \propto f^{a m_1 m_2}
\frac{\omega_{k_2} - \omega_{k_1}}{\sqrt{\omega_{k_1} \omega_{k_2}}}
\ee
The comparison to eq.\,(\ref{eq_eigenval_expression_for_GammaP}) is
striking, especially if one takes into account that in the case of
absence of condensates the eigenvalues of $G^{-1}$ can be interpreted
as single-particle energies, cf. app. \ref{app_RPA_from_tdvp}. The
second quantity needed is 
$1/E^0_N$; $E^0_N$ is obviously the sum of the energies of the
two gluons, i.e. $E^0_N = \omega_{k_1} + \omega_{k_2}$, and therefore
the similarity of $1/E^0_N$ to $(B^{11})^{-1}$ is
obvious, cf. eq.\,(\ref{eq_2nd_term_moi}). Therefore, it appears that the
``quantum part'' of the moment 
of inertia contains only the (albeit generalized) screening component
of the static interaction potential, and the (obviously dominant)
anti-screening component has to be found elsewhere. This has to be
subject to further investigations.
\subsubsection{Evaluation to Leading Order in Perturbation Theory}
After this excursion, we go back to perturbation theory to make an
evaluation of our expressions for the two components needed for the
moment of inertia. We note\footnote{This can be
read off the expressions 
eqs.\,(\ref{expr_Gamma_P}), (\ref{expr_Gamma_p}) together with the
mean-field expectation value of $\Gamma^a({\bf x})$ given in 
eq.\,(\ref{eq_Gauss_law_c-a-decomp}).} that $\Gamma^a_P$ is, in the
perturbative scaling scheme given in app. \,\ref{App_scaling}, of
higher order in $g$ than\footnote{For the purpose of counting powers
of $g$, we take $\bar{\bf A}$ to be of $\mathcal{O}(g^{-1})$,
s.t. $\hat{\bar{\bf D}}$ is completely of $\mathcal{O}(g^0)$.}
$\Gamma^a_p$. Thus, we just compute the leading order piece of the
moment of inertia: 
\bea
(\mathcal{M}^{-1})^{ac}({\bf x}, {\bf y}) \left(  \left(\Gamma^c_{p}({\bf
y})\right)^{b_1}_{n_1}({\bf z}_1)  (F^{-1})^{b_1 b_2}_{n_1 n_2}({\bf
z}_1, {\bf z}_2) (\Gamma^b_{p}({\bf z}))^{b_2}_{n_2}({\bf
z}_2) \right) &=&\nonumber \\  - ({\hat{\bar{\bf D}}} {\hat{\bar{\bf
D}}})^{bc}({\bf z}) (\mathcal{M}^{-1})^{ca}({\bf z}, {\bf x})
&\stackrel{!}{=}& \delta^{ba} \delta_{{\bf x} {\bf z}}.  
\label{eq_leading_order_mom_of_in}
\eea
From this we conclude that the leading order piece of the moment of
inertia in a perturbative expansion is just the Green's function of the
covariant Laplacian in the background field ${\bar{\bf A}}$:
\be \left(\mathcal{M}^{-1} \right)^{ab} ({\bf x},{\bf y}) = -
  G^{ab}_{\Delta}({\bf x},{\bf y}) + \mathcal{O}(g^2)
\ee
with
\be
\hat{\bar{\mathbf{D}}}{}^{ad}_{\mathbf{x},l}
\hat{\bar{\mathbf{D}}}{}^{db}_{\mathbf{x},l}  
G^{bc}_{\Delta}(\mathbf{x},\mathbf{y}) = \delta^{ac} \delta_{\mathbf{xy}}.
\ee
Obviously, in the case of a vanishing background field, we obtain 
Coulomb's law:
\be
\left(\mathcal{M}^{-1} \right)^{ab}({\bf x}, {\bf y}) = 
  \frac{ \delta^{ab}}{4 \pi} \frac{1}{|{\bf x}-{\bf y}|}. 
\ee
Let us now consider the perturbative evaluation of
eq.\,(\ref{eq_2nd_term_moi}). If we solve the mean-field equation in a
perturbative approximation (cf. e.g. \cite{Kerman:1989kb,
Heinemann:2000ja, schroedo:2002a}) for vanishing $\bar{\vec{A}}$, we
obtain for $G^{-1}$ 
\be
(G^{-1})^{ab}_{ij}(\vec{x}, \vec{y}) = 2 \delta^{ab} \int
\frac{d^3k}{(2\pi)^3} e^{i \vec{k} \cdot (\vec{x} - \vec{y})}
|\vec{k}| (\delta_{ij} - \frac{\vec{k}_i \vec{k}_j}{\vec{k}^2}).
\ee
This shows immediately that we have a problem of building a gRPA
treatment on top of the perturbative ground state, since $G^{-1}$ has
zero modes which is (as we have argued before) unacceptable for the
gRPA formalism. Since this section has mainly illustrative purposes, we
proceed by simply ignoring the zero modes in what 
follows. The non-zero eigenvalues of $G^{-1}$ are given by $
\lambda_{2 |\vec{q}|} = 2 |\vec{q}|$, the corresponding projector onto
this eigenspace is given by 
\be
(\PP^{2 | \vec{k}|})^{ab}_{ij} (\vec{x}, \vec{y}) = \int
\frac{|\vec{k}|^2 d \Omega_{\vec{k}}}{(2 \pi)^3} \delta^{ab}
(\delta_{ij} - \frac{\vec{k}_i \vec{k}_j}{\vec{k}^2}) e^{i
\vec{k}\cdot(\vec{x} - \vec{y})}, \label{eq_projector_PT}
\ee
where $d \Omega_{\vec{k}}$ denotes the solid angle in $\vec{k}$-space.
Inserting these expressions for $\lambda_{2 | \vec{k}|}$ and $(\PP^{2
| \vec{k}|})^{ab}_{ij} $ into eq.\,(\ref{eq_2nd_term_moi}), we obtain 
\bea
2 (\Gamma^c_P(\vec{x}))_{i_1 i_2} ((B^{11})^{-1})_{i_1 i_2 m_1 m_2}
(\Gamma^b_P(\vec{y}))_{m_1 m_2} &=& \frac{N g^2}{48 \pi^2}
\ln{(\Lambda^2_{UV}/\Lambda^2_{IR})} \int \frac{d^3p}{(2\pi)^3}
\vec{p}^2 e^{i \vec{p} \cdot (\vec{x} - \vec{y})} \delta^{bc}
\nonumber \\ & & \hspace*{3em} + \text{
finite terms}
\eea
where finite terms are not determined explicitly. $\Lambda_{IR}$ is an
IR-cutoff 
needed in the calculation due to the primitive way used to evaluate
the loop integral. The main point is here, however, that the ``quantum
part'' of the moment of inertia is just the classical part, multiplied
by a divergent constant. Therefore, abbreviating $\frac{N g^2}{48 \pi^2}
\ln{(\Lambda^2_{UV}/\Lambda^2_{IR})}$ by $\alpha$, we can now rewrite
eq.\,(\ref{eq_for_momofin}) as 
 \bea
(- \Delta_{\vec{z}} (\mathcal{M}^{-1})^{ab}(\vec{x}, \vec{z})) ( 1 +
 \alpha) = \delta^{ab} \delta_{\vec{x} \vec{z}},
\eea
where $\Delta_{\vec{z}}$ is the Laplacian w.r.t. the coordinates $\vec{z}$.
Thus,
\be
(\mathcal{M}^{-1})^{ab}(\vec{x}, \vec{z}) = \frac{\delta^{ab}}{1 +
\alpha} \frac{1}{4 \pi} \frac{1}{| \vec{x} - \vec{z}|}.
\ee
If we introduce a renormalized coupling in the standard fashion
\cite{Kerman:1989kb} via
\be
\frac{1}{g^2_R(\mu)} = \frac{1}{g^2} + \beta_0 \ln{(\Lambda^2_{UV}/\mu^2)}
\ee
the expression for $\mathcal{M}$ is made finite only for $\beta_0 =
\alpha$, thereby corroborating our earlier claim that the moment of
inertia can only account for the screening contribution of the static
quark potential. \\
It is interesting to note that the moment of inertia obtained here is
the same as in a cranking type calculation
\cite{Heinemann:2000ja}. This also could have been anticipated, as it
is well-known in nuclear physics that the moments of inertia
determined from RPA and those from cranking are in fact identical in
nuclear physics, cf. \cite{MW69}.
\subsection{Interpretation of Moment of Inertia as Static Quark
Potential \label{sec_interpret_mom_of_in}}
We have to conclude that we \textit{cannot interpret} the moment of
inertia as the static quark potential directly. However, this is not
necessarily fatal for a gRPA treatment of Yang-Mills theory. Due to
the approximations made to the Gauss law operator (e.g. making it
Abelian) it is not necessarily clear that\footnote{The gRPA vacuum $| RPA
  \rangle$ is annihilated by all gRPA annihilation operators $Q^{\nu}_B$ and
  by the gRPA approximated symmetry generators, $\Gamma_B |RPA \rangle =
  0$. Whereas in the case of non-zero modes the excited states are generated
  by the creation operators, in the case of zero modes the excited states
  $|\gamma \rangle$ are given as plane waves and employ the canonically
  conjugate coordinate of the symmetry generator, $|\gamma \rangle = e^{i
    \gamma^b \Theta^b} |RPA \rangle$. These excited states are obviously
  eigenstates of the symmetry generators: $\Gamma^a |\gamma \rangle = \gamma^a
  | \gamma \rangle$. Their energy can be determined easily, too: $H_B | \gamma
  \rangle = \frac{1}{2} (\mathcal{M}^{-1})^{ab} \gamma^a \gamma^b | \gamma
  \rangle $.} $| \gamma \rangle = e^{i \gamma^a \Theta^a} | RPA \rangle$ is a
correctly charged state, and 
thus $\langle \gamma | (\mathcal{M}^{-1})^{ab} \Gamma^a_B \Gamma^b_B |
\gamma \rangle$ is also not necessarily the energy of a state with a
certain prescribed charge. This demonstrates, however, that further
developments are necessary in two areas: First, one needs a systematic
expansion where the quasi-boson approximation is the leading
order. Second, a clear interpretation of gRPA states in terms of real
physical states appears necessary. 
\subsection{Observations in Electrodynamics}
We want to close this section with a few remarks and observations on
electrodynamics. In electrodynamics the (full) Gauss law operator is
given by
\be
\Gamma (\vec{x}) = \nabla_i^{\vec{x}} \vec{\Pi}_i (\vec{x}).
\ee
We guess immediately the form of the corresponding ``angle'' coordinate
$\Theta$ as was discussed before in the context of gRPA; if we take 
\be
\Theta(\vec{x}) = -G_{\Delta} (\vec{x}, \vec{y}) \nabla_i^{\vec{y}}
\vec{A}_i(\vec{y}) \text{ with } \Delta^{\vec{x}} G_{\Delta} (\vec{x},
\vec{y}) = \delta_{\vec{x} \vec{y}} \label{angle_variable_QED_exact}
\ee
then on the one hand
\be
[\Theta(\vec{x}), \Gamma(\vec{y})] = i \delta_{\vec{x} \vec{y}}
\ee
and on the other hand
\be
[\Theta(\vec{x}), H] = - G_{\Delta} (\vec{x}, \vec{y}) \Gamma (\vec{y}),
\ee
thereby identifying the moment of inertia as
\be
(\mathcal{M}^{-1})(\vec{x}, \vec{y}) = - G_{\Delta} (\vec{x}, \vec{y}).
\ee
It seems interesting that the angle coordinate
eq.\,(\ref{angle_variable_QED_exact}) shown here appears to 
be closely related to the \textit{dressing function} used in
\cite{Dirac:1955uv, Dirac:1967}, cf. also \cite{Lavelle:1997ty}, to
construct gauge invariant electron fields,
\be
\psi_c(\vec{x}) = e^{i e G_{\Delta}(\vec{x}, \vec{y})
\nabla^{\vec{y}}_i \vec{A}_i(\vec{y})} \psi(\vec{x}).
\ee  
The Hamiltonian of electrodynamics furthermore allows for a
decomposition very similar to the one made possible by the gRPA
treatment for general bosonic theories,
cf. eq.\,(\ref{RPA_gen_diag_HB_incl_zero_modes}). If we introduce the
longitudinal and transversal projectors $\PP^L_{ij}(\vec{x}, \vec{y}),
\PP^T_{ij}(\vec{x}, \vec{y})$ as
\be
\PP^L_{ij}(\vec{x}, \vec{y}) = \nabla^{\vec{x}}_i
G_{\Delta}(\vec{x}, \vec{y}) \nabla^{\vec{y}}_j \text{ and }
\PP^T_{ij} (\vec{x}, \vec{y}) = \delta_{ij} \delta_{\vec{x} \vec{y}} -
\PP^L_{ij}(\vec{x}, \vec{y}), 
\ee  
we can decompose the Hamiltonian into one depending solely on
transversal degrees of freedom, $\vec{\Pi}^T_i (\vec{x}) =
\PP^T_{ij}(\vec{x}, \vec{y}) \vec{\Pi}_j(\vec{y})$, $\vec{A}^T_i
(\vec{x}) = \PP^T_{ij}(\vec{x}, \vec{y}) \vec{A}_j(\vec{y})$ , and one
depending solely on the Gauss law operator (symmetry generator):
\be
H = \frac{1}{2} \int d^3x\, \left( \vec{\Pi}^T_i(\vec{x})
\vec{\Pi}^T_i(\vec{x}) + \vec{B}^2[\vec{A}^T] \right) + \frac{1}{2}
\int d^3x_1\, d^3x_2 \, G_{\Delta}(\vec{x}_2, \vec{x}_1) \Gamma
(\vec{x}_1) \Gamma (\vec{x}_2). 
\ee
The gRPA treatment, however, differs from the general treatment
discussed above, though not in an unexpected manner. Since
electrodynamics is eventually a free theory, the solution of the full
Schr\"odinger equation gives the same result as the mean-field
treatment, thus we can use as reference state the well-known ground
state of electrodynamics \cite{Greensite:1979}:
\be
\psi[\vec{A}] \propto \exp{ \left\{-\frac{1}{2} \int d^3x\, d^3y
\vec{A}_i(\vec{x}) \left[\int \frac{d^3p}{(2\pi)^3}
|\vec{p}| (\delta_{ij} - \frac{\vec{p}_i
 \vec{p}_j}{\vec{p}^2}) e^{i\vec{p}\cdot(\vec{x} - \vec{y})} \right] \vec{A}_j(\vec{y}) \right\}}.
\ee
This state, however, is annihilated by the Gauss law operator and
consequently we expect $\Gamma_B$ to vanish, and indeed this is
verified by explicit computation. It should be noted in passing,
however, that an application of
eq.\,(\ref{eq_leading_order_mom_of_in}) for the moment of 
inertia gives the correct result (since the zeros in $\Gamma_p$, which
come about since $\nabla$ acts upon the transversal $U^{-1}$, are
canceled by $F^{-1} = G$ which is infinite in the longitudinal
subspace, $G^{-1}$ being proportional to the transversal projector).
\section{Summary and Conclusion \label{sec_summary_and_conclusion}}
Let us shortly summarize what has been achieved in this paper,
and which problems remain to be solved. 
We started by introducing shortly the canonical quantization and
Schr\"odinger picture treatment of general bosonic theories with a
standard kinetic term.
We then considered the operator
formulation of the (generalized) Random Phase Approximation that
is the one common in nuclear physics, and which has recently also been
investigated in the context of pion physics. We demonstrated that
this approach can also be implemented for a general class of bosonic field
theories, but 
that the class of excitation operators to be considered has to be
larger than in  nuclear physics; namely, one has to also allow for
terms linear in creation/annihilation operators of the fundamental
boson fields. It turned out to be possible (at least for a certain
class of Hamiltonians with standard kinetic term) to prove the
equivalence of the operator formulation to the formulation starting
from the time-dependent variational principle.
Then we demonstrated that, in the absence of zero modes,
the gRPA Hamiltonian is just a collection of harmonic oscillators. The
zero modes required special attention, but the problems could be
solved along lines paralleling nuclear physics. We then considered the
question of how conservation laws of the full theory translate into
conservation laws of the theory in generalized Random Phase
Approximation, and saw that, at 
least in the case of symmetries generated by one-body operators,
existence of symmetries in the full theory implies existence of
symmetries in the gRPA-approximated theory, although these symmetries
need not be the same; in Yang-Mills theory, the generalized RPA only
carries an Abelian symmetry. This also opened up a possible way of
improving the gRPA by requiring a certain fulfillment of the
commutation relations by the approximated operators. \\
Then we investigated the
difference between the mean-field energy and the energy of the gRPA
ground state with special emphasis on the energy contribution of the
zero modes. For this purpose we had to calculate the
moment-of-inertia tensor which (at least to lowest order in perturbation
theory) turned out to be the Green's function of the covariant
Laplacian in the background field at the stationary point. \\
Furthermore, we compared the quantum corrections to the moment of
inertia, and found that their structure is very similar to the
\textit{screening} contribution in a Coulomb gauge perturbative
calculation of the static quark potential. \\
As a last point, we made some observations of electrodynamics where
the concepts of moment of inertia and zero mode operators can be
applied to the system without any form of approximation. In the gRPA
treatment of electrodynamics some of the general observations can be
verified explicitly; the most interesting point seems to be that -
even though the Gauss law operator vanishes in the gRPA treatment, due
to the fact that the reference state is gauge invariant, - it is still
possible to determine correctly the moment of inertia. \\
To put the method into perspective, let us summarize the main positive
and negative aspects: \newline On the positive side, we first have to mention
that in the generalized RPA only energy differences w.r.t. the ground state are
computed. This simplifies matters, since that part of  renormalization
that is usually done by normal-ordering is automatically taken care
of. This brings us directly to the second point: energies of excited
states can be computed. This is usually very difficult if one relies upon
e.g. the Rayleigh-Ritz principle.  The most important point, however,
is the effective implementation of the Gauss law constraint. Even
though the gauge symmetry is broken in the mean-field treatment, the
unphysical excitations generated by the (gRPA approximated) Gauss law
operator (more precisely given by plane waves given in the coordinate
conjugate to the Gauss law operator) are
orthogonal to all the other physical excited states, which is
almost as good as if they didn't even exist. \newline 
But there are also a number of drawbacks of the method 
presented. The first has to do with the ground state energy: 
Whereas one does obtain a lowering of the ground state energy
w.r.t. the mean-field energy due to deformation, in a manner which is
even formally quite similar to the lowering obtained e.g. in the second
order Kamlah expansion (at least to the order in perturbation theory
considered) and also in other methods, the point of the calculation at which
the energy correction due to the deformation is considered is fundamentally
different.  In the Kamlah expansion and also in the Thouless-Valatin method,
the corrections 
are subtracted {\it before} the variation is carried out, whereas in
the generalized RPA they are subtracted only {\it after} the
variation. Therefore 
the energy correction does not have any influence on the parameters of
the mean-field. This brings us directly to the next problematic point: 
of course the mean-field vacuum is not the gRPA ground state. In
nuclear physics, one was able, however, to construct the gRPA ground
state from the mean-field ground state; but, in the presence of zero
modes, this state has a divergent norm. This is due to the fact that the
gRPA is a 'small angle' approximation \cite{MW69}, or in other words, the
compact nature of the non-Abelian symmetry is lost. We have seen
explicitly that in the case of Yang-Mills theory, the compact $SU(N)$
symmetry is replaced by the non-compact $U(1)^{N^2-1}$ symmetry. In
nuclear physics, at least in the case of two-dimensional rotations, the
fact that RPA is only a small-angle approximation 
was not so much of a 
problem, since the global dependence on the angle is generally known and
therefore one can extract enough information from the small-angle
approximation to determine the whole wave function \cite{MW69}. In Yang-Mills
theory, we have no comparable knowledge that could be put to use
practically and therefore, we cannot compute the ground-state wave
functional. Lacking this knowledge, however, one needs other methods
to evaluate matrix elements of operators, which up to now we haven't
developed. Therefore, the only quantities we currently can calculate are
the energies of the excited states. As a last though very important point, we
have to mention that we have not dealt with the problems of
renormalization. However, it should be possible to deal with them,
since (at least in the case of 
$\phi^4$ theory) this problem has been faced already by Kerman {\it et
al}\, \cite{Kerman:1995uj, Kerman:1998vt} in the context of the
generalized RPA derived from the time-dependent variational principle.

\begin{acknowledgments}
The authors would like to thank Michael Engelhardt, Jean-Dominique
L\"ange, Markus Quandt, Cecile Martin, and Peter Schuck for continuous
discussion. This work was
supported by the \textit{Deutsche Forschungsgemeinschaft} (DFG) under grants
Mu 705/3, GRK 683, Re 856/4-1, Re 856/5-1 and Schr 749/1-1. This work
is also supported in part by funds provided by the U.S. Department of
Energy (D.O.E.) under cooperative research agreement DF-FC02-94ER40818.  
\end{acknowledgments}

\appendix
%
%
%
\section{Generalized RPA from the Time-Dependent Variational
Principle\label{app_RPA_from_tdvp}} 
In this appendix we present the approach to the generalized Random Phase
Approximation via the time-dependent variational principle as
pioneered by Kerman {\it et al} \cite{Kerman:1976yn, Kerman:1995uj,
Kerman:1998vt}. We 
show that it leads to equations of motion of coupled harmonic oscillators.
The purpose of this appendix is to make this paper reasonably
self-contained, so that the reader can follow our claim that (under
the circumstances outlined in the main text) the operator approach and
the time-dependent approach to the generalized RPA yield identical
equations of motion. \\ 
We
don't want to deal with a specific theory at the moment, 
we will only require the Hamiltonian to be of the form
\be
H = \frac{1}{2}\pi_i^2 + V[\phi_i], \label{tdvpRPAHamiltonian}
\ee
where $\phi_i$ are a set of fields, and $\pi_i$ are the
canonical momenta conjugate to $\phi_i$, in the 
field (coordinate) representation under consideration
\be
\pi_i = \frac{1}{i} \frac{\delta}{\delta \phi_i} 
\ee
and $i$ is a super-index, containing a position variable ${\bf x}$, and all
other indices required (like color, spatial etc). The Einstein
summation convention is adopted, implying sums over all discrete and
integrals over all continuous variables.
$V[\phi]$ is a functional of the field operators, in the following
referred to as 'potential'. \\ The states that we consider as trial
states for the time-dependent variational principle are the most
general time-dependent Gaussian states (we only indicate the
time-variable explicitly, $i,j$ are super-indices and $\mathcal{N}$ is a
normalization constant):
\be
\psi[\phi,t] = \mathcal{N} \exp{\left( -(\phi-\bar{\phi}(t))_{i} (\frac{1}{4}G^{-1}(t) - i
\Sigma(t))_{ij}(\phi-\bar{\phi}(t))_{j} + i \bar{\pi}_i(t)
(\phi-\bar{\phi}(t))_{i}\right)}. \label{RPA_most_general_TD_Gauss}
\ee
The meaning of the parameters becomes clear by considering expectation
values:
\bea
\langle \psi(t) | \phi_i | \psi(t) \rangle &=& \bar{\phi}_i(t) \\ \langle
\psi(t) | \phi_i \phi_j| \psi(t) \rangle &=& \bar{\phi}_i(t) \bar{\phi}_j(t) +
G_{ij}(t) \\
\langle \psi(t) | \pi_i | \psi(t) \rangle &=& \bar{\pi}_i(t) \\
\langle \psi(t) | \pi_i \pi_j  | \psi(t) \rangle &=&  
\bar{\pi}_i \bar{\pi}_j + \frac{1}{4} G^{-1}_{ij}(t) + 4 (\Sigma(t)
G(t) \Sigma(t))_{ij}. \nonumber  \\
\label{eq_expect_pi_squared}
\eea
We can now compute the action of the time-dependent variational
principle \cite{Kerman:1976yn}
\be
S = \int dt \langle \psi(t) | i \partial_t - H | \psi(t) \rangle
\ee
and obtain
\be
S = \int dt  \Bigg\{ \Big[\bar{\pi}_i(t) \dot{\bar{\phi}}_i(t) -
\tr{(\dot{\Sigma} G)} + \frac{i}{4}\tr{(\dot{G}^{-1} G)} \Big] -
{\cal H}(t)[\bar{\phi},\bar{\pi},G,\Sigma] \Bigg\}
\ee
with
\be
{\cal H}(t) =  \langle \psi(t) | H | \psi(t) \rangle. \label{cal_H}
\ee
We can now add a total time derivative\footnote{Its form can e.g. be
found in \cite{Kerman:1995uj} as (here only symbolically)
$\frac{d}{dt}(-\frac{i}{4} \log{(G^{-1})} - \Sigma G)$.} 
 that does not change the
equations of motion, and obtain for the action
\be
S = \int dt (\Sigma_{ij}(t) \dot{G}_{ij}(t) + \bar{\pi}_i(t)
\dot{\bar{\phi}}_i(t) - {\cal H}(t)),
\ee
which shows that $\Sigma$ is to be considered as the canonical
momentum conjugate to $G$, and $\bar{\pi}$ that of $\bar{\phi}$. 
The parameters of the wave functional are now determined via Hamilton's
classical equations of motion:
\bea
\dot{\bar{\phi}}_i(t) = \frac{\delta {\cal H}}{\delta \bar{\pi}_i} &;
& \dot{\bar{\pi}}_i(t) = - \frac{\delta {\cal H}}{\delta
\bar{\phi}_i} \label{canon_eom-RPA1a}\\
\dot{G}_{ij}(t) =\frac{\delta {\cal H}}{\delta \Sigma_{ij}}  &; &
\dot{\Sigma}_{ij}(t) = - \frac{\delta {\cal H}}{\delta
G_{ij}}.\label{canon_eom-RPA1b} 
\eea
However, in general it will be much too complicated to solve these
equations, therefore we now consider a two-step procedure
\begin{enumerate}
\item look for static solutions to the equations of motion
\item consider small fluctuations around these static solutions.
\end{enumerate}
The static solutions are (obviously) determined via
\be
\dot{\bar{\phi}}_i(t) = 0 \hspace*{0.5em}; \hspace*{1.5em} 
\dot{\bar{\pi}}_i(t) = 0 \hspace*{0.5em}; \hspace*{1.5em} 
\dot{G}_{ij}(t) = 0 \hspace*{0.5em}; \hspace*{1.5em} 
\dot{\Sigma}_{ij}(t) = 0.\label{canon_eom-RPA2a} 
\ee
But these are nothing but the equations resulting from the Rayleigh-Ritz
principle:
\be
\frac{\delta {\cal H}}{\delta \bar{\pi}_i} = 0 \hspace*{0.5em};
\hspace*{1.5em}  
\frac{\delta {\cal H}}{\delta \bar{\phi}_i} = 0 \hspace*{0.5em};
\hspace*{1.5em}  
\frac{\delta {\cal H}}{\delta \Sigma_{ij}} = 0 \hspace*{0.5em};
\hspace*{1.5em}
\frac{\delta {\cal H}}{\delta G_{ij}} = 0. \label{canon_eom-RPA2b} 
\ee
Thus, for a static solution of the time-dependent variational
principle, the parameters are those which minimize (or at least extremize) 
the energy. This 
should not really come as a surprise, since for a static state
$\psi[\phi]$, the action reduces just to minus the energy times the
respective time interval under consideration. \\ For the next step,
we decompose the general, time-dependent parameters into the static
solution plus a time-dependent contribution that later on is
considered to be small, e.g. for $\bar{\phi}$:
\be
\bar{\phi}_i(t) = \bar{\phi}_{i,s} + \delta \bar{\phi}_i(t),
\ee
where $\bar{\phi}_{i,s}$ denotes the static solution\footnote{In other
words, if we evaluate the first derivative $\frac{\delta {\cal H}}{\delta
\bar{\pi}_i}$ for $\bar{\pi} = \bar{\pi}_s$, it is zero, and
correspondingly for $\bar{\phi}, G, \Sigma$.}, and $\delta
\bar{\phi}_i(t)$ the 'small' time-dependent part.
We insert this decomposition into the equations of motion
eqs.\,(\ref{canon_eom-RPA1a}), (\ref{canon_eom-RPA1b}), and obtain 
\bea
\delta \dot{\bar{\phi}}_i(t) &=& \frac{\delta {\cal H}}{\delta
\bar{\pi}_i}[\bar{\phi}_{s} + \delta \bar{\phi}(t), \ldots
] \nonumber \\
 &=& \frac{\delta {\cal H}}{\delta
\bar{\pi}_i}[\bar{\phi}_{s},\bar{\pi}_{s},G_{s}, \Sigma_{s}]
 \label{tdvpRPA_smallfluct1} 
\\ & & 
+ \frac{\delta^2 {\cal H}}{\delta \bar{\pi}_j
\delta \bar{\pi}_i}[\bar{\phi}_{s},\bar{\pi}_{s},G_{s}, \Sigma_{s}]
\delta {\bar{\pi}}_j  
+ \ldots + {\cal O} (\delta^2). \nonumber
\eea
Now the meaning of $\delta \bar{\phi}$, etc. being small is clarified:
in the equations of motion terms of higher than linear order are
neglected (in the action, it would be terms of higher than quadratic order).
The first contribution $\frac{\delta {\cal H}}{\delta
\bar{\pi}_i}[\bar{\phi}_{s},\bar{\pi}_{s},G_{s}, \Sigma_{s}]$ vanishes
by virtue of the static equations of motion
eqs.\,(\ref{canon_eom-RPA2a}), (\ref{canon_eom-RPA2b}). The same
construction can be carried out for all four parameter types
($\bar{\phi}, \bar{\pi}, \Sigma, G$), and the
resulting equations of motion can be nicely summarized as follows:
\be
\left(
\begin{array}{c} \delta \dot{\bar{\phi}} \\ \delta \dot{\bar{\pi}}
\\ \delta \dot{G} \\ \delta \dot{\Sigma} \end{array} \right) = \left(
\begin{array}{rrrr}
\frac{\delta^2 {\cal H}}{\delta \bar{\pi} \delta \bar{\phi}} &
\frac{\delta^2 {\cal H}}{\delta \bar{\pi} \delta \bar{\pi} } &
\frac{\delta^2 {\cal H}}{\delta \bar{\pi} \delta G } & 
\frac{\delta^2 {\cal H}}{\delta \bar{\pi} \delta \Sigma } \\
-\frac{\delta^2 {\cal H}}{\delta \bar{\phi} \delta \bar{\phi} } &
-\frac{\delta^2 {\cal H}}{\delta \bar{\phi} \delta \bar{\pi} } &
-\frac{\delta^2 {\cal H}}{\delta \bar{\phi} \delta G } & 
-\frac{\delta^2 {\cal H}}{\delta \bar{\phi} \delta \Sigma } \\
\frac{\delta^2 {\cal H}}{\delta \Sigma\delta \bar{\phi} } &
\frac{\delta^2 {\cal H}}{\delta \Sigma\delta \bar{\pi} } &
\frac{\delta^2 {\cal H}}{\delta \Sigma\delta G } & 
\frac{\delta^2 {\cal H}}{\delta \Sigma \delta \Sigma } \\ 
-\frac{\delta^2 {\cal H}}{\delta G \delta \bar{\phi}} &
-\frac{\delta^2 {\cal H}}{\delta G \delta \bar{\pi}} &
-\frac{\delta^2 {\cal H}}{\delta G \delta G} & 
-\frac{\delta^2 {\cal H}}{\delta G \delta \Sigma} 
\end{array}
\right)
\left(
\begin{array}{c} \delta \bar{\phi} \\ \delta \bar{\pi}
\\ \delta G \\ \delta \Sigma \end{array} \right),
\label{tdvpRPAeom1}
\ee 
where $\frac{\delta^2 {\cal H}}{\delta A \delta B} = \frac{\delta^2
{\cal H}}{\delta A \delta B}[\bar{\phi}_s, \bar{\pi}_s, G_s,
\Sigma_s]$ 
and can be made even more transparent if one introduces an auxiliary
matrix $\Omega$:
\be
\Omega = \left( 
\begin{array}{cccc} 0 & 1 & & \\ -1 & 0 & & \\ & & 0 & 1 \\ & & -1 & 0
\end{array} \right).  
\ee
Then one obtains
\be
\left(
\begin{array}{c} \delta \dot{\bar{\phi}} \\ \delta \dot{\bar{\pi}}
\\ \delta \dot{G} \\ \delta \dot{\Sigma} \end{array} \right) = 
\Omega
\left(
\begin{array}{rrrr}
\frac{\delta^2 {\cal H}}{\delta \bar{\phi} \delta \bar{\phi}} &
\frac{\delta^2 {\cal H}}{\delta \bar{\phi} \delta \bar{\pi} } &
\frac{\delta^2 {\cal H}}{\delta \bar{\phi} \delta G} & 
\frac{\delta^2 {\cal H}}{\delta \bar{\phi} \delta \Sigma} \\
\frac{\delta^2 {\cal H}}{\delta \bar{\pi} \delta \bar{\phi}} &
\frac{\delta^2 {\cal H}}{\delta \bar{\pi} \delta \bar{\pi}} &
\frac{\delta^2 {\cal H}}{\delta \bar{\pi} \delta G} & 
\frac{\delta^2 {\cal H}}{\delta \bar{\pi} \delta \Sigma} \\
\frac{\delta^2 {\cal H}}{\delta G \delta \bar{\phi}} &
\frac{\delta^2 {\cal H}}{\delta G \delta \bar{\pi}} &
\frac{\delta^2 {\cal H}}{\delta G \delta G} & 
\frac{\delta^2 {\cal H}}{\delta G \delta \Sigma } \\
\frac{\delta^2 {\cal H}}{\delta \Sigma \delta \bar{\phi}} &
\frac{\delta^2 {\cal H}}{\delta \Sigma \delta \bar{\pi}} &
\frac{\delta^2 {\cal H}}{\delta \Sigma \delta G} & 
\frac{\delta^2 {\cal H}}{\delta \Sigma \delta \Sigma}
\end{array}
\right)
\left(
\begin{array}{c} \delta \bar{\phi} \\ \delta \bar{\pi}
\\ \delta G \\ \delta \Sigma \end{array} \right).
\label{tdvpRPAeom2}
\ee 
At this point it becomes clear what determines the spectrum of small
fluctuations around a static mean-field solution: it's the {\it
stability matrix} of this static mean-field
solution. Eqs. (\ref{tdvpRPAeom2}) are 
usually called the generalized RPA equations. Up to now, the only
assumption that has been used was the assumption of $\psi$ being a
Gaussian state. 
But we have also restricted the choice of Hamiltonians that we want to
consider by eq.\,(\ref{tdvpRPAHamiltonian}). This restriction will allow to
carry the calculation a bit further. ${\cal H}$ depends only via
$\langle \pi^2 \rangle$ on both $\bar{\pi}$ and $\Sigma$, $V[\phi]$
depends only on 
$\bar{\phi}$ and $G$. Thus for all Hamiltonians that we are
considering, we have the kinetic energy written as
\be
\langle \psi |\frac{1}{2} \pi_i^2 | \psi \rangle = \frac{1}{2} \bar{\pi}_i
\bar{\pi}_i + \frac{1}{8} \Tr{(G^{-1})}  + 2 \Tr{(\Sigma G \Sigma)},
\ee
where $\Tr$ denotes the trace over the super-indices.
The static solutions $\bar{\pi}_{i,s},\Sigma_{ij,s}$ are determined via
\be
\frac{\delta {\cal H}}{\delta \bar{\pi}_i} = 0
\hspace*{0.5em},
\hspace*{1.5em}  
\frac{\delta {\cal H}}{\delta \Sigma_{ij}} = 0
\ee
and result in
\be
\bar{\pi}_{i,s} = 0 \hspace*{0.5em},\hspace*{1.5em} \Sigma_{ij,s} = 0.
\ee
This information, together with the knowledge that $V[\phi]$ does
neither depend on $\bar{\pi}$ nor on $\Sigma$, determines a number of
second derivatives:
\be
\begin{array}{rccrccrccrcc}
\frac{\delta^2 {\cal H}}{\delta \bar{\pi} \delta \bar{\phi}} &=& 0, &
\frac{\delta^2 {\cal H}}{\delta \Sigma \delta \bar{\phi}} &=& 0, &
\frac{\delta^2 {\cal H}}{\delta G \delta \bar{\pi}} &=& 0, & 
\frac{\delta^2 {\cal H}}{\delta \bar{\pi}_i \delta \bar{\pi}_j} &=&
\delta_{ij}, \\
\frac{\delta^2 {\cal H}}{\delta \Sigma \delta \bar{\pi}} &=& 0, &
\frac{\delta^2 {\cal H}}{\delta \bar{\pi} \delta G} &=& 0, &
\frac{\delta^2 {\cal H}}{\delta \Sigma \delta G} &=& 0, &
\frac{\delta^2 {\cal H}}{\delta \Sigma \delta \Sigma} &=& (G \unit), \\
\end{array}
\ee
where we have used the abbreviation\footnote{In this context, one should note
  that $\frac{\delta \Sigma_{ij}}{\delta \Sigma_{kl}} = \frac{1}{2}
  (\delta_{ik} \delta_{jl} + \delta_{il} \delta_{jk}).$}
\bea
(G \unit)_{ij;kl} &=& 2 \frac{\delta}{\delta \Sigma_{ij}} \frac{\delta}
{\delta \Sigma_{kl}} \tr{\left(\Sigma G \Sigma \right)} \nonumber \\ &=& G_{ki}
\delta_{lj} +G_{kj} \delta_{li} + G_{jl} \delta_{ki} + G_{li}
\delta_{kj}. \hspace*{2em}  
\eea
These results can be used to simplify eq.\,(\ref{tdvpRPAeom1}):
\be
\left(
\begin{array}{c} \delta \dot{\bar{\phi}} \\ \delta \dot{\bar{\pi}}
\\ \delta \dot{G} \\ \delta \dot{\Sigma} \end{array} \right) = \left(
\begin{array}{rrrr}
0 & \unit & 0 & 0 \\
-\frac{\delta^2 {\cal H}}{\delta \bar{\phi} \delta \bar{\phi}} &
0 & -\frac{\delta^2 {\cal H}}{\delta \bar{\phi} \delta G } & 0 \\
0 & 0 & 0 & (G\unit) \\ 
-\frac{\delta^2 {\cal H}}{\delta G  \delta \bar{\phi}} &
0 & -\frac{\delta^2 {\cal H}}{\delta G \delta G} & 
0 \\
\end{array}
\right)
\left(
\begin{array}{c} \delta \bar{\phi} \\ \delta \bar{\pi}
\\ \delta G \\ \delta \Sigma \end{array} \right),
\label{tdvpRPAeom3}
\ee 
with $(\unit)_{ij} = \delta_{ij}$.
By 
taking the derivative of eq.\,(\ref{tdvpRPAeom3}) with
respect to time, and reinserting eq.\,(\ref{tdvpRPAeom3}), one obtains
two sets of partially decoupled equations
\bea
\text{(I)} \hspace*{0.5em} 
\left(
\begin{array}{c} 
 \delta \ddot{\bar{\phi}} \\ \delta \ddot{G} \\
\end{array}
\right) &=& 
-
\left(
\begin{array}{rr} 
 \frac{\delta^2 {\cal H}}{\delta \bar{\phi} \delta \bar{\phi}} & 
  \frac{\delta^2 {\cal H}}{\delta \bar{\phi} \delta G} \\ 
  (G \unit) \frac{\delta^2 {\cal H}}{\delta G \delta \bar{\phi}} & (G
  \unit) \frac{\delta^2 {\cal H}}{\delta G \delta G} \\
\end{array}
\right)
\left(
\begin{array}{c} 
\delta \bar{\phi}  \\ \delta G\\
\end{array}
\right), \nonumber \\ && \\ 
\text{(II)} \hspace*{0.5em} 
\left(
\begin{array}{c} 
 \delta \ddot{\bar{\pi}} \\ \delta \ddot{\Sigma} \\
\end{array}
\right) &=& -
\left(
\begin{array}{cc} 
\frac{\delta^2 {\cal H}}{\delta \bar{\phi} \delta \bar{\phi}} &
\frac{\delta^2 {\cal H}}{\delta \bar{\phi} \delta G} (G \unit)
\\\frac{\delta^2 {\cal H}}{\delta G \delta \bar{\phi}}  &
\frac{\delta^2 {\cal H}}{\delta G \delta G}  (G \unit) \\ 
\end{array}
\right)
\left(
\begin{array}{c} 
 \delta {\bar {\pi}}  \\ \delta \Sigma \\
\end{array}
\right). \nonumber \\ && 
\eea
If we now make the {\it ansatz} of a harmonic time-dependence
\bea
\text{(I)} \hspace*{0.5em}
\left( \begin{array}{c} \delta \bar{\phi}_i(t) \\ \delta  G_{ij}(t)
\end{array} \right) = \left( \begin{array}{c} \delta \bar{\phi}_i^{(0)} \\
\delta  G_{ij}^{(0)} \end{array} \right) \cos{(\omega t + \delta_1)}, 
\\
\text{(II)} \hspace*{0.5em}
\left( \begin{array}{c} \delta \bar{\pi}_i(t) \\ \delta  \Sigma_{ij}(t)
\end{array} \right) = \left( \begin{array}{c} \delta \bar{\pi}_i^{(0)} \\
\delta  \Sigma_{ij}^{(0)} \end{array} \right) \cos{(\omega t + \delta_2)}
\eea
we obtain the eigenvalue equations
\bea
\text{(I)} \hspace*{0.5em}
\omega^2
\left(
\begin{array}{c} 
 \delta \bar{\phi} \\ \delta G \\
\end{array}
\right)
&=&
\left(
\begin{array}{rr} 
 \frac{\delta^2 {\cal H}}{\delta \bar{\phi} \delta \bar{\phi}} & 
  \frac{\delta^2 {\cal H}}{\delta \bar{\phi} \delta G} \\ 
  (G \unit) \frac{\delta^2 {\cal H}}{\delta G \delta \bar{\phi}} & (G
  \unit) \frac{\delta^2 {\cal H}}{\delta G \delta G} \\
\end{array}
\right)
\left(
\begin{array}{c} 
\delta \bar{\phi}  \\ \delta G\\
\end{array}
\right), \nonumber \\
\label{tdvp_final0} \\
\text{(II)} \hspace*{0.5em}
 \omega^2
\left(
\begin{array}{c} 
 \delta \bar{\pi} \\ \delta \Sigma \\
\end{array}
\right) & = &
\left(
\begin{array}{cc} 
\frac{\delta^2 {\cal H}}{\delta \bar{\phi} \delta \bar{\phi}} &
\frac{\delta^2 {\cal H}}{\delta \bar{\phi} \delta G} (G \unit)
\\\frac{\delta^2 {\cal H}}{\delta G \delta \bar{\phi}}  &
\frac{\delta^2 {\cal H}}{\delta G \delta G}  (G \unit) \\ 
\end{array}
\right)
\left(
\begin{array}{c} 
 \delta {\bar {\pi}}  \\ \delta \Sigma \\
\end{array}
\right). \nonumber \\
\label{tdvp_final1}
\eea
In analyzing properties of the generalized RPA equations
eqs.\,(\ref{tdvp_final0}), (\ref{tdvp_final1}) it is often simpler to
study the Hamiltonian 
they are derived from than to study the equations themselves.
In this
context one can imagine eq.\,(\ref{tdvpRPAeom3}) to originate from a
Hamiltonian\footnote{They originate by the canonical equations of
motion, e.g. $\delta \dot{\bar{\pi}} = - \delta H/\delta (\delta
\bar{\phi})$.} $H$: 
\be
H = \frac{1}{2} (\delta \bar{\pi} ~\delta \Sigma) \left( \begin{array}{cc}
\unit & 0 \\ 0 & (G \unit) \end{array} \right) \left(
\begin{array}{c} \delta \bar{\pi} \\ \delta \Sigma \end{array} \right)
+ \frac{1}{2} (\delta \bar{\phi} ~\delta G) \left( \begin{array}{cc}
\frac{\delta^2 {\cal H}}{\delta \bar{\phi} \delta \bar{\phi}} &
\frac{\delta^2 {\cal H}}{\delta \bar{\phi} \delta G}
 \\\frac{\delta^2 {\cal H}}{\delta G \delta \bar{\phi}}  &
 \frac{\delta^2 {\cal H}}{\delta G \delta G}   \end{array} \right)
 \left( \begin{array}{c} \delta \bar{\phi} \\ \delta G \end{array} \right).
\label{tdvpHamiltonian1}
\ee
This evidently is the Hamiltonian of a set of coupled oscillators.
The important point is that the signs of the eigenvalues of the
decoupled oscillators  are determined by the reduced stability
matrix\footnote{The problem of coupled oscillators is well-known,
cf. \cite{Go91, FK92}. 
If the reduced stability matrix is a positive matrix (and thus our mean
field vacuum is indeed a minimum), all the eigenvalues will be
positive, and thus all oscillator frequencies will be real.} containing only
second derivatives w.r.t. to $\bar{\phi}$ and $G$ (We assume here that
$(G \unit)$ is positive definite and since $(G \unit)$ is only
multiplied by objects symmetric in their two indices, e.g. $\delta
\Sigma_{ij} = \delta \Sigma_{ji}$, this boils down to assuming that
$G$ is positive definite. This is a sensible assumption connected to
the normalizabilty of $\psi[\phi]$ and is 
discussed extensively in sec.\,\ref{sec_on_creat_annihil_op}). An interesting
observation can be made immediately: usually $\frac{\delta^2 {\cal
H}}{\delta \bar{\phi} \delta G}$ will be only non-zero if there is a
condensate (i.e. $\bar{\phi} \neq 0$) in the system. Thus if we don't have a condensate, the  
equations for the one- and two-particle content\footnote{We see
in sec.\,\ref{pqformulationofRPA} that in a creation/annihilation
operator formalism 
$\delta \bar{\phi}, \delta \bar{\pi}$ are connected to the amplitudes
of operators containing one creation/annihilation operator, whereas
$\delta G, \delta \Sigma$ are connected to the amplitudes containing
two creation/annihilation operators.} decouple,
 and, since in
a system without condensate $G$ will usually be translation invariant,
we can see by considering the Fourier transformed
quantities that $1/\tilde{g}({\bf p})$ 
just describes the energy spectrum of
single-particle excitations with momentum ${\bf p}$, where
$\tilde{g}({\bf p})$ is defined by the following procedure.
We introduce the Fourier transform $\tilde{G}({\bf p}, {\bf q})$ by
\be
G({\bf x},{\bf y}) = \int \frac{d^3p}{(2 \pi)^3} \frac{d^3q}{(2\pi)^3}
e^{i {\bf p}.{\bf x}} \tilde{G}({\bf p}, {\bf q}) e^{-i {\bf q}.{\bf y}}. 
\ee
If we now require that $G$ shall be translation invariant,
i.e. $G({\bf x},{\bf y}) = G({\bf x}-{\bf y})$, we obtain for
$\tilde{G}({\bf p}, {\bf q})$ 
\be
\tilde{G}({\bf p}, {\bf q}) = (2 \pi)^3 \delta({\bf p} - {\bf q})
\tilde{g}({\bf p}).
\ee
This defines $\tilde{g}({\bf p})$.

%
%
%
\section{Hamiltonian Treatment of Yang-Mills Theory \label{app_YMT}}
\subsection{Yang-Mills Theory}
There is not much to be said about the Hamiltonian treatment of
Yang-Mills theory in particular, since there is an excellent
treatment in \cite{Jackiw:1980ur}. The main points
to be kept in mind are: \textit{First}, in order to work in the
Hamiltonian formalism at all one chooses the \textit{Weyl
gauge}, i.e. $A_0 = 0$. The price to be paid in this gauge is that the
classical constraint equation $\hat{\vec{D}}^{ab}_i \vec{E}^b_i = 0$
($\vec{E}$ is the color-electric field, for the definition of $\hat{\vec{D}}$
cf. eq.\,(\ref{def_cov_deriv_adj}) below),
i.e. the Gauss law equation, has to be implemented as a constraint on
states: $| \psi \rangle$ is a physical state if it is annihilated
by the Gauss law operator (which is constructed from $\hat{\vec{D}}^{ab}_i
\vec{E}^b_i$ upon quantization). Even in the
presence of this constraint the canonical pair of $\vec{A},
\vec{\Pi}$ can be quantized straightforwardly without the appearance of
non-trivial metric tensors \cite{Christ:1980ku}. \\
To put our treatment of Yang-Mills theory in a nutshell: We consider
it to be described as a canonical system, defined in terms of
coordinates $\vec{A}_i^a(\vec{x})$ and conjugate momenta
$\vec{\Pi}^a_i(\vec{x})$ which satisfy ordinary commutation relations:
\be
\ [\vec{A}^a_i(\vec{x}), \vec{\Pi}^b_j(\vec{y})] = i \delta^{ab}
\delta_{ij} \delta_{\vec{x} \vec{y}}.
\ee 
The states in the physical subspace have to satisfy Gauss' law, i.e.
\be
\Gamma^a(\vec{x}) | \psi \rangle
= 0,
\ee
where we have introduced both the Gauss law operator
\be
\Gamma^a(\vec{x}) =  \hat{\vec{D}}^{ab}_{i}(\vec{x}) \vec{\Pi}^b_i(\vec{x}) 
\label{eq_def_Gauss_law_op}
\ee
and the covariant derivative in the adjoint representation
\be
\hat{\vec{D}}^{ab}_{i}(\vec{x}) = \delta^{ab} \nabla_i - (g) f^{acb}
\vec{A}^c_i(\vec{x}). \label{def_cov_deriv_adj}
\ee
We have put the coupling constant in brackets since its appearance
depends on whether we have chosen ``perturbative'' or
``non-perturbative'' scaling, cf. app. \ref{App_scaling}. The $SU(N)$ structure
constants are denoted as $f^{abc}$. One should note that the Gauss law
operator used here may differ from those in other publications by some
proportionality factors, as it is not directly the generator of (small) gauge
transformations. The reason for choosing this form of the Gauss law operator 
is that the only factors of $g$ appearing during a change from perturbative to
non-perturbative scaling appear inside the covariant derivative. \\ 
The wave functional of the reference state taken for Yang-Mills
theory is of Gaussian form 
\bea 
\psi[\vec{A}] &=& \mathcal{N} \exp{\left\{ - \left(\vec{A}^a_i (\vec{x}) -
\bar{\vec{A}}^a_i (\vec{x}) \right) \left[\frac{1}{4}
(G^{-1})^{ab}_{ij}(\vec{x}, 
\vec{y}) - i \Sigma ^{ab}_{ij}(\vec{x}, \vec{y}) \right] \left(\vec{A}^b_j
(\vec{y}) - \bar{\vec{A}}^b_j (\vec{y})\right)\right\}} \nonumber \\
 & & \times \exp{ \left\{ i \bar{\bm \pi}^a_i
(\vec{x}) \left(\vec{A}^a_i (\vec{x}) - \bar{\vec{A}}^a_i
(\vec{x})\right)\right\}}, 
\eea
falling into the class of reference states taken in the main text for
generic bosonic theories. The remaining defintions, like the
Hamiltonian, can be found in the next section, app.\,\ref{App_scaling}.
\subsection{Factors of g\label{App_scaling}}
In Yang-Mills theory one has basically two options concerning where
one wants to put the 
coupling constant, either in front of the action (here called '{\it
non-perturbative scaling}'), or in front of the 
commutator term in the field strength (here called '\textit{perturbative
scaling}'). In table \ref{tab_pert_nonpert_scaling} we give a short  
list concerning which convention leads to which placing of factors of g in
other quantities of interest.
\begin{longtable*}[c]{|l|c|c|}
 \hline \hline  
  &  non-perturbative scaling &  perturbative \ scaling \\ \hline
 \hline  \endhead 
  covariant derivative & $ D_{\mu} = \partial_{\mu} - i
 A_{\mu}  $  & $
 D_{\mu} = \partial_{\mu} - i \mathbf{g} A_{\mu} $ \\ \hline
  field strength &  $ F_{\mu \nu} = \partial_{\mu} A_{\nu} -
 \partial_{\nu} A_{\mu} - i [A_{\mu}, A_{\nu}]  $  &
  $ F_{\mu \nu} = \partial_{\mu} A_{\nu} - \partial_{\nu} A_{\mu} - i
 \mathbf{g}[A_{\mu}, A_{\nu}] $  \\   
 & $ = i[D_{\mu}, D_{\nu}] $ & $ =  \frac{i}{\mathbf{g}} [D_{\mu},
 D_{\nu}] $ \\  
 \hline 
  action & $ S = - \frac{1}{4 \mathbf{g}^2} F^a_{\mu \nu}(x)
 F^{a \, \mu \nu}(x) $ & $ S = - \frac{1}{4} F^a_{\mu
 \nu}(x) F^{a \,  \mu \nu}(x)$ \\ \hline
  electrical field & $ \mathbf{E}^a_{i} = F^a_{0i} $ & $ \mathbf{E}^a_{i} =
 F^a_{0i}$ \\ \hline 
  magnetic field & $  \mathbf{B}^a_{i} = -\frac{1}{2} \epsilon_{ijk} F^{a
 \, jk}  $ & $  \mathbf{B}^a_{i} = -\frac{1}{2} \epsilon_{ijk} F^{a \,
 jk} $ \\  
 & $ = (\nabla \times \mathbf{A})^a_i - \frac{1}{2} f^{abc}(\mathbf{A}^b
 \times \mathbf{A}^c)_i$ & $ = (\nabla \times \mathbf{A})^a_i -
 \frac{\mathbf{g}}{2} 
 f^{abc}(\mathbf{A}^b  \times \mathbf{A}^c)_i$ \\
 \hline
  momenta $\quad \mathbf{\Pi}^a_i = \frac{\partial \mathcal{L}}{\partial
 \dot{\mathbf{A}}^{a}_i} $ & $ \mathbf{\Pi}^a_i =
 \frac{1}{\mathbf{g}^2} F^a_{i0} =  -\frac{1}{\mathbf{g}^2} 
 \mathbf{E}^a_{i} $ & $ \mathbf{\Pi}^a_i =  F^a_{i0} =  -
 \mathbf{E}^a_{i} $ \\ \hline   
  Hamiltonian & $ H = \frac{\mathbf{g}^2}{2}
  \mathbf{\Pi}^a_i(\mathbf{x}) \mathbf{\Pi}^a_i(\mathbf{x}) 
 + \frac{1}{2 \mathbf{g}^2} \mathbf{B}^a_i(\mathbf{x})
 \mathbf{B}^a_i(\mathbf{x}) $ & $ H = 
 \frac{1}{2} \mathbf{\Pi}^a_i(\mathbf{x}) \mathbf{\Pi}^a_i(\mathbf{x})
  + \frac{1}{2} 
 \mathbf{B}^a_i(\mathbf{x}) \mathbf{B}^a_i(\mathbf{x}) $ \\  
 & $ =  \frac{1}{2 \mathbf{g}^2}( \mathbf{E}^a_i(\mathbf{x})
  \mathbf{E}^a_i(\mathbf{x}) + \mathbf{B}^a_i(\mathbf{x})
  \mathbf{B}^a_i(\mathbf{x})) $ & $ = 
 \frac{1}{2 }( \mathbf{E}^a_i(\mathbf{x}) \mathbf{E}^a_i(\mathbf{x}) +
  \mathbf{B}^a_i(\mathbf{x}) \mathbf{B}^a_i(\mathbf{x})) $ \\ 
 \hline
  wave functional of & $ \psi[\mathbf{A}] \sim
 e^{-\frac{1}{\mathbf{g}^2} \mathbf{A} G^{-1} \mathbf{A}}
 $ & $  \psi[\mathbf{A}] \sim e^{- \mathbf{A} G^{-1} \mathbf{A}} $ \\ 
  'free' theory & &  \\ \hline
generators of time-in-& $ [(-\Gamma^a_\mathbf{x}), (-\Gamma^b_\mathbf{y})] = i
 \delta_{\mathbf{x}\mathbf{y}} f^{abc} (-\Gamma^c_\mathbf{x}) \qquad$ & $
 [\frac{1}{\mathbf{g}} (-\Gamma^a_\mathbf{x}), \frac{1}{\mathbf{g}}
 (-\Gamma^b_\mathbf{y})] =
 \frac{\mathbf{g}}{\mathbf{g}} i \delta_{\mathbf{x} \mathbf{y}} f^{abc}
 \frac{1}{\mathbf{g}} (-\Gamma^c_\mathbf{x}) $ \\ 
dependent gauge trafos & & \\ \hline
finite gauge trafos & $  e^{i \int \phi^a \Gamma^a} $ & $  
  e^{i \int \mathbf{g} \phi^a \frac{1}{\mathbf{g}} \Gamma^a} $ 
\\ 
(gluonic part) & &  \\ \hline \hline 
\caption{Placement of the coupling constant $g$ in the
'non-perturbative' and the 'perturbative' scaling
scheme. We have used the shorter form $\Gamma^a_{\mathbf{x}}$ for
$\Gamma^a(\mathbf{x})$. \label{tab_pert_nonpert_scaling}}   
\end{longtable*}
%
%
%
\section{Explicit Expressions \label{app_explicit_expressions}}
\subsection{Matrices A-F \label{app_expl_A-F}}
In this appendix, we want to give explicit expressions that are valid
for the theories that have a Hamiltonian of the form
eq.\,(\ref{gRPAHamiltonian}).  
As we have already mentioned, terms in the Hamiltonian that contain more
than four c/a operators do not contribute to any of the matrix
elements due to the second gRPA approximation\footnote{One can use
Wick's theorem to compute the double commutators that appear in the
definition of the matrices $A,\ldots,F$. Then one realizes
that for those terms of the Hamiltonian that contain more than four c/a
operators the evaluated double commutators still contain at least one c/a
operator. Taking the vacuum expectation values - where according to the second
gRPA approximation we use the mean-field vacuum state - of these expressions
sets these terms then to zero.}. Then the computation is straightforward: we
simply insert 
eq.\,(\ref{H_in_ca_rep}) into eq.\,(\ref{defA-F}), and compute the double
commutators. The computation is simplified by the observation that
only those terms of $H$ contribute to the matrices where the number of
creation operators together with the number of creation operators in
the definition of the matrix elements match the respective number of
annihilation operators. This is the reason for the matrices $A,C,D,E$
having contributions from one term of $H$ only. $B$ obtains
contributions from two terms of $H$ as is to be expected, whereas $F$
obtains only one.
Individually, the matrices read:
\begin{itemize}
\item Matrix A
\be
A_{minj} = - 24 H^{ \{minj\} }_{04}
\ee
\item Matrix B
\be
B_{minj} = B^{11}_{minj} + B^{22}_{minj}
\ee
with
\be
B^{11}_{minj} = (H_{11}^{ji} \delta_{nm} + H_{11}^{ni} \delta_{jm} +
H_{11}^{jm} \delta_{ni} + H_{11}^{nm} \delta_{ji}) \hspace*{1em} ;
\hspace*{1em}  B^{22}_{minj} = 4 H_{22}^{\{nj \} \{mi\} } 
\ee
\item Matrix C
\be
C_{jmi} = -6 H_{03}^{\{jmi\}}
\ee
\item Matrix D
\be
D_{jmi} = 2 H_{12}^{i \{jm\} }
\ee
\item Matrix E
\be
E_{mi} = -2 H_{02}^{ \{mi\} }
\ee
\item Matrix F
\be
F_{mi} = H^{im}_{11},
\ee
\end{itemize}
where $ ^{\{ij\} }$ means: symmetrize in the indices $i,j$ (i.e. add
all permutations and divide by the number of permutations) and 
$H_{ab}^{\ldots}$ means 'that factor in the Hamiltonian that multiplies
$a$ creation and $b$ annihilation operators'.
We see that at the stationary point they simplify considerably:
\bea 
A_{minj} & = & - 4 U_{m m_1} U_{i i_1}  U_{n n_1} U_{j j_1}
\frac{\delta^2}{\delta G_{m_1 i_1} \delta G_{n_1 j_1}}
\langle V \rangle, \label{eq_A_at_stationary_point}\\ 
B_{minj} &=&  \hspace*{1em} B^{11}_{minj} + B^{22}_{minj}, \\ 
B^{11}_{minj} &=& \hspace*{1em} (F_{mn} \delta_{ij} +F_{mj} \delta_{in} + 
F_{in} \delta_{mj} + F_{ij} \delta_{mn}), \\ 
B^{22}_{minj} &=& \hspace*{1em} 4 U_{m m_1} U_{i i_1}  U_{n n_1} U_{j j_1}
\frac{\delta^2}{\delta G_{m_1 i_1} \delta G_{n_1 j_1}}
\langle V \rangle,  \\
C_{jmi} \hspace*{0.45em} & = & - 2 U_{j j_1} U_{m m_1} U_{i i_1}
\frac{\delta^2}{\delta \bar{\phi}_{j_1} \delta G_{m_1 i_1} }
\langle H \rangle,  \\ 
D_{jmi} \hspace*{0.45em} & = & - C_{jmi}, \\
E_{mi} \hspace*{0.9em}   & = & 0,  \\
F_{mi} \hspace*{0.9em} &=& \frac{1}{2}
G^{-1}_{mi}. \label{eq_F_at_stationary_point} 
\eea
\subsection{Gauss Law Operator \label{app_Gauss_law}}
The Gauss law operator was defined in
eq.\,(\ref{eq_def_Gauss_law_op}). Here, we give its decomposition into
creation/annihilation operators. 
After normal ordering, we obtain the result
\bea
\hat{\mathbf{D}}{}^{ab}_{\mathbf{x},i} \mathbf{\Pi}^{b}_{i \mathbf{x}}  = & &
\hat{\bar{\mathbf{D}}}{}^{ab}_{\mathbf{x},i} 
\bar{\bm \pi }^{b}_{i \mathbf{x}} + 2g \tr{(\hat{T}^a
  \Sigma_{\mathbf{x}\mathbf{x}_1} G_{\mathbf{x}_1 \mathbf{x}} )} \nonumber \\  
&+ & \left( \frac{\delta}{\delta \bar{\mathbf{A}}^{a_1}_{l_1 \mathbf{x}_1}}
  \langle \hat{\mathbf{D}}{}^{ab}_{\mathbf{x},i} \mathbf{\Pi}^{b}_{i
    \mathbf{x}} \rangle \right) U^{a_1 b_1,l_1 n_1}_{\mathbf{x}_1 \mathbf{z}_1}
  (a^{\dagger b_1 n_1}_{\mathbf{z}_1} + a^{b_1 n_1}_{\mathbf{z}_1}) 
\nonumber  \\   
&+ & 2 \left( \frac{\delta}{\delta \bar{\bm \pi}^{a_1}_{l_1 \mathbf{x}_1}}
  \langle \hat{\mathbf{D}}{}^{ab}_{\mathbf{x},i} \mathbf{\Pi}^{b}_{i
    \mathbf{x}}  \rangle \right)  
\Big(\frac{i}{4}(U^{-1})^{a_1 b_1,l_1 n_1}_{\mathbf{x}_1
  \mathbf{z}_1}(a^{\dagger b_1 n_1}_{\mathbf{z}_1} -  
a^{b_1 n_1}_{\mathbf{z}_1}) \nonumber \\ && \hspace*{10em} + \Sigma^{a_1 c
_1,l_1 k_1}_{\mathbf{x}_1 \mathbf{y}_1} U^{c_1 b_1,k_1 
  n_1}_{\mathbf{y}_1 \mathbf{z}_1}(a^{\dagger b_1 n_1}_{\mathbf{z}_1} + a^{b_1
  n_1}_{\mathbf{z}_1})\Big) 
\nonumber \\ 
&+ &  \left( \frac{\delta}{\delta G^{a_1 a_2,l_1 l_2}_{\mathbf{x}_1
 \mathbf{x}_2}} \langle 
 \hat{\mathbf{D}}{}^{ab}_{\mathbf{x},i} \mathbf{\Pi}^{b}_{i \mathbf{x}}
 \rangle \right) U^{a_1 b_1, l_1 
  n_1}_{\mathbf{x}_1 \mathbf{z}_1} U^{a_2 b_2, l_2 n_2}_{\mathbf{x}_2
  \mathbf{z}_2} \nonumber \\ & & 
\hspace*{10em} \times (a^{\dagger b_1 n_1}_{\mathbf{z}_1}  
a^{\dagger b_2 n_2}_{\mathbf{z}_2}  + a^{b_1 n_1}_{\mathbf{z}_1} a^{b_2
 n_2}_{\mathbf{z}_2} +2 
a^{\dagger b_1 n_1}_{\mathbf{z}_1} a^{b_2 n_2}_{\mathbf{z}_2}) \nonumber \\
&+ &  \left( \frac{\delta}{\delta \Sigma^{a_1 a_2,l_1 l_2}_{\mathbf{
 x}_1 \mathbf{x}_2}} 
  \langle \hat{\mathbf{D}}{}^{ab}_{\mathbf{x},i} \mathbf{\Pi}^{b}_{i
    \mathbf{x}} \rangle \right) \frac{i}{4} (U^{-1})^{a_1  
  b_1, l_1 n_1}_{\mathbf{x}_1 \mathbf{z}_1} (U^{-1})^{a_2 b_2, l_2
 n_2}_{\mathbf{x}_2 \mathbf{z}_2} (a^{\dagger 
  b_1 n_1}_{\mathbf{z}_1} a^{\dagger b_2 n_2}_{\mathbf{z}_2} - a^{b_1
 n_1}_{\mathbf{z}_1} a^{b_2 n_2}_{\mathbf{z}_2}) \nonumber \\
&+ & g \frac{i}{2} (\hat{T}^a)^{a_1 a_2} (U^{a_1 b_1, l n_1}_{\mathbf{x}
  \mathbf{z}_1} (U^{-1})^{a_2 b_2, l  n_2}_{\mathbf{x} \mathbf{z}_2} +
(U^{-1})^{a_1  b_1, l  n_1}_{\mathbf{x} \mathbf{z}_1} U^{a_2 b_2, l 
  n_2}_{\mathbf{x} \mathbf{z}_2}) a^{\dagger b_1 n_1}_{\mathbf{z}_1} a^{b_2
 n_2}_{\mathbf{z}_2}.  \label{eq_Gauss_law_c-a-decomp}
\eea
Here we have employed the notation
\be
(\hat{T}^a)^{a_1 a_2} = f^{a_1 a a_2}
\ee
where $f^{a_1 a a_2}$ denote the $SU(N)$ structure constants.
%
%
%
\section{Commutation relations of normal modes \label{gRPA_comm_rel}}
In this appendix we want to consider the question 'what are the
conditions under which the eigenmodes $Q_{B\, \nu}^{\dagger}$ can be treated as
harmonic oscillators ?' We start with the observation that
eq.\,(\ref{RPA-SHO}) implies by hermitian conjugation
\be
\ [ H_B, Q_{B\, \nu} ] = -\Omega_{\nu} Q_{B\, \nu}. \label{RPA-SHOadj}
\ee
Thus, to every eigenfrequency, the negative eigenfrequency also belongs
to the spectrum. Thus, without loss of generality\footnote{We have seen
in sec.\,\ref{pqformulationofRPA} that the {\it assumption} that the 
eigenenergies of the modes are {\it real} already implies that we are
dealing with a stable mean-field solution.}
in the following we will assume that
\be
\Omega_{\nu} \ge 0,
\ee
otherwise we just exchange the respective $Q_{B\, \nu},
Q_{B\, \nu}^{\dagger}$. 
The next point is that all the commutators 
\be
\ [Q^{\dagger}_{B\, \nu}, Q^{\dagger}_{B\, \mu}] \hspace*{1em} [Q_{B\, \nu},
Q^{\dagger}_{B\, \mu}] \hspace*{1em} [Q_{B\, \nu}, Q_{B\, \mu}]  
\ee
are pure numbers. This is due to the fact that all operators
considered in this context are by construction linear in $\BB, \BBd,
a, a^{\dagger}$.  
We denote these numbers as
\be
\ [Q^{\dagger}_{B\, \mu}, Q^{\dagger}_{B\, \nu}] = M_{\mu \nu}; \hspace*{1em}
[Q_{B\, \mu}, Q^{\dagger}_{B\, \nu}] = N_{\mu \nu}; \hspace*{1em} 
[Q_{B\, \mu}, Q_{B\, \nu}] = O_{\mu \nu}.
\ee
Up to now these numbers are arbitrary; one can, however, put the
equations of motion to some good use. Consider
\bea
\Omega_{\nu} [Q_{B\, \mu}, Q^{\dagger}_{B\, \nu}] &
\stackrel{eq.\,(\ref{RPA-SHO})}{=} 
&  [Q_{B\, \mu}, [H_B, Q^{\dagger}_{B\, \nu}]] \\ 
& \stackrel{Jacobi~id.}{=} & - [H_{B}, [ Q^{\dagger}_{B\, \nu}, Q_{B\, \mu}]] -
[Q^{\dagger}_{B\, \nu}, [Q_{B\, \mu}, H_B]] \\
&=& \underbrace{[H_B, N_{\mu \nu}]}_{=0} + [Q^{\dagger}_{B\, \nu},
\underbrace{[H_B, Q_{B\, \mu}]}_{= - \Omega_{\mu} Q_{B\, \mu}}] \\
 & = & \Omega_{\mu} [Q_{B\, \mu}, Q^{\dagger}_{B\, \nu}].
\eea
In other words:
\be
(\Omega_{\nu} - \Omega_{\mu}) N_{\mu \nu} = 0 \ \ ({\rm no ~sum ~over
~} \mu, \nu). \label{Ndet}
\ee
If {\it all eigenvalues are distinct and non-zero} then
eq.\,(\ref{Ndet}) implies that $N_{\mu \nu}$ is diagonal. Since the
gRPA
equations are homogeneous equations, we may now normalize the
amplitudes in such a way that\footnote{\label{CR_footnote}In
\cite{Ring:1980} it is stated
that one can show that $N_{\mu \nu}$ is positive if one is considering
a positive definite Hessian at the stationary point of the mean-field
problem; a possible proof of this statement  works as follows: only if the
Hessian is a positive matrix it is guaranteed that all eigenvalues
$\Omega^2$ are positive. This has been assumed so far (e.g. how we
concluded that for every positive frequency there is also a negative
one etc.); thus we know that the system is stable. Knowing this, we can
argue as follows: we know that $[H, Q_{B\, \nu}^{\dagger} ] = \Omega_{\nu}
Q_{B\, \nu}^{\dagger}, [H, Q_{B\, \nu} ] = -\Omega_{\nu} Q_{B\, \nu}$ and
$[Q_{B\, \nu}, Q^{\dagger}_{B\, \nu'}] = {\cal N}_{\nu} \delta_{\nu
\nu'}$. From this we conclude that the Hamiltonian has to look like
$ H = \sum_{\nu} (\Omega_{\nu}/{\cal N}_{\nu}) Q^{\dagger}_{B\, \nu} Q_{B\,
  \nu}$. 
Now we can study two different scenarios, namely ${\cal N}_{\nu}$ can
be positive (we call the set of $\nu$ for which this is true $\nu_+$)
or negative (correspondingly $\nu_{-}$). In order to have the usual
creation/annihilation commutation relations we have - for $\nu \in
\nu_{-}$ - to interchange 'creation' and 'annihilation' operators; for
clarity, we introduce new letters for them, i.e. for $\nu \in \nu_{-},
Q_{B\, \nu} \rightarrow P^{\dagger}_{B\, \nu}, Q^{\dagger}_{B\, \nu}
\rightarrow P_{B\, \nu}$. Now we normalize $Q_B$s and $P_B$s s.t. for
$\nu \in \nu_+, {\cal N}_{\nu} = 1$ and for  $\nu \in \nu_{-}, {\cal
N}_{\nu} = -1$. The Hamiltonian then reads $ H = 
 \sum_{\nu \in \nu_{+}} \Omega_{\nu} Q^{\dagger}_{B\, \nu} Q_{B\, \nu} +
 \sum_{\nu \in \nu_{-}} (-\Omega_{\nu}) P^{\dagger}_{B\, \nu} P_{B\, \nu} +
{\rm const}$. But here we see that we are dealing with a rather unstable
system: the more modes are generated by $P^{\dagger}_B$, the lower the
energy becomes. This cannot be true, however, since we know that we
started from a stable system (with all frequencies real) ! Thus
$\nu_{-}$ has to be empty.}${}^,$\footnote{This relation has also as
its consequence that the states generated by $Q^{\dagger}_{B\, \mu},
Q^{\dagger}_{B\, \nu}$ from the gRPA vacuum (defined to be annihilated by
all gRPA annihilation operators, $Q_{B\, \nu} | RPA \rangle = 0 \forall
\nu$) 
are indeed orthogonal: $\langle \mbox{RPA} | Q_{B\, \mu} Q^{\dagger}_{B\, \nu}
| \mbox{RPA} \rangle = \langle \mbox{RPA} | [Q_{B\, \mu}, Q^{\dagger}_{B\, \nu}]
| \mbox{RPA} \rangle = \langle
 | [Q_{B\, \mu}, Q^{\dagger}_{B\, \nu}] | \rangle = \delta_{\mu \nu}$ where we
have used the second gRPA approximation in the next to last
step\,\cite{Ring:1980}. } 
\be
N_{\mu \nu} = \delta_{\mu \nu}.
\ee
The same procedure can be carried out for $M_{\mu \nu}, O_{\mu
\nu}$. We derive in analogy to eq.\,(\ref{Ndet}):
\be
(\Omega_{\nu} + \Omega_{\mu}) M_{\mu \nu} = 0, \hspace*{2em}
(\Omega_{\nu} + \Omega_{\mu}) O_{\mu \nu} = 0, 
\ee
where again no sum over double indices is performed. From this we
conclude, again if there are no zero modes, that 
\be
M_{\mu \nu} = 0, \hspace*{2em} O_{\mu \nu} = 0.
\ee
Thus, under the aforementioned conditions, the {\it normal modes
satisfy ordinary c/a commutation relations}. To arrive at this, we have indeed
used both gRPA approximations, since the basic ingredient was that
$Q^{\dagger}$ is a one-particle operator (first gRPA approximation) and
that all the commutators of $a, a^{\dagger}, \BB, \BBd$ are pure
numbers (quasi-boson approximation). \\
%
%
%
\section{General Potential\label{app_pot}}
In this appendix we will demonstrate that a potential that is an
arbitrary polynomial in the field operators, can - once one decomposes
the field operator into creation and annihilation operators - be
written in normal ordered form s.t. all the coefficients appearing in
front of the normal ordered products of creation/annihilation
operators can be written as functional derivatives w.r.t. G and
$\bar{\phi}$ of the vacuum expectation value of the potential (the
creation/annihilation operators are defined with respect to that
vacuum). \\ For simplicity we take the potential of the form
\be
V[\phi] = {\cal M}_{x_1 \ldots x_n} \phi_{x_1} \cdots
\phi_{x_n}, \label{Vpoly1}  
\ee
where $x_i$ are super-indices.
It is clear that, if the claim holds for this potential, it will
hold for an arbitrary polynomial since it will be a sum of terms of
type (\ref{Vpoly1}).\\ Since the field operators commute, it is
sufficient to consider an ${\cal M}_{x_1 \ldots x_n}$ that is symmetric
in all indices. We have seen in section \ref{sec_on_creat_annihil_op}
that one can write 
\be
\phi_x = \bar{\phi}_x + U_{xy}(b^{\dagger}_{y} + b_y),
\ee
where $x,y$ are super-indices, $U^2 = G$, and $[b_x, b^{\dagger}_y] =
\delta_{xy}$. For our purposes, it will be more useful to define
rescaled operators
\be
a^{\dagger}_x = U_{xy} b^{\dagger}_y \hspace*{2em}; \hspace*{2em}
a_x = U_{xy} b_y \hspace*{1em} \mbox{with } \hspace*{0.5em} [a_x,
a^{\dagger}_y] = G_{xy}. 
\ee
Sometimes we find it also useful to write
\be
\phi_x = \bar{\phi}_x + \varphi_x \mbox{ with } \varphi_x =
a^{\dagger}_x + a_x.
\ee
After all this notational introduction, let's come to the proof. We
note that, similar to $\phi$, both $\bar{\phi}$ and $\varphi$
commute; thus we can write equally\footnote{It is understood,
obviously, that the index of $x$ increases from left to right; if
it should ever decrease, as in the case $m=0$, the $\bar{\phi}$s are
to be considered absent.} instead of eq.\,(\ref{Vpoly1}):
\be V[\phi] =
{\cal M}_{x_1 \ldots x_n} \sum_{m=0}^{n} \left(\begin{array}{c}
\displaystyle{n} \\ \displaystyle{m} 
\end{array} \right) \bar{\phi}_{x_1} \cdots \bar{\phi}_{x_m}
\varphi_{x_{m+1}} \cdots \varphi_{x_n}. \label{Vpoly2}
\ee
Now we normal order the terms with a fixed $m$:
\bea
& & \varphi_{x_{m+1}} \cdots \varphi_{x_n} \nonumber \\   &= &(a_{x_{m+1}} +
a^{\dagger}_{x_{m+1}}) \cdots (a_{x_n} + a^{\dagger}_{x_n}) \nonumber \\
  & = & a^{\dagger}_{x_{m+1}}a^{\dagger}_{x_{m+2}} \cdots
a^{\dagger}_{x_n} \nonumber \\
& + &  a_{x_{m+1}} a^{\dagger}_{x_{m+2}}   a^{\dagger}_{x_{m+3}} \cdots
a^{\dagger}_{x_n} 
+ \cdots +
a^{\dagger}_{x_{m+1}}a^{\dagger}_{x_{m+2}}   a^{\dagger}_{x_{m+3}}
\cdots a_{x_n} \nonumber \\
& + &  a_{x_{m+1}} a_{x_{m+2}}   a^{\dagger}_{x_{m+3}} \cdots
a^{\dagger}_{x_n} 
+ \cdots +
a^{\dagger}_{x_{m+1}}a^{\dagger}_{x_{m+2}}   a^{\dagger}_{x_{m+3}}
\cdots a_{x_n} \nonumber \\
& \vdots & \nonumber \\
& + &  a_{x_{m+1}} a_{x_{m+2}} \cdots a_{x_n}. 
\eea
We see that every row contains a fixed number of creation/annihilation
operators and that it contains all possible permutations of types
(creation or annihilation) among the possible indices. Especially for
every term 
\be
\cdots a^{\dagger}_{x_p} \cdots a_{x_q} \cdots \label{partner_0}
\ee
there also exists a term 
\be
\cdots a_{x_p} \cdots a^{\dagger}_{x_q} \cdots \label{partner_1}
\ee
with all undenoted operators identical. We now use Wick's theorem,
cf. e.g. \cite{Ring:1980}, to
put every line into normal order. It is practical to deal with the
whole line for the following reason: if we use Wick's theorem naively
we obtain the normal ordered expression plus the normal ordered
expression of two less operators times their contraction etc.\,. However,
a lot of these contractions are zero, since (we denote the contraction
of two operators by ${\cal C}$)
\be
{\cal C}(a_i^{\dagger} a_j) = 0.
\ee
However, if we deal with the complete line at once, we can use that,
since for every arrangement eq.\,(\ref{partner_0}) there also exists a
partner eq.\,(\ref{partner_1}), we will always obtain contractions
\be
{\cal C}(a_i^{\dagger} a_j + a_i a^{\dagger}_j) = G_{ij}.
\ee
It is clear that for $k$ contractions present we need $2^k$ partners
to obtain a non-vanishing contribution. The important point is that
they exist if we deal with the whole line at once. Remember that the
contractions are multiplied by normal ordered terms, and that $[a, a] =
[a^{\dagger}, a^{\dagger}] = 0$. Since in addition ${\cal M}_{x_1
\ldots x_n}$ is symmetric in all its indices, all terms with a fixed
number of contractions, creation, and annihilation operators, will give
an identical contribution. Thus the question appears: assume we start
out from the line where every term contains p creation operators, q
annihilation operators and consider now terms with k contractions; how
many terms will we obtain ? The answer is simple,
\be
\mbox{number of terms} = \frac{1}{2^k} \times \frac{(p+q)!}{p! q!}
\times \left( \frac{1}{k !} \frac{p !}{(p-k)!} \frac{q !}{(q-k)!} 
 \right), \label{no_of_perm1}
\ee
and comes about as follows:\footnote{In the following, we abbreviate
'creation operators' as c's and 'annihilation operators' as a's.}
\begin{enumerate}
\item the line containing p c's and q a's contains $\frac{(p+q)!}{p!
q!}$ terms; 
\item out of the p c's and q a's we take k each, and put them together
to form contractions: there are obviously $\frac{p !}{(p-k)!} \frac{q
!}{(q-k)!}$ ways to do this.
\item However, it does not matter in which order we perform the
contractions since the contractions commute; thus in the step before
we have overcounted by a factor of $k!$.
\item We now have to take into account what was said above: On the
one hand, a lot of contractions are zero. On the other hand we can
form pairs - this gives as argued above an additional factor of $2^{-k}$.
\end{enumerate}
With this formula at hand, we can at first answer the following
important question: assume that we have started from an expression
with $n$ field operators; upon normal ordering we obtain expressions with
$n,n-2,\ldots,n-2k$ c/a operators each; the question now is: is the
{\it relative} number of $P$ c's and $Q$ a's (with $P+Q$ fixed) always the
same, no matter from which n one starts and how many contractions one
needs\footnote{As an example: no matter where one starts - if one can
get to the expression with, in total, two c's and a's, will they always
come as $a a + a^{\dagger} a^{\dagger} + 2 a^{\dagger} a$ ?} (provided
$n-(P+Q)$ is even) ? This
question can be answered in the affirmative in the following way:
We start out with an expression that contains $p$ c's and $q$ a's; after k
contractions we will end up with $P=p-k$ c's and $Q=q-k$ a's. Since in
the beginning $p+q=n$ was fixed, and we end up with $P+Q=n'$ fixed we
need for every term the same number of contractions, namely $2k=
n-n'$. With this we can rewrite eq.\,(\ref{no_of_perm1}) as
\be
\frac{n!}{2^k k!} \frac{1}{P ! (n' - P)!} = \frac{n!}{2^k k!}
\frac{1}{P ! Q!}. 
\ee
Thus, we have decomposed the number of terms into a factor that depends on
$n,n'$ which is a constant for $P+Q=n'$ fixed and $P$ varying, and a factor
that depends on the number of creation and annihilation operators. If we go
back to eq.\,(\ref{Vpoly2}) we 
see that we have different possibilities to end up with $P$ c's and
$Q$ a's ($P+Q=n'$): either start from $n$ $\varphi$s, and perform $k$
contractions, or start from two $\bar{\phi}$s, $(n-2) \varphi$s and
perform $k-1$ contractions etc. Thus the contribution to ${\cal
M}_{x_1 \ldots x_n} \phi_{x_1} \cdots \phi_{x_n}$ containing $n'$ c's
and a's can be written as
\footnote{Note that the factors of the
contractions and of the binomial decomposition of eq.\,(\ref{Vpoly2})
can be put together in a practical manner: $ \frac{(n'+2k)!}{2^k k!}
 \binom{n}{n+1 - (n'+2k+1)} = \frac{n!}{2^k k! (n-n'-2k)!}$. Note
 also that we have arranged here $\varphi, \bar{\phi}$ opposite to
 eq.\,(\ref{Vpoly2}).} 
\bea
& & {\cal M}_{x_1 \ldots x_n} \left[\sum_{P=0}^{n'} \frac{1}{P! (n'-P)!} a^{\dagger}_{x_1} \cdots
a^{\dagger}_{x_P} a_{x_{P+1}} \cdots a_{x_{n'}} \right] \nonumber \\ &
\times & \sum_{k=0}^{[\frac{1}{2} (n-n')]} \Big( \frac{n!}{2^k k! (n-n'
- 2k)!}
G_{x_{n'+1} x_{n'+2}} \cdots \times \nonumber \\ & & \hspace*{2em}
\times G_{x_{n'+(2k-1)} x_{n'+2k}} 
\bar{\phi}_{x_{n'+2k+1}} \cdots \bar{\phi}_{x_n}
\Big), \label{addendP+Q}
\eea
where the latter sum runs to $\frac{1}{2}(n-n')$ if $n-n'$ is even,
and to $\frac{1}{2}(n-n'-1)$ if it is odd - we will deal with these
details below.
This expression allows us to write down the vacuum expectation value (VEV)
of $V[\phi]$ since it corresponds to the case $n'=0$. Distinguish
\begin{itemize}
\item $n=2N$: the VEV reads
\bea
\langle V[\phi] \rangle = 
{\cal M}_{x_1 \ldots x_{2N}}
 \sum_{k=0}^{N} \Big( && \hspace*{-1em} \frac{n!}{2^k k! (n-2k)!}
G_{x_{1} x_{2}}  \cdots \label{VEVeven} \\ &&
 G_{x_{(2k-1)} x_{2k}} \bar{\phi}_{x_{2k+1}} \cdots \bar{\phi}_{x_{2N}}
\Big) \nonumber
\eea
\item $n=2N+1$: the VEV reads
\bea
\langle V[\phi] \rangle = 
{\cal M}_{x_1 \ldots x_{2N+1}} \sum_{k=0}^{N} \Big( && \hspace*{-1em}
\frac{n!}{2^k k! (n-2k)!} 
G_{x_{1} x_{2}} \cdots \label{VEVodd} \\ && \hspace*{-2em}
G_{x_{(2k-1)} x_{2k}} 
\bar{\phi}_{x_{2k+1}} \cdots \bar{\phi}_{x_{2N}} \Big) \bar{\phi}_{x_{2N+1}}.
\nonumber
\eea
\end{itemize}
For the following treatment, we can treat both cases with the same
formula if we realize that ${\cal M}_{x_1 \ldots x_{2N+1}}
\bar{\phi}_{x_{2N+1}}$ has the same properties as ${\cal M}_{x_1
\ldots x_{2N}}$ and that in the sum in eq.\,(\ref{VEVodd}) always at
least one $\bar{\phi}$ survives. Thus we only have to treat the case
with $n=2N$. We now consider 
\bea
 & & \frac{\delta}{\delta G_{y_1 y_2}} \langle V[\phi]
\rangle \nonumber \\
& = & \frac{\delta}{\delta G_{y_1 y_2}} 
{\cal M}_{x_1 \ldots x_{2N}}
 \sum_{k=0}^{N} \left( \left. \frac{n!}{2^k k! (n-n'-2k)!} \right|_{n'=0}
G_{x_{1} x_{2}} \cdots G_{x_{(2k-1)} x_{2k}}
\bar{\phi}_{x_{2k+1}} \cdots \bar{\phi}_{x_{2N}}
\right) \nonumber \\
& \stackrel{(*)}{=} & {\cal M}_{y_1 y_2 x_3 \ldots x_{2N}}
 \sum_{k=1}^{N} \left( \left. \frac{n!}{2^k k! (n-n'-2k)!} \right|_{n'=0}
k ~G_{x_{3} x_{4}} \cdots G_{x_{(2k-1)} x_{2k}}
\bar{\phi}_{x_{2k+1}} \cdots \bar{\phi}_{x_{2N}}
\right) \nonumber \\
&=& \frac{1}{2^{N'}}{\cal M}_{y_1 y_2 x_{n'+1} \ldots x_{2N}}
 \nonumber \\ & & \hspace*{2em} \times
 \sum_{k=0}^{N-N'} \left( \left. \frac{n!}{2^k k! (n-n'-2k)!} 
G_{x_{n'+1} x_{n'+2}} \cdots G_{x_{(2k-1)} x_{2k}}
\bar{\phi}_{x_{2k+1}} \cdots \bar{\phi}_{x_{2N}} \right|_{n'=2 = 2N'}
\right). \nonumber \\
\eea
In (*) we have used the symmetry of ${\cal M}$.
Obviously, as has been indicated by the suggestive notation, one is
not restricted to one functional derivative but one can also perform
$N'$ of them, and then the restriction $N'=1$ in the last line is
rendered unnecessary. We see clearly that, apart from the factor
$2^{-N'}$ the outcome of $N'$ derivatives of the vacuum expectation
value w.r.t. $G$ is identical to the prefactor of the addend
containing $P+Q= n'=2N'$ c's and a's in eq.\,(\ref{addendP+Q}). Thus we
can rewrite eq.\,(\ref{addendP+Q}) as
\bea
2^{N'} \left[\sum_{P=0}^{n'} \frac{1}{P! (n'-P)!} a^{\dagger}_{y_1} \cdots
a^{\dagger}_{y_P} a_{y_{P+1}} \cdots a_{y_{n'}} \right]
\frac{\delta}{\delta G_{y_1 y_2}} \cdots \frac{\delta}{\delta
G_{y_{n'-1} y_{n'}}} \langle V[\phi] \rangle.  \label{addendP+Qder}
\eea
The treatment we have presented up to here is valid for $P+Q$ even. If
$P+Q$ is odd, we have to perform derivatives w.r.t. $\bar{\phi}$ and thus we
have to treat the $n$ even/odd cases individually. Let's start with n
even:
\bea
& & \frac{\delta}{\delta \bar{\phi}_{y_1}} \langle V[\phi]
\rangle \nonumber \\ & = & 
\frac{\delta}{\delta \bar{\phi}_{y_1}}
{\cal M}_{x_1 \ldots x_{n-1} x_n}
 \sum_{k=0}^{N} \left( \left. \frac{n!}{2^k k! (n-n'-2k)!} \right|_{n'=0}
G_{x_{1} x_{2}} \cdots G_{x_{(2k-1)} x_{2k}}
\bar{\phi}_{x_{2k+1}} \cdots \bar{\phi}_{x_{n-1}} \bar{\phi}_{x_n}
\right) \nonumber \\
&\stackrel{(*)}{=}& {\cal M}_{x_1 \ldots x_{n-1} y_1}
 \sum_{k=0}^{N-1} \left( \left. \frac{n!}{2^k k! (n-n'-2k)!} \right|_{n'=0}
G_{x_{1} x_{2}} \cdots G_{x_{(2k-1)} x_{2k}}
\bar{\phi}_{x_{2k+1}} \cdots \bar{\phi}_{x_{n-1}} (n-2k)
\right) \nonumber \\
&=& {\cal M}_{x_1 \ldots x_{n-1} y_1}
 \sum_{k=0}^{N-1} \left( \left. \frac{n!}{2^k k! (n-n'-2k)!} \right|_{n'=1}
G_{x_{1} x_{2}} \cdots G_{x_{(2k-1)} x_{2k}}
\bar{\phi}_{x_{2k+1}} \cdots \bar{\phi}_{x_{n-1}} 
\right) \nonumber \\
&=& {\cal M}_{y_1 x_2 \ldots x_{n}}
 \sum_{k=0}^{N-1} \left( \left. \frac{n!}{2^k k! (n-n'-2k)!}
G_{x_{n'+1} x_{n'+2}} \cdots G_{x_{n'+(2k-1)} x_{n'+2k}}
\bar{\phi}_{x_{n'+2k+1}} \cdots \bar{\phi}_{x_{n}} \right|_{n'=1}
\right). \nonumber \\
\eea
The main point happened in line $(*)$ where the fact that we started
from even n played a role. If n is even, $k=N$ means that this addend
doesn't contain a single factor of $\bar{\phi}$, thus its derivative
vanishes. This is different in case of n odd, there the upper boundary
is not affected by the first differentiation. It is different if one
performs two derivatives w.r.t. $\bar{\phi}$ since then the upper limit
of the sum changes once altogether independent of whether one starts
from n even or odd. Thus one obtains a relation
between derivatives w.r.t. $G$ and to $\bar{\phi}$
\be
\frac{1}{2} \frac{\delta^2}{\delta \bar{\phi}_x \delta \bar{\phi}_y}
\langle V[\phi] \rangle = \frac{\delta}{\delta G_{xy}} \langle V[\phi] \rangle,
\label{trade_phiphi_forG}
\ee
which will be very useful in proving the equivalence between
the generalized RPA using the operator approach and
generalized RPA as derived from the time-dependent variational principle. \\ 
To put this appendix in a nutshell, we have proved that a potential
that is an arbitrary polynomial in the field operators can be
decomposed into creation and annihilation operators, s.t. upon normal
ordering one obtains a sum of subsums where each subsum contains a
fixed number of c's and a's. The subsum consisting of 
the terms containing $n'$ c's and a's can be written
as a standard polynomial in c's and a's
\be
\left[\sum_{P=0}^{n'} \frac{1}{P! (n'-P)!} a^{\dagger}_{x_1} \cdots
a^{\dagger}_{x_P} a_{x_{P+1}} \cdots a_{x_{n'}} \right]
\ee
multiplied by
\begin{itemize}
\item if $n'$ is even, $n'/2$ derivatives w.r.t. $G$ times a factor
$2^{n'/2}$ 
\item if $n'$ is odd, one derivative w.r.t. $\bar{\phi}$ and
$(n'-1)/2$ derivatives w.r.t. $G$ times a factor $2^{(n'-1)/2}$.
\end{itemize}
We have also shown that each derivative w.r.t. $G$ may be traded
for two derivatives w.r.t. $\bar{\phi}$.
%
%
%
%
%
%
%
%
%
%
%
%

\end{document}